\documentclass[
aps,
showpacs,
nofootinbib,
superscriptaddress,
floatfix
]{revtex4}

\usepackage{amsmath}
\usepackage{amssymb} 

\usepackage{graphicx}

\usepackage[utf8]{inputenc} 

\usepackage{enumerate}
\usepackage{enumitem}

\usepackage[colorlinks=true, urlcolor=blue, linkcolor=blue, citecolor=blue]{hyperref}

\usepackage{definition_AT}
\usepackage{definition_CL}
\begin{document}

\title{Revisiting the mechanical properties of the nucleon}

\author{C\'edric Lorc\'e}
\affiliation{Centre de Physique Th\'eorique, \'Ecole polytechnique, 
	CNRS, Universit\'e Paris-Saclay, F-91128 Palaiseau, France}

\author{Herv\'e Moutarde}
\affiliation{IRFU, CEA, Universit\'e Paris-Saclay, F-91191 Gif-sur-Yvette, France}

\author{Arkadiusz P. Trawi\'nski}
\affiliation{Centre de Physique Th\'eorique, \'Ecole polytechnique, 
	CNRS, Universit\'e Paris-Saclay, F-91128 Palaiseau, France}
\affiliation{IRFU, CEA, Universit\'e Paris-Saclay, F-91191 Gif-sur-Yvette, France}


\begin{abstract}
We discuss in detail the distributions of energy, radial pressure and tangential pressure inside the nucleon. In particular, this discussion is carried on in both the instant form and the front form of dynamics. Moreover we show for the first time how these mechanical concepts can be defined when the average nucleon momentum does not vanish. We express the conditions of hydrostatic equilibrium and stability in terms of these two and three-dimensional energy and pressure distributions. We briefly discuss the phenomenological relevance of our findings with a simple yet realistic model. In the light of this exhaustive mechanical description of the nucleon, we also present several possible connections between hadronic physics and compact stars, like e.g. the study of the equation of state for matter under extreme conditions and stability constraints.
\end{abstract}

\pacs{11.30.Cp,13.88.+e,
Published in \href{https://doi.org/10.1140/epjc/s10052-019-6572-3}{Eur.Phys.J. C79 (2019) no.1, 89}
}
\maketitle

\section{Introduction}

Understanding how quarks and gluons bind together to form nucleons is a challenging and formidable open problem. Since a couple of decades, one of the main focuses of hadronic physics consists in the study of the mass and spin structure of the nucleon. This information is encoded in the energy-momentum tensor (EMT) which can be probed in various high-energy exclusive scattering experiments~\cite{Ji:1996ek,Ji:1996nm}. Since the corresponding cross-sections are small, these processes are studied in high-luminosity setups like Jefferson Laboratory or COMPASS at CERN, and constitute an important aspect of the experimental program to be conducted in a near future in a U.S.-based Electron-Ion Collider~\cite{Accardi:2012qut}. The nucleon EMT is also intensively studied using Lattice QCD, see~\cite{Yang:2018bft,Alexandrou:2018xnp,Yang:2018nqn,Shanahan:2018pib,Shanahan:2018nnv} for recent developments.

Beside mass and spin, the EMT is also a fundamental object encoding mechanical properties of the nucleon like stress~\cite{Pais:1949vdk, Polyakov:2002yz, Polyakov:2018zvc}. Unlike ordinary media at equilibrium, the stress inside the nucleon is not isotropic. Indeed, some theoretical investigations~\cite{Ruderman:1972aj, Canuto:1974ft} already showed that the nuclear matter itself may become anisotropic at very high densities ($>10^{15}$ g/cm$^3$), where the nuclear interactions must be treated relativistically. Such conditions are typically met inside compact stars which cannot be explained properly in terms of an ordinary equation of state (EoS)~\cite{Alcock:1986hz, Haensel:1986qb, Weber:2004kj, PerezGarcia:2010ap, Rodrigues:2011zza}. Stress anisotropy in self-gravitating systems has been studied in~\cite{Bowers:1974tgi, Herrera:1997plx, Mak:2001eb} and has been shown to affect the physical properties, stability and structure of stellar matter~\cite{Dev:2000gt, Dev:2003qd, Silva:2014fca}. In particular, anisotropic stellar objects can be much more compact than the isotropic ones~\cite{Boehmer:2006ye}. In the chiral quark-soliton model~\cite{Goeke:2007fp}, it has been found that the energy density at the center of the nucleon is about $3\times10^{15}$ g/cm$^3$, i.e. 13 times higher than the average density of nuclear matter. It is therefore no wonder that stress anisotropy plays a significant role in the mechanical structure of the nucleon.

So far, the mechanical properties of the nucleon have been studied in the Breit frame based on the symmetric Belinfante-Rosenfeld form of the EMT~\cite{Polyakov:2002yz, Goeke:2007fp}. It is known however that distributions defined in the Breit frame are subject to relativistic corrections~\cite{Burkardt:2002hr} and that the spin of the constituents makes the nucleon EMT asymmetric~\cite{Leader:2013jra, Lorce:2017wkb}. In view of this, we propose a detailed revisit of the mechanical structure of the nucleon addressing the above shortcomings. 
\newline

The paper is organized as follows. In Section~\ref{sec:2} we explain how to describe a relativistic quantum system localized in phase space and present the matrix elements of the general asymmetric EMT. Mechanical properties of the nucleon defined in the instant form of dynamics are discussed in Section~\ref{sec:3}. After a quick review of the proper decomposition of the nucleon EMT into quark and gluon contributions, we present the three-dimensional spatial distributions defined in the Breit frame where the system is in average at rest. We extend the study to the more general class of elastic frames introduced in~\cite{Lorce:2017wkb}, where two-dimensional spatial distributions can be defined in the plane transverse to the average motion of the system. In Section~\ref{sec:4}, we discuss for the first time the mechanical structure of the nucleon using the front form of dynamics. In such a framework, we define two-dimensional spatial distributions free of relativistic corrections and compare them with the corresponding distributions in instant form. The questions of hydrostatic equilibrium and stability conditions are discussed in Section~\ref{sec:5}. Finally, we conclude the paper with Section~\ref{sec:6} where we summarize our results.

\section{Matrix elements of the energy-momentum tensor}\label{sec:2}

In order to define the distribution of a physical quantity inside a system, one should first localize the system in both position and momentum space. In a quantum theory, this can be achieved in the Wigner sense~\cite{Hillery:1983ms}. A system with average position $R$ and average momentum $P$ is described by the covariant phase-space density operator
\begin{equation}\label{eq:RPcov}
\rho_{R,P}=
\int\frac{\ud P^2}{2\pi}\int\frac{\ud^4\Delta}{(2\pi)^4}\,2\pi\,\delta(2P\cdot\Delta)\,2\pi\,\delta(P^2+\tfrac{\Delta^2}{4}-M^2)\,e^{-i\Delta\cdot R}\,\ket{P-\tfrac{\Delta}{2}}\bra{P+\tfrac{\Delta}{2}}\,,
\end{equation}
where the delta functions\footnote{For notational convenience, we omit to write the theta functions which select the positive mass shells.} ensure that initial and final states have the same mass $M$. 
It follows from the relativistic normalization for momentum eigenstates $\langle p'|p\rangle=2p^0(2\pi)^3\,\delta^{(3)}(\uvec p'-\uvec p)$ that $\uTr[\rho_{R,P}]=1$. Defining the covariant ``position'' states as
\begin{equation}
\ket{x}=\int\frac{\ud^4p}{(2\pi)^4}\,2\pi\,\delta(p^2-M^2)\,e^{ip\cdot x}\,\ket{p}\,,
\end{equation} 
with normalization $\langle x'|x\rangle=\int\frac{\ud^3p}{(2\pi)^32p^0}\,e^{-ip\cdot(x'-x)}$, the covariant phase-space density operator can alternatively be expressed as
\begin{equation}
\rho_{R,P}=\int\frac{\ud P^2}{2\pi}\int\ud^4Z\,e^{-iP\cdot Z}\,\ket{R+\tfrac{Z}{2}}\bra{R-\tfrac{Z}{2}}\,.
\end{equation}
If we integrate the covariant phase-space density operator over the position $\uvec R$, we recover the density operator in momentum space
\begin{equation}
\int\ud^3 \uvec R\,\rho_{R,P}=\int\frac{\ud P^2}{2\pi}\,2\pi\,\delta(P^2-M^2)\,\frac{\ket{P}\bra{P}}{2P^0}\,,
\end{equation}
and if we integrate over the momentum $\uvec P$, we recover the density operator in position space
\begin{equation}
\int\frac{\ud^3\uvec P}{(2\pi)^3 2P^0}\,\rho_{R,P} =\ket{R}\bra{R}\,.
\end{equation}
Matrix elements of position-dependent operators are then given by
\begin{equation}
\langle O(X)\rangle_{R,P}=\uTr[O(X)\rho_{R,P}]\,,
\end{equation}
and translation invariance implies that
\begin{equation}
\langle O(X)\rangle_{R,P}=\langle O(x)\rangle_{0,P}\,,
\end{equation}
where $x=X-R$ is the relative average position.

In the present work, we are interested in matrix elements of the (renormalized) nucleon EMT $T^{\mu\nu}(X)$. The latter can be expressed in terms of gravitational (or energy-momentum) form factors (GFFs), first introduced\footnote{Note that a tensor decomposition of the electron EMT disregarding polarization appeared in an earlier work by Villars~\cite{Villars:1950pkp}.} by Kobzarev and Okun~\cite{Kobzarev:1962wt,Kobzarev:1962wt2} and later by Pagels~\cite{Pagels:1966zza}. In the general case of a local, gauge-invariant asymmetric EMT for a spin-$\frac{1}{2}$ target, the standard parametrization reads~\cite{Bakker:2004ib,Leader:2013jra,Lorce:2015lna,Lorce:2017wkb}
\begin{align}
\label{eq:TmunuUU}
\bra{p',\uvec s'} T^{\mu\nu}_a(0)\ket{p,\uvec s}
= \bar u(p',\uvec s') \Bigg\{&
\frac{P^\mu P^\nu}{M}\,A_a(t)
+ \frac{\Delta^\mu\Delta^\nu - \eta^{\mu\nu}\Delta^2}{M}\, C_a(t)
+ M \eta^{\mu\nu}\bar C_a(t)\\
&+ \frac{P^{\{\mu} i\sigma^{\nu\}\Delta}}{4M}\left[A_a(t)+B_a(t)\right]+ \frac{P^{[\mu} i\sigma^{\nu]\Delta}}{4M}\,D_a(t)\Bigg\} u(p,\uvec s)\nn\,,
\end{align}
where $p$ ($p'$) and $\uvec s$ ($\uvec s'$) are the four-momentum and canonical polarization of the initial (final) nucleon of a mass $M$, $\eta^{\mu\nu} = \text{diag}(+,-,-,-)$ is the Minkowski metric, and $t=\Delta^2$ with $\Delta=p'-p$ and $P=(p'+p)/2$. For convenience, we introduced the symmetrizer $a^{\{\mu}b^{\nu\}}  = a^\mu b^\nu + a^\nu b^\mu$, the antisymmetrizer $a^{[\mu}b^{\nu]} = a^\mu b^\nu - a^\nu b^\mu$, and the notation $i\sigma^{\mu\Delta}= i\sigma^{\mu\lambda}\Delta_\lambda$. The label $a$ in Eq.~\eqref{eq:TmunuUU} refers to the contribution from a particular type of constituents, typically $a=q$ for quarks and $a=G$ for gluons, here defined in the $\overline{\text{MS}}$ scheme . The total EMT is then simply obtained by summing over all the constituent types $T^{\mu\nu}=\sum_a T^{\mu\nu}_a$.

The generic matrix element~\eqref{eq:TmunuUU} is parametrized in terms of five GFFs, namely $A_a(t)$, $B_a(t)$, $C_a(t)$, $\bar C_a(t)$ and $D_a(t)$. Beside their $t$-dependence, the GFFs are also usually scale and scheme-dependent. Except for $D_a(t)$, these additional dependences drop out when summing over all quark and gluon contributions. In particular, three of the GFFs associated with the symmetric part of the EMT satisfy the sum rules
\begin{equation}\label{eq:A}
\sum_{a=q,G} A_a(0)= 1,\qquad
\sum_{a=q,G} B_a(0)= 0\,,\qquad
\sum_{a=q,G} \bar C_a(t)= 0\,,
\end{equation}
which arise from the Poincar\'e invariance of the theory~\cite{Teryaev:1999su,Leader:2013jra,Lowdon:2017idv}. The vanishing of the total anomalous gravitomagnetic moment $\sum_aB_a(0)=0$ is related to the equivalence principle in General Relativity~\cite{Kobzarev:1962wt,Kobzarev:1962wt2,Teryaev:1999su,Teryaev:2016edw} and holds separately for each Fock component of the state~\cite{Brodsky:2000ii}. The GFF $\bar C_a(t)$ accounts for the non-conservation of the partial EMT $\langle p',\uvec s'|\partial_\mu T^{\mu\nu}_a(0)|p,\uvec s\rangle=i\Delta^\nu M\, \bar u(p',\uvec s')u(p,\uvec s)\,\bar C_a(t)$ and should naturally vanish when summed over all the constituents. The GFFs have been studied in a variety of theoretical approaches, see e.g.~\cite{Kim:2012ts} and references therein, and also using Lattice QCD simulations~\cite{Hagler:2007xi,Bali:2016wqg,Alexandrou:2017oeh,Yang:2018bft,Alexandrou:2018xnp,Yang:2018nqn,Shanahan:2018pib}.
\newline

Although a direct measurement of nucleon scattering by a gravitational field is not realistic with the current technology, it is remarkable that the GFFs can in principle be extracted from experimental data. Ji~\cite{Ji:1996ek,Ji:1998pc} showed that the three GFFs $A_a(t)$, $B_a(t)$ and $C_a(t)$ can be obtained from the second Mellin moment of leading-twist generalized parton distributions (GPDs)~\cite{Mueller:1998fv, Ji:1996nm, Radyushkin:1996nd}, which are accessible in several exclusive processes, like deeply virtual Compton scattering~\cite{Kumericki:2016ehc} and meson production~\cite{Favart:2015umi}. Recently, the corresponding GFFs for a pion target have been extracted from Belle data on $\gamma^\star\gamma\rightarrow\pi^0\pi^0$~\cite{Kumano:2017lhr}. The GFF $\bar C_a(t)$, which can formally be obtained from higher-twist GPDs~\cite{Leader:2012ar,Leader:2013jra,Tanaka:2018wea}, is related to the $\sigma_{\pi N}$ term extracted from $\pi N$ scattering amplitudes~\cite{Alarcon:2011zs,Hoferichter:2015dsa}, and to the trace anomaly which can be studied through the production of heavy quarkonia at threshold~\cite{Kharzeev:2002fm, Krein:2017usp, Joosten:2018gyo, Joosten:2018fql, Hatta:2018ina}. Finally, the GFF associated with the antisymmetric part of the EMT is directly related to the axial-vector form factor~\cite{Lorce:2017wkb}
\begin{equation}
D_q(t)= - G^q_A(t),\qquad D_G(t) =0\,,
\end{equation}
and hence can be obtained from quasi-elastic neutrino scattering and pion electroproduction processes~\cite{Bernard:2001rs}. 

For illustrative purposes, we will adopt in this work a simple multipole Ansatz for the GFFs
\begin{equation}
\label{eq:model}
F_a(t)=\frac{F_a(0)}{\left(1-t/\Lambda^2_{F_a}\right)^{n_F}}\,,
\end{equation}
which is supported by model calculations for $|t|<1$ GeV$^2$~\cite{Goeke:2007fp}, together with the parameters given in Table~\ref{tab:model}, in the $\overline{\text{MS}}$ scheme with renormalization scale $\mu = 2~$GeV. We adopt a standard dipole Ansatz (i.e. $n_F=2$) for $A_a$, $\bar C_a$ and $D_a$, but for $B_a$ and $C_a$ we choose a tripole Ansatz (i.e. $n_F=3$) in order for the energy and pressure distributions to be realistic, see discussion in Section~\ref{sec:5}. 
\begin{table}[t!]
\begin{center}
\caption{\footnotesize{Parameters for the multipole parametrization~\eqref{eq:model} of the GFFs, in the $\overline{\text{MS}}$ scheme with renormalization scale $\mu = 2~$GeV.}}
\label{tab:model}
\begin{tabular}{@{\quad\!}c@{\quad}c@{\quad}|@{\quad}c@{\quad}c@{\quad}c@{\quad}c@{\quad\!}}
$F_a$&$n_F$&$F_q(0)$&$\Lambda_{F_q}\,[$GeV$]$&$F_G(0)$&$\Lambda_{F_G}\,[$GeV$]$\\
\hline
$A_a$&$2$&$~~0\,.55$&$0\,.91$&$~~0\,.45$&$0\,.91$\\
$B_a$&$3$&$-0\,.07$&$0\,.8~~$&$~~0\,.07$&$0\,.8~~$\\
$C_a$&$3$&$-0\,.32$&$0\,.8~~$&$-0\,.56$&$0\,.8~~$\\
$\bar C_a$&$2$&$-0\,.11$&$0\,.91$&$~~0\,.11$&$0\,.91$\\
$D_a$&$2$&$-0\,.33$&$1.74$&--&--
\end{tabular}
\end{center}
\end{table}
The normalization $A_q(0)\approx 0\,.55$ is taken from the recent MMHT2014 analysis~\cite{Harland-Lang:2014zoa}. We set $B_q(0)\approx-0\,.07$ as suggested by the AdS/QCD correspondence~\cite{Mondal:2015fok,Kumar:2017dbf} and which agrees in magnitude with recent estimates from Lattice QCD~\cite{Deka:2013zha,Shanahan:2018pib}. We also use the values $C_q(0)=d^q_1(0)/5$ with $d^q_1(0)\approx-1.59$ obtained in a dispersive analysis of deeply virtual Compton scattering~\cite{Pasquini:2014vua} which is close to a recent experimental extraction reported in~\cite{Burkert:2018bqq}, $\bar C_q(0)\approx -0\,.11$ given by a phenomenological estimate~\cite{Lorce:2017xzd} supported by a recent Lattice calculation~\cite{Yang:2018nqn}, and $D_q(0)=-G^q_A(0)\approx -0\,.33$ obtained from a leading twist NNLO analysis by the HERMES collaboration~\cite{Airapetian:2006vy}. The quark multipole masses $\Lambda_{F_q}$ with $F=A,B,C,\bar C$ are motivated\footnote{Only a dipole fit of the GFFs has been reported in that model. We estimated the tripole mass $\Lambda_{C_q}$ by multiplying the reported dipole mass $0\,.65$ GeV with $\sqrt{3/2}$ so as to leave the quantity $\ud C_q(t)/\ud t|_{t=0}$ unchanged. No multipole mass for $B_q(t)$ has been reported, so we simply choose $\Lambda_{B_q}=\Lambda_{C_q}$.} by results obtained in the chiral quark-soliton model~\cite{Goeke:2007fp,Polyakov:2018exb}, and $\Lambda_{D_q}=\Lambda_{G^q_A}$ is taken from a recent Lattice estimate~\cite{Alexandrou:2017hac}. In the gluon sector, the normalizations $A_G(0)$, $B_G(0)$ and $\bar C_G(0)$ are determined by the sum rules~\eqref{eq:A}. As suggested in~\cite{Hatta:2018ina}, we use the simple relation $C_G(0)=\frac{16}{3n_f}\,C_q(0)$ with $n_f=3$. Since we lack information about the gluon GFFs, we simply set $\Lambda_{F_G} = \Lambda_{F_q}$ for $F= A,\,B,\,C,\,\bar C$. 

This simple parametrization should be considered only as a naive model with the sole aim of allowing us to illustrate the various distributions discussed in the rest of the paper. In particular, one of its advantages is to permit an evaluation of the two and three-dimensional Fourier transforms of the multipole distributions in closed forms
\begin{align}
\int\frac{\ud^2\uvec\Delta_\perp}{(2\pi)^2}\,\frac{e^{-i\uvec\Delta_\perp\cdot\uvec b_\perp}}{(1+\frac{\uvec\Delta^2_\perp}{\Lambda^2})^n}&=\frac{\Lambda}{b\pi}\left(\frac{\Lambda b}{2}\right)^n\,\frac{K_{n-1}(\Lambda b)}{(n-1)!}\,,\\
\int\frac{\ud^3\uvec\Delta}{(2\pi)^3}\,\frac{e^{-i\uvec\Delta\cdot\uvec r}}{(1+\frac{\uvec\Delta^2}{\Lambda^2})^2}&=\frac{\Lambda^3}{8\pi}\,e^{-\Lambda r}\,,\\
\int\frac{\ud^3\uvec\Delta}{(2\pi)^3}\,\frac{e^{-i\uvec\Delta\cdot\uvec r}}{(1+\frac{\uvec\Delta^2}{\Lambda^2})^3}&=(1+\Lambda r)\,\frac{\Lambda^3}{32\pi}\,e^{-\Lambda r}\,,
\end{align}
where $b=|\uvec b_\perp|$ with $\uvec b_\perp$ a two-dimensional vector, and $r=|\uvec r|$ with $\uvec r$ a three-dimensional vector.

\section{Distributions in instant form}\label{sec:3}

Performing the integrals over $P^2$ and $\Delta^0$ in the covariant phase-space density operator~\eqref{eq:RPcov}  leads to
\begin{equation}\label{RPnoncov}
\rho_{R,P}=\int\frac{\ud^3\uvec\Delta}{(2\pi)^3\,2P^0}\,e^{-i\Delta\cdot R}\,\ket{P-\tfrac{\Delta}{2}}\bra{P+\tfrac{\Delta}{2}}\,,
\end{equation}
where the initial and final target energies are given by
\begin{equation}\label{eq:Delta0}
p^0=\sqrt{(\uvec P-\tfrac{\uvec\Delta}{2})^2+M^2}\,,\qquad p'^0=\sqrt{(\uvec P+\tfrac{\uvec\Delta}{2})^2+M^2}\,.
\end{equation}
The non-explicitly covariant form~\eqref{RPnoncov} coincides with that of Appendix B in Ref.~\cite{Lorce:2018zpf} if we replace the normalization factor $2P^0$ by $2\sqrt{p'^0p^0}$. The difference in the normalization comes from the fact that ``position'' states in Ref.~\cite{Lorce:2018zpf} were defined with the non-relativistic normalization $\langle x'|x\rangle=\delta^{(3)}(\uvec x'-\uvec x)$ at equal times. This difference will however not concern us since we will essentially be interested in the case $\Delta^0=0$, when $p^0 = p'^0$.

Setting the origin at the average position of the system $\uvec R=\uvec 0$ and denoting $x^\mu=(0,\uvec r)$, the static EMT encoding the distribution of energy and momentum inside the system with canonical polarization $\uvec s$ and average momentum $\uvec P$, is defined as the following Fourier transform
\begin{equation}\label{TFT}
\mathcal T^{\mu\nu}_a(\uvec r;\uvec P)=\langle T^{\mu\nu}_a( x)\rangle_{0,P}=\int\frac{\ud^3\uvec\Delta}{(2\pi)^3}\,e^{-i\uvec\Delta\cdot\uvec r}\,\langle\!\langle T^{\mu\nu}_a(0)\rangle\!\rangle
\end{equation} 
of the off-forward amplitude~\cite{Lorce:2017wkb} 
\begin{equation}\label{off-forward}
\langle\!\langle T^{\mu\nu}_a(0)\rangle\!\rangle=\frac{\bra{p',\uvec s} T^{\mu\nu}_a(0)\ket{p,\uvec s}}{2P^0}\,.
\end{equation}
Note that $x^0=0$ because the positions $\uvec X$ and $\uvec R$ are considered at the same time $X^0=R^0$. If we allow $X^0$ and $R^0$ to be different, the above static EMT $\mathcal T^{\mu\nu}_a(\uvec r;\uvec P)$ can be recovered by considering the time average $\int\ud X^0/(2\pi\,\delta(0))$ in a frame where the energy transfer vanishes $\Delta^0=0$~\cite{Polyakov:2002yz}.

The GFFs in Eq.~\eqref{eq:TmunuUU} are multiplied by two types of Dirac bilinears, namely $\bar u'u$ and $\bar u'i\sigma^{\mu\Delta} u$. Using the same canonical polarization for both initial and final states, these bilinears can be expressed in instant form as~\cite{Lorce:2017isp}
\begin{align}
\bar u'u&=\mathcal N^{-1}\left[2(P^2+MP^0)+i\epsilon^{0P\Delta S}\right],\\
\bar u'i\sigma^{\mu\Delta} u&=\mathcal N^{-1}\left\{P^\mu\Delta^2+M(\eta^{\mu 0}\Delta^2-\Delta^\mu\Delta^0)+2\left[(P^0+M)\,i\epsilon^{\mu\Delta P S}-\frac{\Delta^2}{4}\,i\epsilon^{\mu\Delta S0}-(P\cdot S)\,i\epsilon^{\mu\Delta P0}\right]\right\},
\end{align}
where $S^\mu=(0\,,\uvec s)$ and $\mathcal N=\sqrt{p'^0+M}\sqrt{p^0+M}$. In the present work, we will restrict ourselves to the case of an unpolarized target, which amounts to setting $S^\mu=0$ in the above expressions. The unpolarized off-forward amplitude then reads
\begin{align}\label{Tred}
\langle\!\langle T^{\mu\nu}_a(0)\rangle\!\rangle&=\frac{P^2+MP^0}{P^0\mathcal N}\left\{\frac{P^\mu P^\nu}{M}\,A_a(t)
+ \frac{\Delta^\mu\Delta^\nu - \eta^{\mu\nu}\Delta^2}{M}\, C_a(t)
+ M \eta^{\mu\nu}\bar C_a(t)\right\}\\
&+\frac{\Delta^2}{4P^0\mathcal N}\left\{\left[\frac{2P^\mu P^\nu}{M}+P^{\{\mu}\eta^{\nu\}0}\right]\frac{A_a(t)+B_a(t)}{2}+P^{[\mu}\eta^{\nu]0}\,\frac{D_a(t)}{2}\right\}\nn\\
&-\frac{\Delta^0}{4P^0\mathcal N}\left\{P^{\{\mu}\Delta^{\nu\}}\,\frac{A_a(t)+B_a(t)}{2}+P^{[\mu}\Delta^{\nu]}\,\frac{D_a(t)}{2}\right\}\nn.
\end{align}
Except the global normalization factor, the first line is identical to the standard parametrization for a spin-$0$ state. The second and third lines can be interpreted as the polarization-independent distortion arising due to the spin of the target. Indeed, it has been shown in~\cite{Polyakov:2002yz,Lorce:2017wkb} that the combinations $[A_a(t)+B_a(t)]/2$ and $-D_a(t)/2$ provide the information about the spatial distribution of total angular momentum and spin associated with constituent type $a$. 

The static EMT can receive a (quasi-)probabilistic interpretation only when no energy is transferred to the system $\Delta^0=0$~\cite{Lorce:2017wkb}. Since the onshell conditions impose that $\Delta^0=\uvec P\cdot\uvec\Delta/P^0$, we will consider the following cases:
\begin{enumerate}[label=(\alph*)]
\item $\uvec \Delta = \uvec 0$ --- forward limit (FL);
\item $\uvec P = \uvec 0$ --- Breit frame (BF);
\item $\uvec P \cdot \uvec \Delta = 0$ --- elastic frame (EF);
\item $\uvec P \cdot \uvec \Delta = 0$ and $|\uvec P| \to \infty$ --- infinite-momentum frame (IMF).
\end{enumerate}
Note that when $\Delta^0=0$ the normalization factor appearing in Eq.~\eqref{Tred} reduces to $\mathcal N=P^0+M$.

\subsection{Forward limit}

Since $\uvec\Delta$ is the Fourier conjugate variable to relative position $\uvec r$, the forward limit (FL) $\uvec\Delta=\uvec 0$ is obtained by integrating the static EMT over $\uvec r$
\begin{align}\label{eq:TmunuFL}
\int\ud^3 \uvec r\,\mathcal T^{\mu\nu}_a(\uvec r;\uvec P)&=\frac{\bra{P,\uvec s} T^{\mu\nu}_a(0)\ket{P,\uvec s}}{2E_P}\\
&=\frac{P^\mu P^\nu}{E_P}\,A_a(0) + \frac{M^2}{E_P}\, \eta^{\mu\nu}\bar C_a(0)\nn\,,
\end{align}
where $E_P=\sqrt{\uvec P^2+M^2}$.
Note that the dependence on the nucleon spin disappears in the FL. This is expected because the EMT is the Noether current associated with invariance under translations and not Lorentz transformations.

Focusing on the $\mathcal T^{00}_a$ component in the rest frame $\uvec P=\uvec 0$, Ji~\cite{Ji:1994av,Ji:1995sv} proposed a decomposition of the nucleon mass based on Eq.~\eqref{eq:TmunuFL}. Recently, a covariant treatment of $\mathcal T^{\mu\nu}_a$ revealed that the gravitational charges $A_a(0)$ and $\bar C_a(0)$ can be interpreted in terms of partial internal energy density and isotropic pressure~\cite{Lorce:2017xzd}. Indeed, denoting the proper volume\footnote{The proper volume of the nucleon can typically be taken to be $V = \frac{4}{3} \pi R_M^3$ with the mass radius $R_M$ defined in Eq.~\eqref{eq:R_M}. Note however that the precise definition of $V$ is somewhat arbitrary and does not affect our results for the average densities as they are always expressed in units of $M/V$.} by $V$ and the boost factor by $\gamma=E_P/M$, one finds that the average density
\begin{equation}\label{eq:identFL}
\frac{\gamma}{V}\int\ud^3 \uvec r\,\mathcal T^{\mu\nu}_a(\uvec r;\uvec P)=\left[\frac{P^\mu P^\nu}{M^2}\,A_a(0) +\eta^{\mu\nu}\bar C_a(0)\right]\frac{M}{V}
\end{equation}
has the same structure as the EMT of an element of perfect fluid~\cite{Eckart:1940te}
\begin{align}
\theta^{\mu\nu}(\uvec r)&=
(\bar\varepsilon +\bar p)\,u^\mu u^\nu - \bar p\, \eta^{\mu\nu}.
\end{align}
The four-velocity of the nucleon being given in the FL by $u^\mu={P^\mu}/{M}$, this suggests that the following combinations
\begin{equation}
\bar\varepsilon_a=\left[A_a(0)+\bar C_a(0)\right]\frac{M}{V},\qquad \bar p_a=-\bar C_a(0)\,\frac{M}{V}\,,
\end{equation}
can be interpreted\footnote{Note that the nucleon is by no means assimilated to a perfect fluid. We are only interested in the mechanical interpretation of particular components of the static EMT, defined in a quantum theory as the expectation value of the EMT operator in a specified state~\cite{Pais:1949vdk,Polyakov:2002yz}.} as the spatial average of partial energy density and isotropic pressure associated with constituent type $a$. The contributions to proper internal energy and pressure-volume work are then given by
\begin{equation}\label{eq:epsilon_p}
U_a=\left[A_a(0)+\bar C_a(0)\right]M\,,\qquad W_a=-\bar C_a(0)\,M\,.
\end{equation}
The nucleon being a stable system with mass $M$, one obtains a mass sum rule and a stability constraint 
\begin{equation}
\sum_{a=q,G}U_a=M\,, \qquad \sum_{a=q,G}W_a=0\,,
\end{equation}
consistent with Eq.~\eqref{eq:A}. While $A_q(0)$ is well determined~\cite{Harland-Lang:2014zoa}, $\bar C_q(0)$ is poorly known. Based on the phenomenological estimates in~\cite{Gao:2015aax}, $\bar C_q(0)$ seems to be negative and sizeable~\cite{Lorce:2017xzd,Hatta:2018sqd}, in agreement with the MIT Bag Model prediction~\cite{Ji:1997gm} and recent Lattice estimates~\cite{Yang:2018nqn}. In contrast, it has been suggested based on the instanton picture of the QCD vacuum that $\bar C_q(0)$ at low scale may be small and positive~\cite{Polyakov:2018exb}.

\subsection{Breit frame}

Since the work of Sachs on the electromagnetic form factors~\cite{Sachs:1962zzc}, the Breit frame (BF) defined by $\uvec P=\uvec 0$ became a popular frame for the physical analysis of form factors in instant form. The phase-space perspective adopted in the present work shows that working in the Breit frame amounts to looking at the system which is in average at rest and sitting in average at the origin. 

The unpolarized off-forward amplitude~\eqref{Tred} reduces in the BF to
\begin{equation}\label{BFFA}
\langle\!\langle T^{\mu\nu}_a(0)\rangle\!\rangle\big|_\text{BF}=M\left\{\eta^{\mu 0}\eta^{\nu 0}\left[A_a(t)+\frac{t}{4M^2}\,B_a(t)\right]+\eta^{\mu\nu}\left[\bar C_a(t)-\frac{t}{M^2}\,C_a(t)\right]+\frac{\Delta^\mu\Delta^\nu}{M^2}\,C_a(t)\right\},
\end{equation}
where we used $P^\mu=\eta^{\mu 0}P^0$. Interestingly, the GFF $D_a(t)$ does not contribute in the BF. We therefore recover the case of a symmetric EMT studied by Polyakov \emph{et al.}~\cite{Polyakov:2002yz,Goeke:2007fp,Polyakov:2018zvc}. After Fourier transform, we find the following unpolarized static EMT
\begin{align}\label{eq:TmunuRF}
\mathcal T^{\mu\nu}_a(\uvec r;\uvec 0)=M&\left\{\eta^{\mu0}\eta^{\nu0}\left[\mathcal A_a(r)+\frac{1}{4M^2}\left(\frac{1}{r^2}\frac{\ud}{\ud r}\!\left(r^2\,\frac{\ud\mathcal B_a(r)}{\ud r}\right)-\frac{4}{r}\frac{\ud\mathcal C_a(r)}{\ud r}\right)\right]\right.\\
&\left. +\eta^{\mu\nu} \left[ \bar{\mathcal{C}}_a(r) - \frac{1}{M^2}\frac{1}{r}\frac{\ud}{\ud r}\!\left(r\,\frac{\ud\mathcal C_a(r)}{\ud r}\right)\right]-\frac{x^\mu\, x^\nu}{r^2}\, \frac{1}{M^2}\,r\,\frac{\ud}{\ud r}\!\left(\frac{1}{r}\frac{\ud\mathcal C_a(r)}{\ud r}\right)\right\}\nn,
\end{align}
where $x^\mu=(0,\uvec r)$ and the three-dimensional Fourier transforms of  GFFs are denoted by
\begin{equation}
\mathcal F_a(r)=\int\frac{\ud^3\uvec\Delta}{(2\pi)^3}\,e^{-i\uvec\Delta\cdot\uvec r}\,F_a(t)
\end{equation}
with $t=-\uvec\Delta^2$.

\begin{figure}[t!]
    \centering
        \includegraphics[width=0.33\textwidth,trim={2.7cm 2.3cm 1.1cm 0.9cm}, clip]{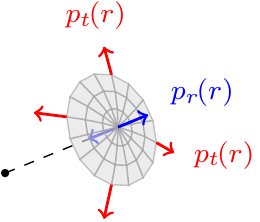}
    \caption{The radial pressure $p_r(r)$ and the tangential pressure $p_t(r)$ at a distance $r$ from the center of the system~\eqref{eq:TmunuAnisotropy}. Spherical symmetry imposes only the equality of the two tangential pressures.}
    \label{fig:cap}
\end{figure}

We observe that the unpolarized static EMT in the BF~\eqref{eq:TmunuRF} has the same structure as the EMT of an anisotropic spherically symmetric compact star~\cite{Bayin:1985cd}
\begin{equation}
\label{eq:TmunuAnisotropy}
\Theta^{\mu\nu}(\uvec r)=[\varepsilon(r)+p_t(r)]\,u^\mu u^\nu - p_t(r)\eta^{\mu\nu}+[p_r(r)-p_t(r)]\,\chi^\mu \chi^\nu,
\end{equation}
where $u^\mu$ and $\chi^\mu=x^\mu/r$ are unit timelike and spacelike four-vectors orthogonal to each other. The functions $\varepsilon(r)$, $p_r(r)$ and $p_t(r)$ represent the energy density, radial pressure and tangential pressure, respectively. As noticed by Einstein and developed first by Lemaitre in 1933~\cite{Lemaitre:1933gd2,Lemaitre:1933gd}, spherical symmetry requires only the equality of the two tangential pressures, see Fig~\ref{fig:cap}. The tensor~\eqref{eq:TmunuAnisotropy} can alternatively be written as
\begin{equation}
\label{eq:TmunuAnisotropy2}
\Theta^{\mu\nu}(\uvec r)=[\varepsilon(r)+p(r)]\,u^\mu u^\nu - p(r)\eta^{\mu\nu}+s(r)\left(\chi^\mu \chi^\nu-\frac{1}{3}\,h^{\mu\nu}\right)
\end{equation}
with $h^{\mu\nu}=u^\mu u^\nu-\eta^{\mu\nu}$. Isotropic pressure $p(r)$ and pressure anisotropy $s(r)$ are related to radial and tangential pressures as follows
\begin{equation}
p(r)=\frac{p_r(r)+2\,p_t(r)}{3}\,,\qquad s(r)=p_r(r)-p_t(r)\,.
\end{equation}

The comparison of the unpolarized static EMT in the BF~\eqref{eq:TmunuRF} with the EMT of an anisotropic spherically symmetric compact star~\eqref{eq:TmunuAnisotropy} or~\eqref{eq:TmunuAnisotropy2} with $u^\mu=\eta^{\mu 0}$ suggests that the following combinations
\begin{align}
\label{eq:epsilon}
\varepsilon_a(r) &= M\left\{\mathcal A_a(r)+\bar{\mathcal C}_a(r) + \frac{1}{4M^2}\frac{1}{r^2}\frac{\ud}{\ud r}\!\left(r^2\frac{\ud}{\ud r}\!\left[\mathcal B_a(r)-4\mathcal C_a(r)\right]\right)\right\},\\
\label{eq:pr}
p_{r,a}(r) &= M\left\{-\bar{\mathcal{C}}_a(r)+ \frac{1}{M^2}\frac{2}{r}\frac{\ud\mathcal C_a(r)}{\ud r} \right\},
\\
\label{eq:pt}
p_{t,a}(r) &= M\left\{-\bar{\mathcal{C}}_a(r) + \frac{1}{M^2}\frac{1}{r}\frac{\ud}{\ud r}\!\left( r\,\frac{\ud\mathcal C_a(r)}{\ud r}\right) \right\},\\
\label{eq:p}
p_a(r) &= M\left\{-\bar{\mathcal{C}}_a(r)+ \frac{2}{3}\frac{1}{M^2}\frac{1}{r^2}\frac{\ud}{\ud r}\!\left(r^2\frac{\ud\mathcal C_a(r)}{\ud r}\right) \right\},
\\
\label{eq:s}
s_a(r) &= M\left\{-\frac{1}{M^2}\,r\,\frac{\ud}{\ud r}\!\left(\frac{1}{r}\frac{\ud\mathcal C_a(r)}{\ud r}\right) \right\},
\end{align}
can be interpreted as the partial energy density, radial pressure, tangential pressure, isotropic pressure, and pressure anisotropy associated with constituent type $a$, respectively. They can alternatively be written as
\begin{align}
\label{eq:epsilonbis}
\varepsilon_a(r) &= M\int\frac{\ud^3\uvec\Delta}{(2\pi)^3}\,e^{-i\uvec\Delta\cdot\uvec r}\left\{A_a(t)+\bar{C}_a(t) + \frac{t}{4M^2}\left[B_a(t)- 4C_a(t)\right]\right\},\\
\label{eq:prbis}
p_{r,a}(r) &=M\int\frac{\ud^3\uvec\Delta}{(2\pi)^3}\,e^{-i\uvec\Delta\cdot\uvec r}\left\{-\bar C_a(t)-\frac{4}{r^2}\frac{t^{-1/2}}{M^2}\frac{\ud}{\ud t}\!\left(t^{3/2}\,C_a(t)\right)\right\},
\\
\label{eq:ptbis}
p_{t,a}(r) &= M\int\frac{\ud^3\uvec\Delta}{(2\pi)^3}\,e^{-i\uvec\Delta\cdot\uvec r}\left\{-\bar C_a(t)+\frac{4}{r^2}\frac{t^{-1/2}}{M^2}\frac{\ud}{\ud t}\!\left[t\frac{\ud}{\ud t}\!\left(t^{3/2}\,C_a(t)\right)\right]\right\},\\
\label{eq:pbis}
p_a(r) &= M\int\frac{\ud^3\uvec\Delta}{(2\pi)^3}\,e^{-i\uvec\Delta\cdot\uvec r}\left\{-\bar{C}_a(t) +\frac{2}{3} \frac{t}{M^2}\,C_a(t)\right\},
\\
\label{eq:sbis}
s_a(r) &=M\int\frac{\ud^3\uvec\Delta}{(2\pi)^3}\,e^{-i\uvec\Delta\cdot\uvec r}\left\{-\frac{4}{r^2}\frac{t^{-1/2}}{M^2}\frac{\ud^2}{\ud t^2}\!\left(t^{5/2}\,C_a(t)\right)\right\},
\end{align}
As indicated by the presence of $B_a(t)$, the non-zero spin of the target affects only the energy distribution in the BF. Classically, we indeed expect angular momentum to push matter away from the center. 
\begin{figure}[t!]
    \centering
    \includegraphics[width=0.49\textwidth]{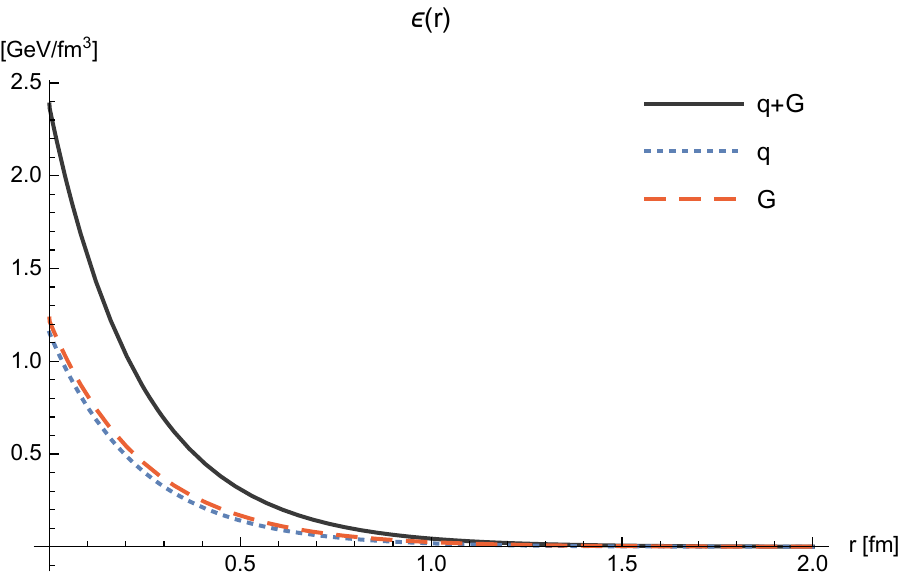}
    \hspace{0.5em}
    \includegraphics[width=0.49\textwidth]{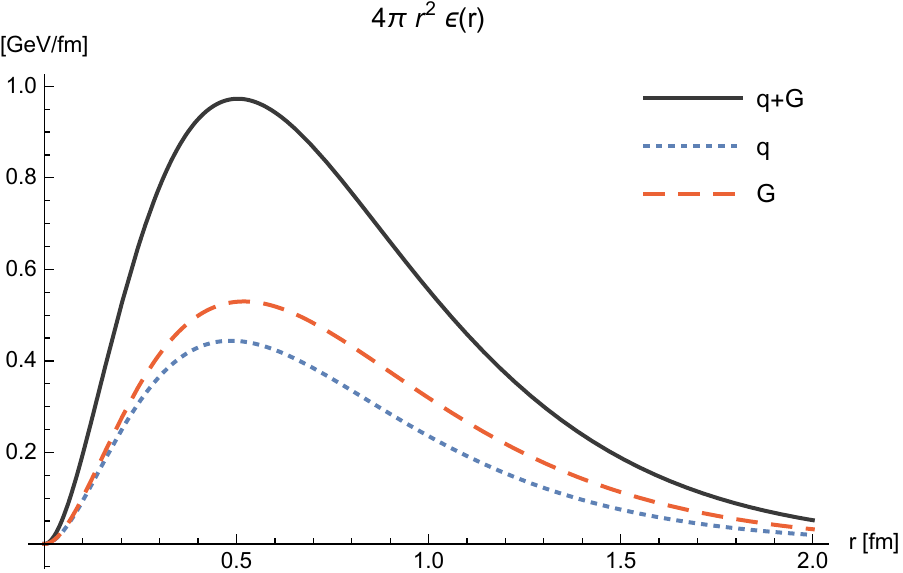}
    \\
	(a)\hspace*{0.45\linewidth}(b)
    \caption{\label{fig:epsilon}
    Plots of the energy density, (a) $\epsilon(r)$ and (b) $4\pi\,r^2\,\epsilon(r)$,
    using the multipole model~\eqref{eq:model} with parameters given in Table~\ref{tab:model}, see Eq.~\eqref{eq:epsilon} or Eq.~\eqref{eq:epsilonbis} for the definition in terms of GFFs.
    }
\end{figure}

\begin{figure}[t!]
    \centering
        \includegraphics[width=0.49\textwidth]{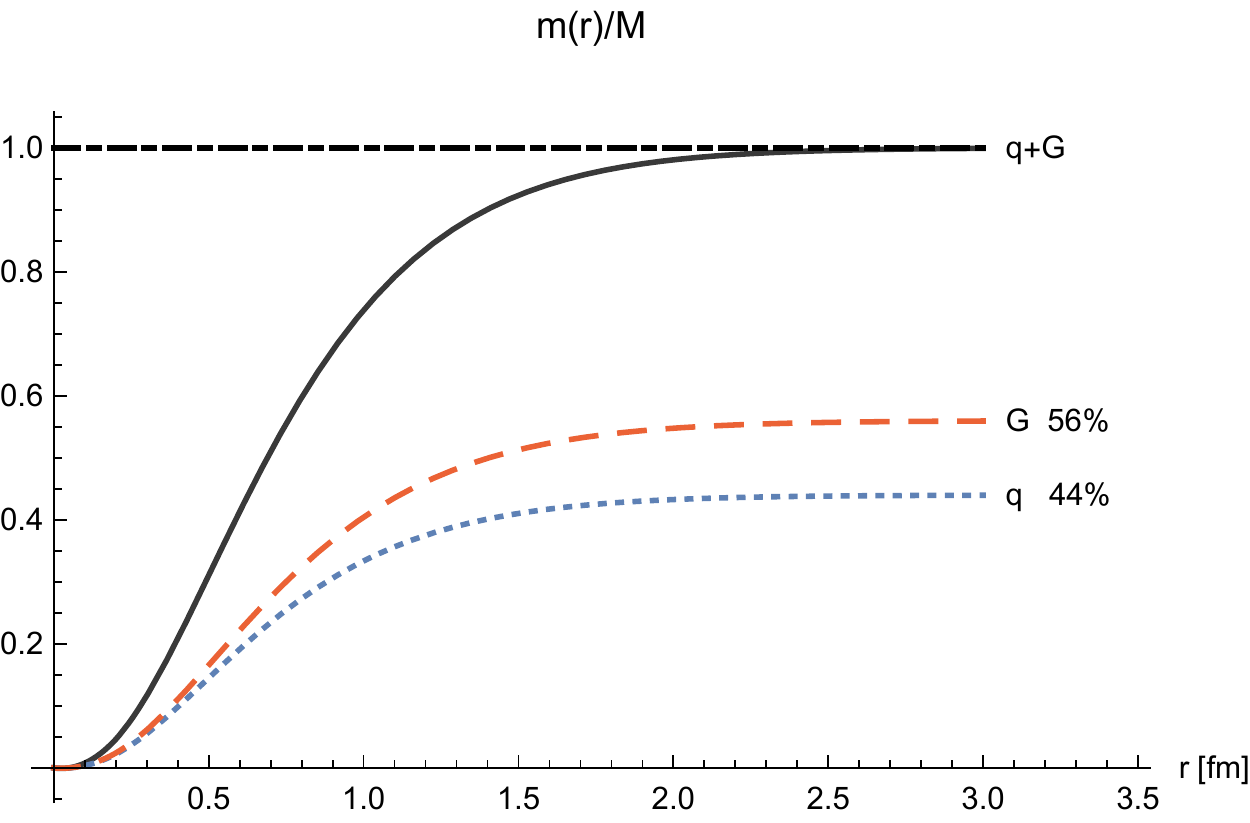}
    \caption{Quark, gluon and total mass functions,
     computed using the multipole model~\eqref{eq:model} with parameters given in Table~\ref{tab:model}, see Eq.~\eqref{massfunc} for the definition.}
    \label{fig:m}
\end{figure}

\begin{figure}[t!]
    \centering
    \includegraphics[width=0.49\textwidth]{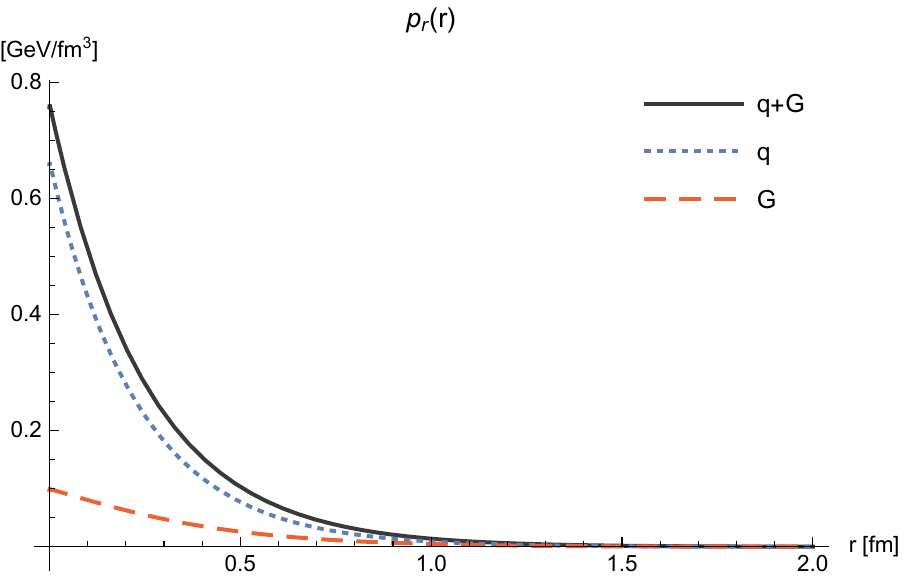}
    \hspace{0.5em}
    \includegraphics[width=0.49\textwidth]{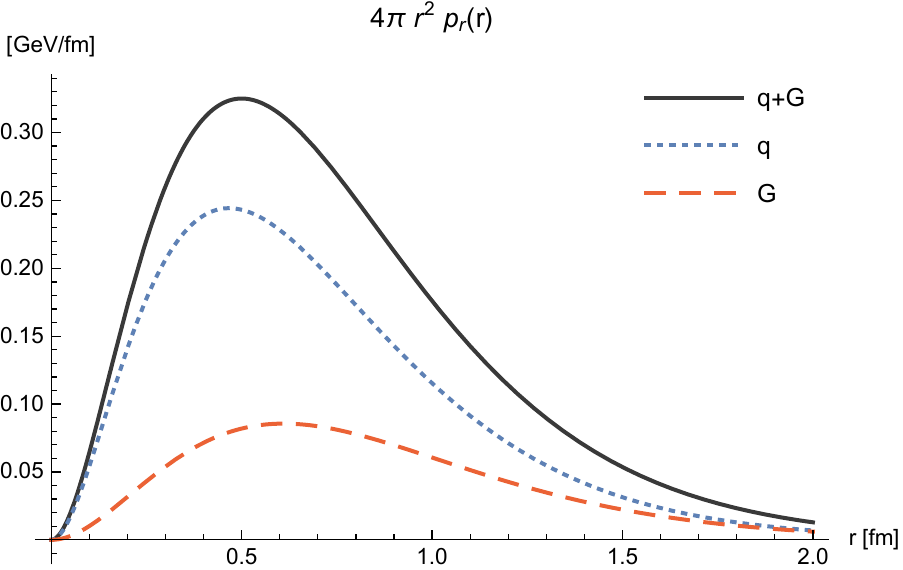}
    \\
	(a)\hspace*{0.45\linewidth}(b)
	\caption{\label{fig:pr}
    Plots of the radial pressure, (a) $p_r(r)$ and (b) $4\pi\,r^2\,p_r(r)$,
    using the multipole model~\eqref{eq:model} with parameters given in Table~\ref{tab:model}, see Eq.~\eqref{eq:pr} or Eq.~\eqref{eq:prbis} for definitions in terms of GFFs.
    }
\end{figure}

\begin{figure}[t!]
    \centering
    \includegraphics[width=0.49\textwidth]{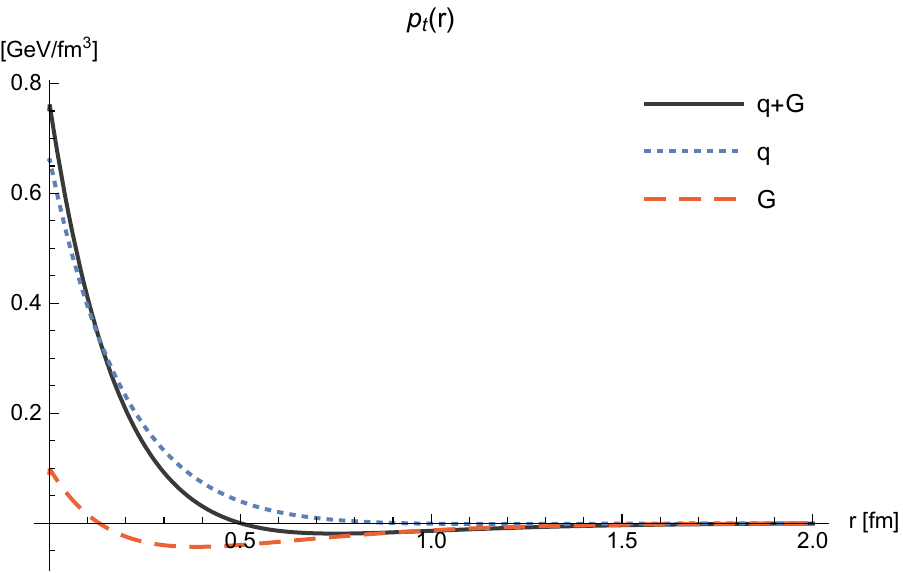}
    \hspace{0.5em}
    \includegraphics[width=0.49\textwidth]{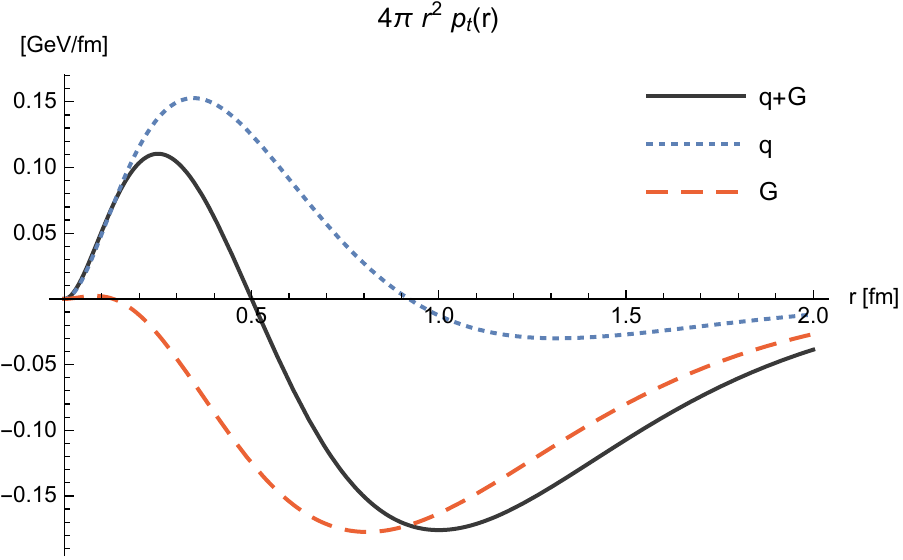}
    \\
	(a)\hspace*{0.45\linewidth}(b)
    \caption{\label{fig:pt}
    Plots of the tangential pressure, (a) $p_t(r)$ and (b) $4\pi\,r^2\,p_t(r)$,
    using the multipole model~\eqref{eq:model} with parameters given in Table~\ref{tab:model}, see Eq.~\eqref{eq:pt} or Eq.~\eqref{eq:ptbis} for definitions in terms of GFFs.
    }
\end{figure}

\begin{figure}[t!]
    \centering
    \includegraphics[width=0.49\textwidth]{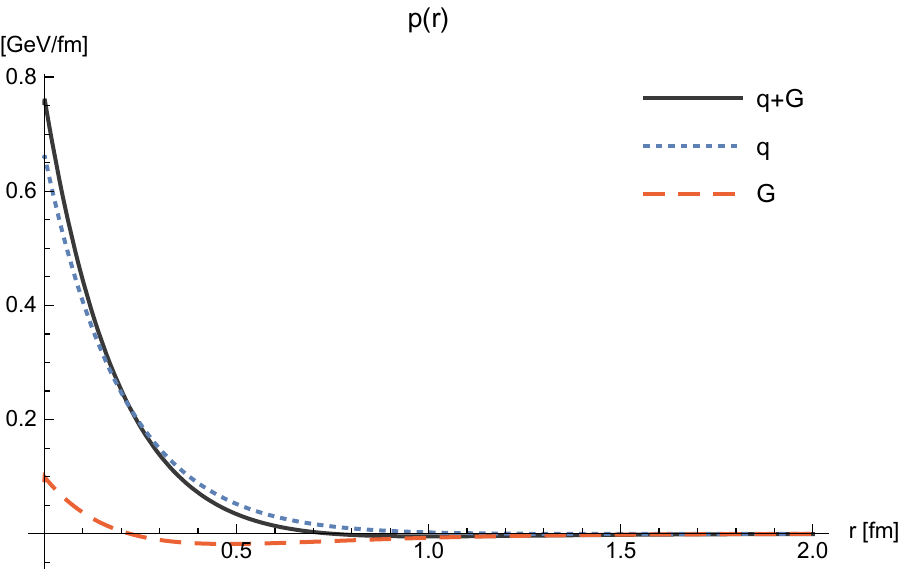}
    \hspace{0.5em}
    \includegraphics[width=0.49\textwidth]{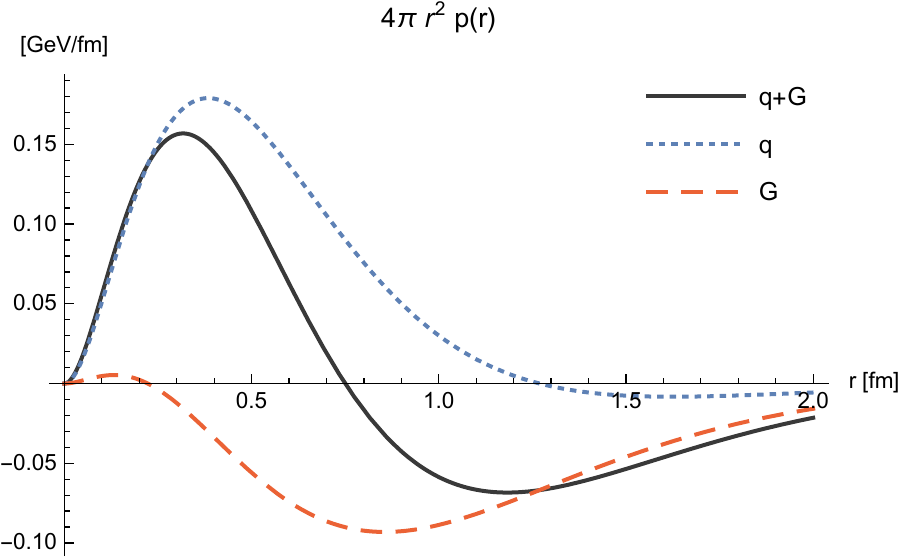}
    \\
	(a)\hspace*{0.45\linewidth}(b)
\caption{\label{fig:p}
    Plots of the isotropic pressure, (a) $p(r)$ and (b) $4\pi\,r^2\,p(r)$,
    using the multipole model~\eqref{eq:model} with parameters given in Table~\ref{tab:model}, see Eq.~\eqref{eq:p} or Eq.~\eqref{eq:pbis} for definitions in terms of GFFs.
    }
\end{figure}

\begin{figure}[t!]
    \centering
    \includegraphics[width=0.49\textwidth]{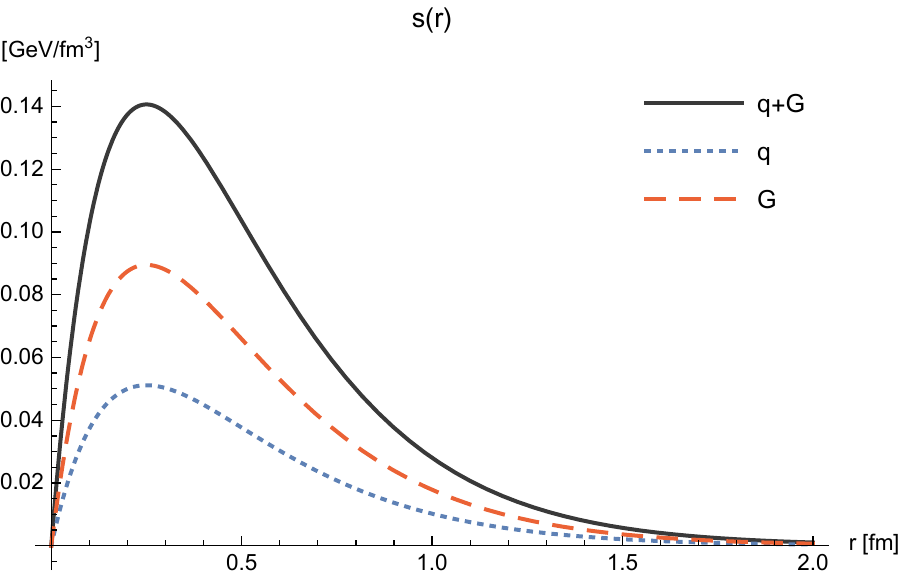}
    \hspace{0.5em}
    \includegraphics[width=0.49\textwidth]{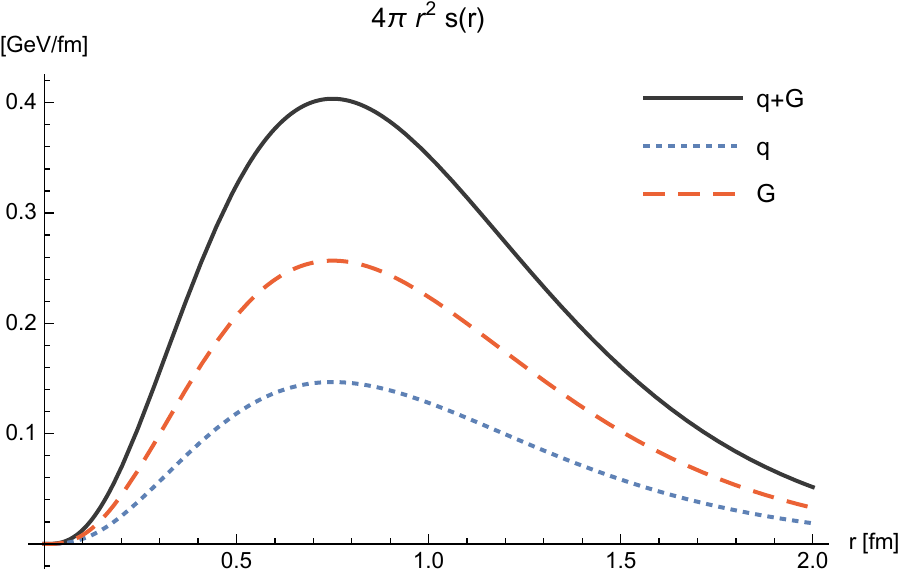}
    \\
	(a)\hspace*{0.45\linewidth}(b)
    \caption{\label{fig:s}
    Plots of the pressure anisotropy, (a) $s(r)$ and (b) $4\pi\,r^2\,s(r)$,
    using the multipole model~\eqref{eq:model} with parameters given in Table~\ref{tab:model}, see Eq.~\eqref{eq:s} or Eq.~\eqref{eq:sbis} for definitions in terms of GFFs.
    }
\end{figure}

The above distributions are illustrated in Figs.~\ref{fig:epsilon}-\ref{fig:s} in units of GeV/fm$^3=1.7827\times 10^{15}$ g/cm$^3$ using the multipole model~\eqref{eq:model} with parameters given in Table~\ref{tab:model}. The energy density in Fig.~\ref{fig:epsilon} is always positive and is approximately shared equally between quark and gluon contributions. One defines the corresponding average squared mass radius as
\begin{equation}
R^2_M=\frac{1}{M}\int\ud^3 \uvec r\,r^2\,\varepsilon(r)=6\left[\frac{\ud A(t)}{\ud t}\Big|_{t=0}-\frac{1}{M^2}\,C(0)\right].
\end{equation}
In our simple model, we find $R_M=0\,.905$ fm which is a bit larger than the charge radius $R_Q=0\,.841$ fm extracted from muonic hydrogen spectroscopy~\cite{Pohl:2010zza,Pohl1:2016xoo} and $R_Q=0\,.879$ fm extracted from electron-proton scattering~\cite{Arrington:2015yxa}. Knowing the distribution of energy density, it is also easy to derive the standard mass function widely used in General Relativity
\begin{equation}\label{massfunc}
m(r)=4\pi\int_0^r\ud r'\,r'^2\,\varepsilon(r')
\end{equation}
which represents the mass contained within a sphere of radius $r$, see Fig.~\ref{fig:m}. 

While the total radial pressure in Fig.~\ref{fig:pr} is always positive and largely dominated by the quark contribution, the total tangential and isotropic pressures in Figs.~\ref{fig:pt} and \ref{fig:p} switch from positive sign at the center of the nucleon (where it is dominated by the quark contribution) to negative sign at the periphery (where it is dominated by the gluon contribution). The pressure anisotropy in Fig.~\ref{fig:s} vanishes at the center of the nucleon, as required by spherical symmetry, and is positive anywhere else, indicating that the radial pressure is always larger than the tangential one. Looking at the separate contributions, we see that the quark and gluon radial forces are both repulsive and of similar range. For the tangential forces, the quark contribution appears to be mostly repulsive and short range whereas the gluon contribution appears to be mostly attractive and long range.

If we integrate the energy density and the isotropic pressure over the whole volume, we naturally recover the FL~\eqref{eq:epsilon_p} discussed in the former section
\begin{equation}
\int\ud^3 \uvec r\,\varepsilon_a(r)=\left[A_a(0)+\bar C_a(0)\right]M,
\qquad
\int\ud^3 \uvec r\,p_a(r)=-\bar C_a(0)\,M.
\end{equation}
One can also relate the value of the GFF $C_a(t)$ at $t=0$ to a weighted integral of the pressure anisotropy~\eqref{eq:sbis}~\cite{Polyakov:2002yz,Goeke:2007fp,Polyakov:2018zvc}
\begin{equation}
\label{eq:sCrel}
\int\ud^3 \uvec r\,r^2\,s_a(r)=-\frac{15}{M}\,C_a(0)\,.
\end{equation}
Summing over the constituents, one obtains the following additional relations~\cite{Goeke:2007fp,Perevalova:2016dln}
\begin{equation}
\label{eq:C0}
\int\ud^3 \uvec r\,r^2 p_r(r)=-\frac{6}{M}\,C(0)\,,\qquad \int\ud^3 \uvec r\,r^2 p_t(r)=\frac{9}{M}\,C(0)\,,\qquad\int\ud^3 \uvec r\,r^2 p(r)=\frac{4}{M}\,C(0)\,.
\end{equation}
Several hints suggest that $C(0)$ is likely negative~\cite{Goeke:2007fp, Hudson:2017xug, Hudson:2017oul, Polyakov:2018zvc}. Hudson and Schweitzer~\cite{Hudson:2017xug} observed that while adding total derivatives to the EMT leaves the Poincar\'e generators unaffected (and hence the particle mass and spin), it does change $C(0)$. Since this term can be extracted from experimental data, one should not be allowed to add these divergence terms without changing some scheme prescriptions, contrary to the common belief. The same conclusion is reached when one consistently treats intrinsic angular momentum at the level of spatial distributions~\cite{Lorce:2017wkb,Lorce:2018zpf}. 



\begin{figure}[t!]
    \centering
    \includegraphics[width=0.49\textwidth]{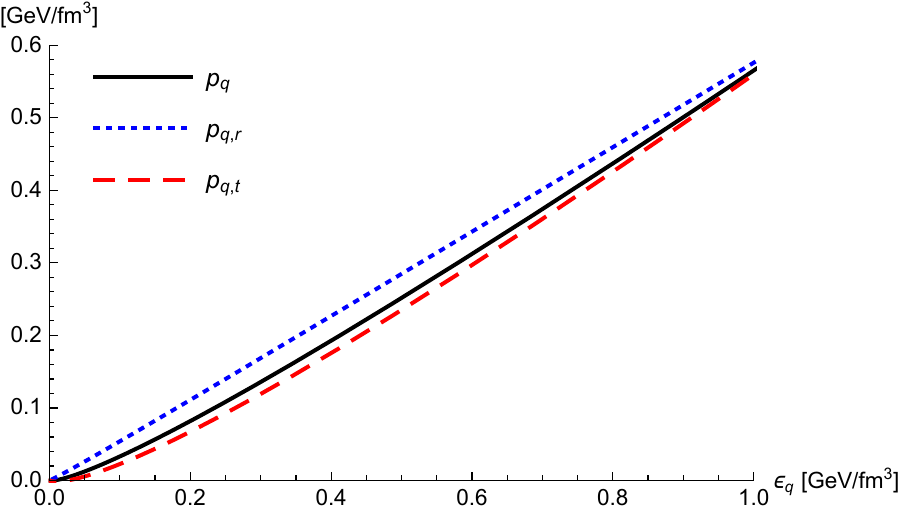}
    \hspace{0.5em}
    \includegraphics[width=0.49\textwidth]{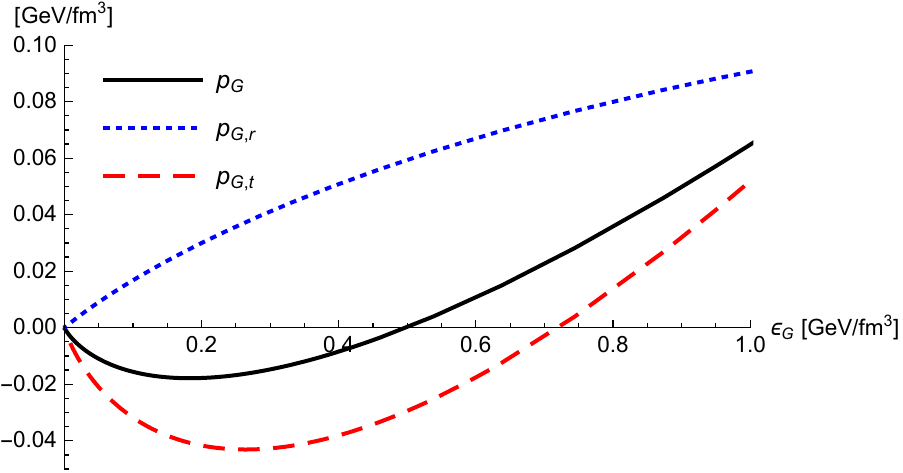}
     \\
	(a)\hspace*{0.45\linewidth}(b)
    \caption{Radial $p_r(\varepsilon)$, tangential $p_t(\varepsilon)$, and isotropic $p(\varepsilon)$ EoS for (a) quarks and (b) gluons computed using the multipole model~\eqref{eq:model} with parameters given in Table~\ref{tab:model}, see Eqs.~\eqref{eq:epsilon}-\eqref{eq:p} for definitions.}
    \label{fig:EoS_q_G}
\end{figure}

\begin{figure}[t!]
    \centering
    \includegraphics[width=0.49\textwidth]{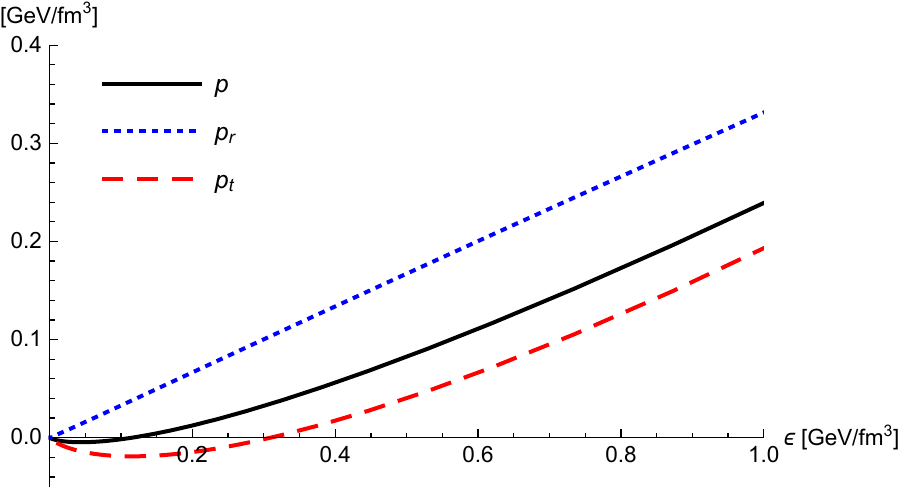}
    \caption{Radial $p_r(\varepsilon)$, tangential $p_t(\varepsilon)$, and isotropic $p(\varepsilon)$ EoS of the nucleon computed using the multipole model~\eqref{eq:model} with parameters given in Table~\ref{tab:model}, see Eqs.~\eqref{eq:epsilon}-\eqref{eq:p} for definitions.}
    \label{fig:EoS}
\end{figure}

Interestingly, in view of the energy density and pressure conditions encountered in a nucleon, one may conjecture that studies of the nucleon EMT will shed some light on the EoS inside compact stars~\cite{Lorce:2017xzd}, which so far remains largely unknown~\cite{Ozel:2016oaf}, and will therefore complement efforts based on heavy-ion collisions~\cite{Danielewicz:2002pu} and gravitational wave observations~\cite{Rezzolla:2016nxn,Annala:2017llu,Paschalidis:2017qmb,Most:2018hfd,Nandi:2018ami,Urbano:2018nrs}. Eliminating the radial variable $r$ in Eqs.~\eqref{eq:epsilon}-\eqref{eq:p}, we obtain the nucleon EoS for radial pressure $p_r(\varepsilon)$, tangential pressure $p_t(\varepsilon)$, and isotropic pressure $p(\varepsilon)$. The results plotted in Figs.~\ref{fig:EoS_q_G} and \ref{fig:EoS} show a pretty stiff behavior compatible with the observation of supermassive ($\sim 2\,M_\odot$) compact stars~\cite{Demorest:2010bx,Antoniadis:2013pzd,Annala:2017llu} well above the Chandrasekhar mass limit $1.44\,M_\odot$~\cite{Chandrasekhar:1931ih}. Although our multipole model is very naive, it supports the idea of an exciting crosstalk between hadronic physics and compact stars. An example of such a connection is given by the use of hadronic models to study the EoS of potential quark matter inside compact stars~\cite{Baym:1976yu,Haensel:1986qb,Alcock:1986hz,Drago:1995ah,RikovskaStone:2006ta,Stone:2010jt,Motta:2018rxp,Deb:2018,Abhishek:2018xml,Jokela:2018ers}.

\subsection{Elastic frame}

\begin{figure}[t!]
	\centering
	\includegraphics[trim={0 1.5cm 0 0}, clip,width=0.5\linewidth]{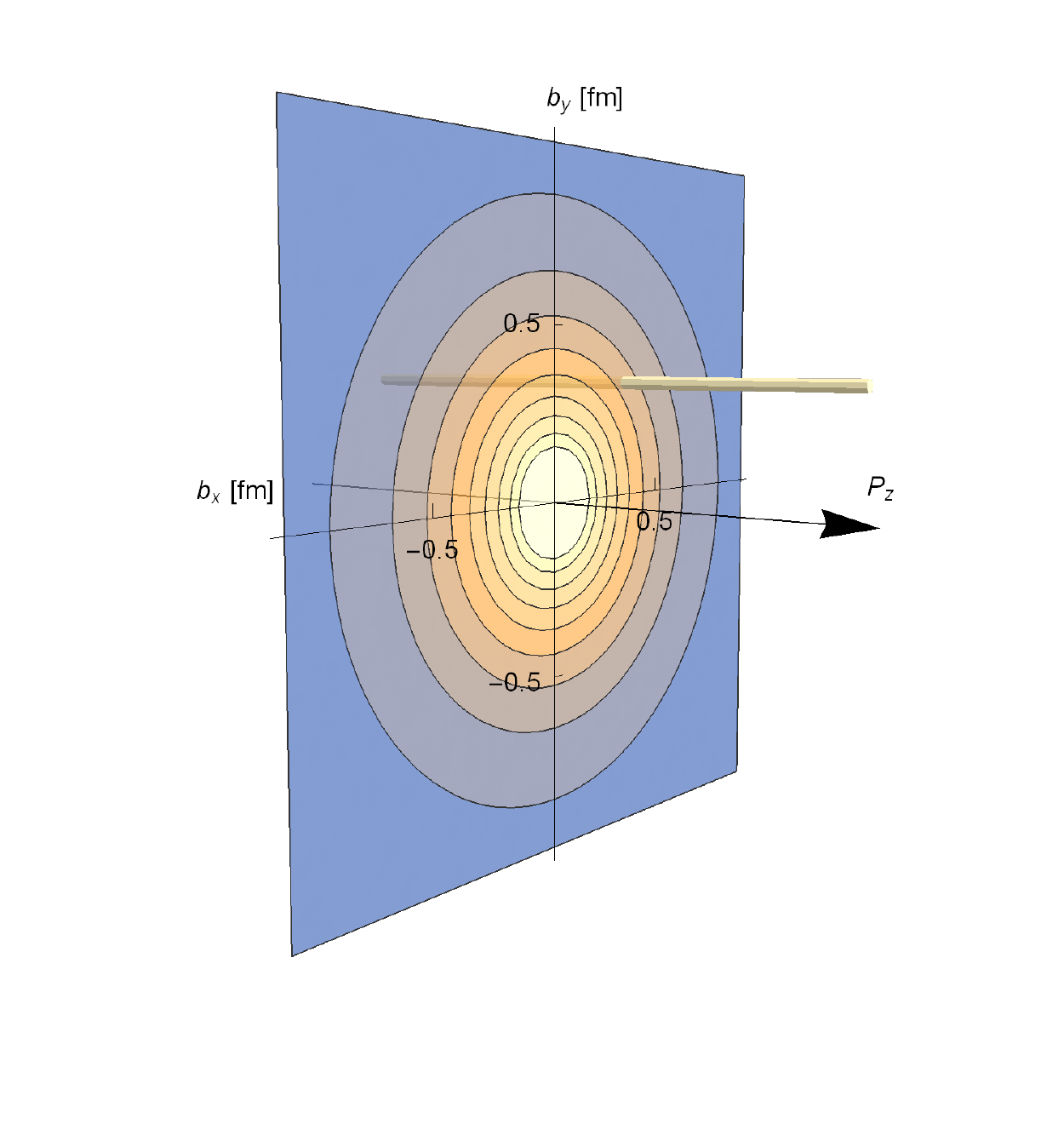}
	\caption{
	Density plot of the two-dimensional energy density $\rho(\uvec b_\perp)$
    using the multipole model~\eqref{eq:model} with parameters given in Table~\ref{tab:model}, see Eq.~\eqref{eq:rho} or Eq.~\eqref{eq:rhobis} for the definition in terms of GFFs.
	The cylinder parallel to $\uvec P=P_z\,\uvec e_z$ illustrates the integration
	over $r^\parallel$, see Eq.~\eqref{eq:int3}. }
	\label{fig:plain}
\end{figure}

Spatial distributions with quasi-probabilistic interpretation can also be introduced when $\uvec P\neq \uvec 0$~\cite{Lorce:2017wkb}. In order to maintain the condition $\Delta^0=0$, we have to restrict ourselves to the set of elastic frames (EF) defined by $\uvec P\cdot\uvec\Delta=0$. They can be obtained by integrating the static EMT over the longitudinal coordinate $r^\parallel=\uvec r\cdot\uvec P/|\uvec P|$
\begin{equation}
\label{eq:int3}
\mathsf T^{\mu\nu}_a(\uvec b_\perp;\uvec P)=\int\ud r^\parallel\,\mathcal T^{\mu\nu}_a(\uvec r;\uvec P)=\int\frac{\ud^2\uvec\Delta_\perp}{(2\pi)^2}\,e^{-i\uvec\Delta_\perp\cdot\uvec b_\perp}\,\langle\!\langle T^{\mu\nu}_a(0)\rangle\!\rangle\big|_\text{EF},
\end{equation}
where $\uvec\Delta_\perp$ and $\uvec b_\perp=\uvec r_\perp$ are vectors lying in the two-dimensional plane orthogonal to $\uvec P$, see Fig.~\ref{fig:plain}.
The unpolarized off-forward amplitude~\eqref{off-forward} in the EF takes the form
\begin{align}\label{eq:EMT_EF}
\langle\!\langle T^{\mu\nu}_a(0)\rangle\!\rangle\big|_\text{EF}&=\left[1-\frac{\uvec P^2}{P^0(P^0+M)}\right]\!\left\{\frac{P^\mu P^\nu}{M}\,A_a(t)
+ \frac{\Delta^\mu_\perp\Delta^\nu_\perp +\eta^{\mu\nu}\uvec \Delta^2_\perp}{M}\, C_a(t)
+ M \eta^{\mu\nu}\bar C_a(t)\right\}\\
&-\frac{\uvec\Delta^2_\perp}{4P^0(P^0+M)}\left\{\left[\frac{2P^\mu P^\nu}{M}+P^{\{\mu}\eta^{\nu\}0}\right]\frac{A_a(t)+B_a(t)}{2}+P^{[\mu}\eta^{\nu]0}\,\frac{D_a(t)}{2}\right\}\,\nn,
\end{align}
where $P^0=\sqrt{\uvec P^2+\frac{\uvec\Delta^2_\perp}{4}+M^2}$ and $\Delta^0_\perp=\Delta^\parallel_\perp=0$. If we further integrate the static EMT over the impact-parameter $\uvec b_\perp$, which amounts to setting $\uvec\Delta_\perp=\uvec 0_\perp$ in Eq.~\eqref{eq:EMT_EF}, we recover the FL expression~\eqref{eq:TmunuFL}. If we set $\uvec P=\uvec 0$ in Eq.~\eqref{eq:EMT_EF}, we recover the BF expression~\eqref{BFFA} integrated over $r^\parallel$, i.e. with $\Delta^\parallel=0$. 

Since $P^0$ depends on $\uvec\Delta_\perp$ in the EF, we could not find simple expressions for the spatial distributions $\mathsf T^{\mu\nu}_a(\uvec b_\perp;\uvec P)$ in terms of Fourier transforms of GFFs. Moreover, these spatial distributions will be $|\uvec P|$-dependent. Let us choose for convenience the $z$-axis along $\uvec P$. If we restrict ourselves to the (1+2)-dimensional (or transverse) static EMT\footnote{Since the longitudinal coordinate is integrated over, it follows that the total transverse EMT is itself conserved $\partial_\alpha \mathsf T^{\alpha\beta}(\uvec b_\perp;\uvec P)=0$.} with components $\alpha,\beta\in\{0,1,2\}$, its structure looks the same\footnote{Note that the proper identification $t^{\alpha\beta}\sim\gamma\,\mathsf T^{\alpha\beta}$ involves a boost factor  accounting for the Lorentz contraction of the volume just like in the FL, see Eq.~\eqref{eq:identFL}.} as that of an anisotropic axially symmetric compact star in two dimensions 
\begin{equation}\label{2DEF}
t^{\alpha\beta}(\uvec b_\perp;P_z)=[\gamma^2\rho(b,P_z)+\sigma_t(b,P_z)]\,v^\alpha v^\beta - \sigma_t(b,P_z)\eta^{\alpha\beta}+[\sigma_r(b,P_z)-\sigma_t(b,P_z)]\,\chi^\alpha \chi^\beta\,,
\end{equation}
where $v^\alpha=(1,\uvec 0_\perp)$ and $\chi^\alpha=(0\,,\uvec b_\perp/b)$. The functions $\rho(b,P_z)$, $\sigma_r(b,P_z)$ and $\sigma_t(b,P_z)$ represent the two-dimensional version of energy density, radial pressure and tangential pressure, respectively. Note that we have included explicitly a factor of $\gamma^2$ in the energy density component that comes from the longitudinal Lorentz boost $t^{00}(\uvec b_\perp;P_z)=\rho(b,P_z) \,u^0u^0=\gamma^2\rho(b,P_z)\, v^0v^0$, allowing us to compare directly $\rho(b,P_z)$ for different values of $P_z$. In particular, for the total EMT we have $\int\ud^2b_\perp\,\rho(b,P_z)=M$. The tensor $t^{\alpha\beta}$ can alternatively be written as
\begin{equation}\label{2DEFbis}
t^{\alpha\beta}(\uvec b_\perp;P_z)=[\gamma^2\rho(b,P_z)+\sigma(b,P_z)]\,v^\alpha v^\beta - \sigma(b,P_z)\eta^{\alpha\beta}+\Pi(b,P_z)\left(\chi^\alpha \chi^\beta-\frac{1}{2}\,l^{\alpha\beta}\right)
\end{equation}
with $l^{\alpha\beta}=v^\alpha v^\beta-\eta^{\alpha\beta}$. The two-dimensional isotropic pressure $\sigma(b,P_z)$ and pressure anisotropy $\Pi(b,P_z)$ are related to radial and tangential pressures as follows
\begin{equation}
\sigma(b,P_z)=\frac{\sigma_r(b,P_z)+\sigma_t(b,P_z)}{2}\,,\qquad \Pi(b,P_z)=\sigma_r(b,P_z)-\sigma_t(b,P_z)\,.
\end{equation}

The particular case $P_z=0$ is simple and corresponds to the BF. We find
\begin{align}\label{eq:TmunuEF}
\mathsf T^{\alpha\beta}_a(\uvec b_\perp;\uvec 0)=M&\left\{\eta^{\alpha0}\eta^{\beta0}\left[\mathsf A_a(b)+\frac{1}{4M^2}\frac{1}{b}\frac{\ud}{\ud b}\!\left(b\,\frac{\ud\mathsf B_a(b)}{\ud b}-4\mathsf C_a(b)\right)\right]\right.\\
&\left. +\eta^{\alpha\beta} \left[ \bar{\mathsf{C}}_a(b) - \frac{1}{M^2}\frac{\ud^2\mathsf C_a(b)}{\ud b^2}\right]-\frac{x^\alpha x^\beta}{b^2}\, \frac{1}{M^2}\,b\,\frac{\ud}{\ud b}\!\left(\frac{1}{b}\frac{\ud\mathsf C_a(b)}{\ud b}\right)\right\}\nn\,,
\end{align}
where $x^\alpha=(0,\uvec b_\perp)$ and the two-dimensional Fourier transforms of GFFs are denoted by
\begin{equation}
\mathsf F_a(b)=\int\frac{\ud^2\uvec\Delta_\perp}{(2\pi)^2}\,e^{-i\uvec\Delta_\perp\cdot\uvec b_\perp}\,F_a(t)
\end{equation}
with $t=-\uvec\Delta^2_\perp$. Comparing Eq.~\eqref{eq:TmunuEF} with the EMT of an anisotropic axially symmetric compact star in two dimensions~\eqref{2DEF} and~\eqref{2DEFbis} suggests that the following combinations
\begin{align}
\label{eq:rho}
\rho_a(b,0) &= M\left\{\mathsf A_a(b)+\bar{\mathsf C}_a(b) + \frac{1}{4M^2}\frac{1}{b}\frac{\ud}{\ud b}\!\left(b\,\frac{\ud}{\ud b}\!\left[\mathsf B_a(b)-4\mathsf C_a(b)\right]\right)\right\},\\
\label{eq:sigmar}
\sigma_{r,a}(b,0) &= M\left\{-\bar{\mathsf{C}}_a(b)+ \frac{1}{M^2}\frac{1}{b}\frac{\ud\mathsf C_a(b)}{\ud b} \right\},
\\
\label{eq:sigmat}
\sigma_{t,a}(b,0) &= M\left\{-\bar{\mathsf{C}}_a(b) + \frac{1}{M^2}\frac{\ud^2\mathsf C_a(b)}{\ud b^2} \right\},\\
\label{eq:sigma}
\sigma_a(b,0) &= M\left\{-\bar{\mathsf{C}}_a(b)+\frac{1}{2M^2}\frac{1}{b}\frac{\ud}{\ud b}\!\left(b\,\frac{\ud\mathsf C_a(b)}{\ud b}\right) \right\},
\\
\label{eq:Pi}
\Pi_a(b,0) &= M\left\{-\frac{1}{M^2}\,b\,\frac{\ud}{\ud b}\!\left(\frac{1}{b}\frac{\ud\mathsf C_a(b)}{\ud b}\right) \right\},
\end{align}
can be interpreted as the two-dimensional partial energy density, radial pressure, tangential pressure, isotropic pressure, and pressure anisotropy associated with constituent type $a$. They can alternatively be written as
\begin{align}
\label{eq:rhobis}
\rho_a(b,0) &= M\int\frac{\ud^2\uvec\Delta_\perp}{(2\pi)^2}\,e^{-i\uvec\Delta_\perp\cdot\uvec b_\perp}\left\{A_a(t)+\bar{C}_a(t) + \frac{t}{4M^2}\left[B_a(t)-4 C_a(t)\right]\right\}\,,\\
\label{eq:sigmarbis}
\sigma_{r,a}(b,0) &=M\int\frac{\ud^2\uvec\Delta_\perp}{(2\pi)^2}\,e^{-i\uvec\Delta_\perp\cdot\uvec b_\perp}\left\{-\bar C_a(t)-\frac{2}{b^2}\frac{1}{M^2}\frac{\ud}{\ud t}\!\left[t\,C_a(t)\right]\right\}\,,
\\
\label{eq:sigmatbis}
\sigma_{t,a}(b,0) &= M\int\frac{\ud^2\uvec\Delta_\perp}{(2\pi)^2}\,e^{-i\uvec\Delta_\perp\cdot\uvec b_\perp}\left\{-\bar C_a(t)+\frac{1}{b^2}\frac{4}{M^2}\frac{\ud}{\ud t}\!\left[t^{1/2}\frac{\ud}{\ud t}\!\left(t^{3/2}\,C_a(t)\right)\right]\right\}\,,\\
\label{eq:sigmabis}
\sigma_a(b,0) &= M\int\frac{\ud^2\uvec\Delta_\perp}{(2\pi)^2}\,e^{-i\uvec\Delta_\perp\cdot\uvec b_\perp}\left\{-\bar{C}_a(t) +\frac{1}{2}\frac{t}{M^2}\,C_a(t)\right\}\,,
\\
\label{eq:Pibis}
\Pi_a(b,0) &=M\int\frac{\ud^2\uvec\Delta_\perp}{(2\pi)^2}\,e^{-i\uvec\Delta_\perp\cdot\uvec b_\perp}\left\{-\frac{1}{b^2}\frac{4}{M^2}\frac{\ud^2}{\ud t^2}\!\left[t^2C_a(t)\right]\right\}\,.
\end{align}
Integrating Eq.~\eqref{eq:TmunuRF} over $z$ allows us to alternatively express these quantities in terms of the three-dimensional energy density and pressures as follows
\begin{align}
\rho_a(b,0)&=\int\ud z\,\varepsilon_a(r)\,,\\
\sigma_{r,a}(b,0)&=\int\ud z\,\frac{b^2p_{r,a}(r)+z^2p_{t,a}(r)}{r^2}\,,\label{2Dsigmar}\\
\sigma_{t,a}(b,0)&=\int\ud z\,p_{t,a}(r)\,,\\
\sigma_a(b,0)&=\int\ud z\left[p_a(r)+\frac{b^2-2z^2}{6 r^2}\,s_a(r)\right],\\
\Pi_a(b,0)&=\int\ud z\,\frac{b^2}{r^2}\,s_a(r)\,,
\end{align}
with $r=\sqrt{b^2+z^2}$.

\begin{figure}[t!]
    \centering
    \includegraphics[width=0.49\textwidth]{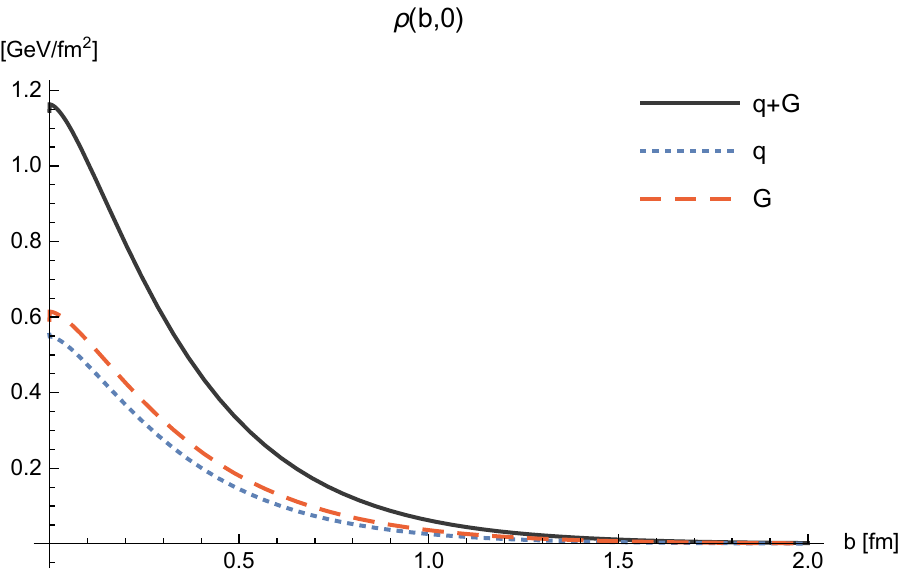}
    \hspace{0.5em}
    \includegraphics[width=0.49\textwidth]{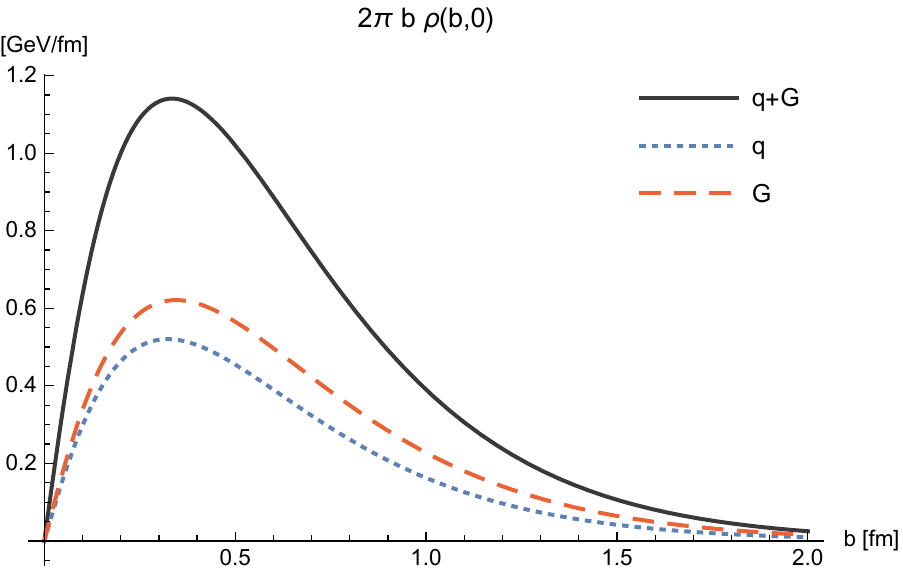}
    \\
	(a)\hspace*{0.45\linewidth}(b)
    \caption{\label{fig:rho}
    Plots of the two-dimensional energy density, (a) $\rho(b,0)$ and (b) $2\pi\,b\,\rho(b,0)$, using the multipole model~\eqref{eq:model} with parameters given in Table~\ref{tab:model}, see Eq.~\eqref{eq:rho} or Eq.~\eqref{eq:rhobis} for the definition in terms of GFFs.
    }
\end{figure}

\begin{figure}[t!]
    \centering
    \includegraphics[width=0.49\textwidth]{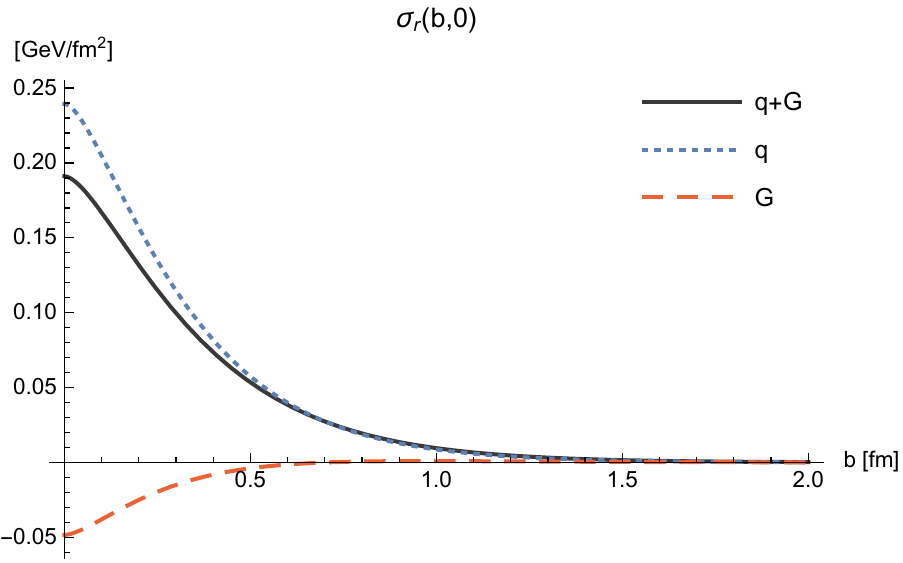}
    \hspace{0.5em}
    \includegraphics[width=0.49\textwidth]{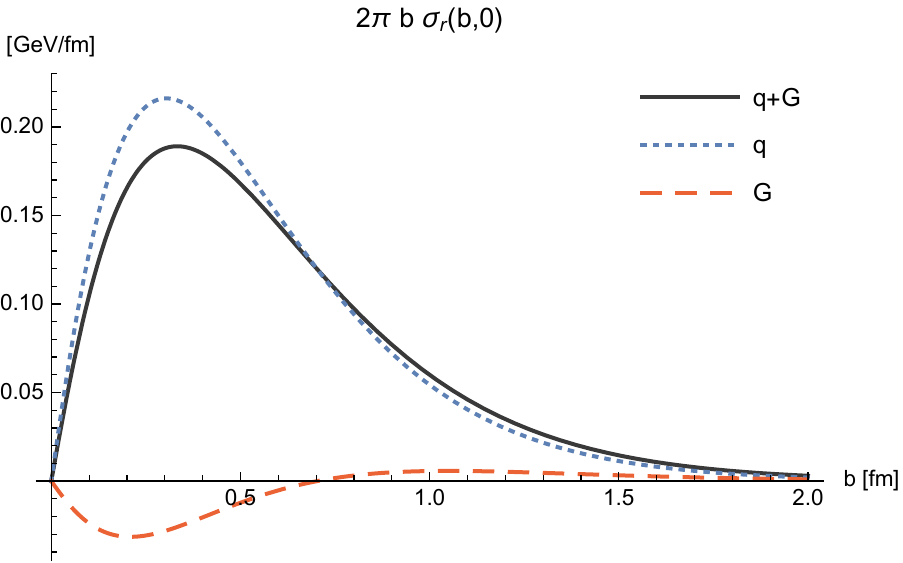}
    \\
	(a)\hspace*{0.45\linewidth}(b)
    \caption{\label{fig:sigmar}
    Plots of the two-dimensional radial pressure, (a) $\sigma_r(b,0)$ and (b) $2\pi\,b\,\sigma_r(b,0)$, using the multipole model~\eqref{eq:model} with parameters given in Table~\ref{tab:model}, see Eq.~\eqref{eq:sigmar} or Eq.~\eqref{eq:sigmarbis} for the definition in terms of GFFs.
    }
\end{figure}

\begin{figure}[t!]
    \centering
    \includegraphics[width=0.49\textwidth]{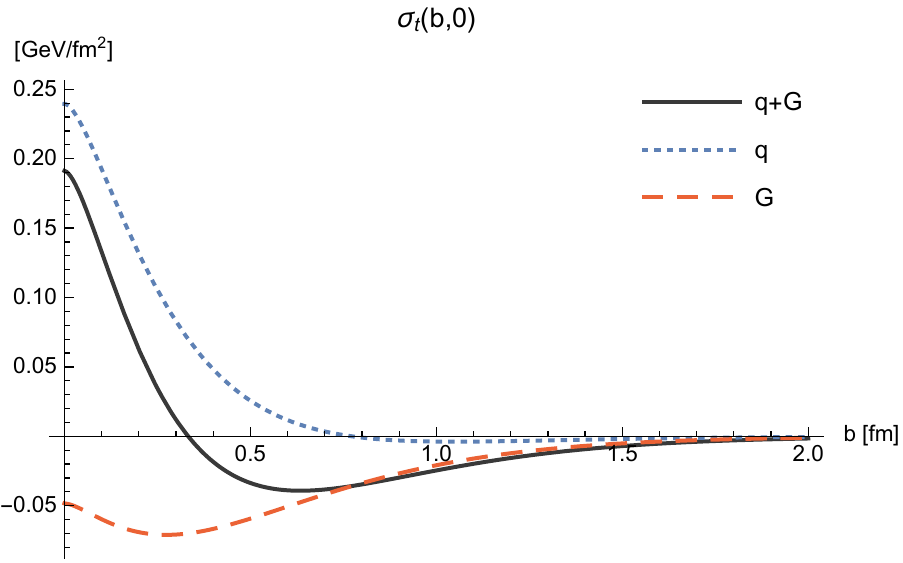}
    \hspace{0.5em}
    \includegraphics[width=0.49\textwidth]{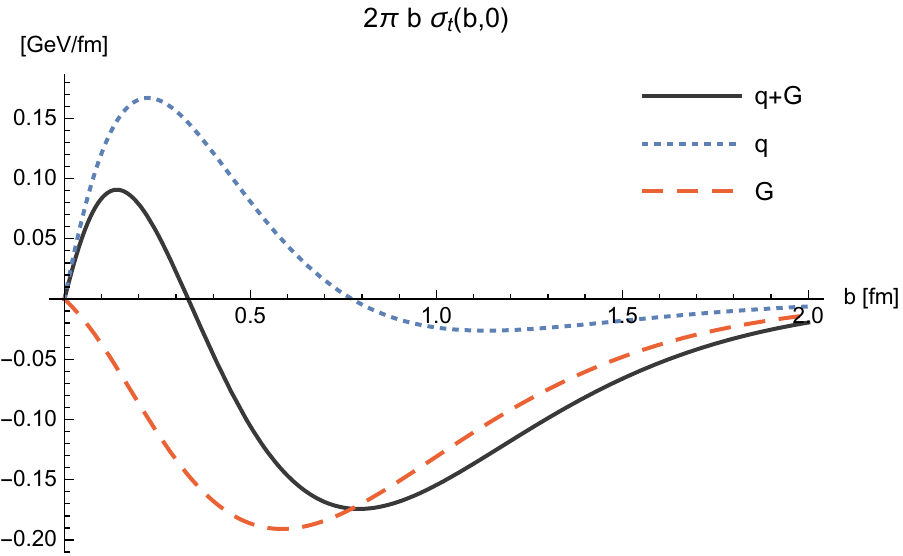}
    \\
	(a)\hspace*{0.45\linewidth}(b)
	\caption{\label{fig:sigmat}
    Plots of the two-dimensional tangential pressure, (a) $\sigma_t(b,0)$ and (b) $2\pi\,b\,\sigma_t(b,0)$, using the multipole model~\eqref{eq:model} with parameters given in Table~\ref{tab:model}, see Eq.~\eqref{eq:sigmat} or~\eqref{eq:sigmatbis} for the definition in terms of GFFs.
    }
    \vspace{2em} 
\end{figure}

\begin{figure}[t!]
    \centering
    \includegraphics[width=0.49\textwidth]{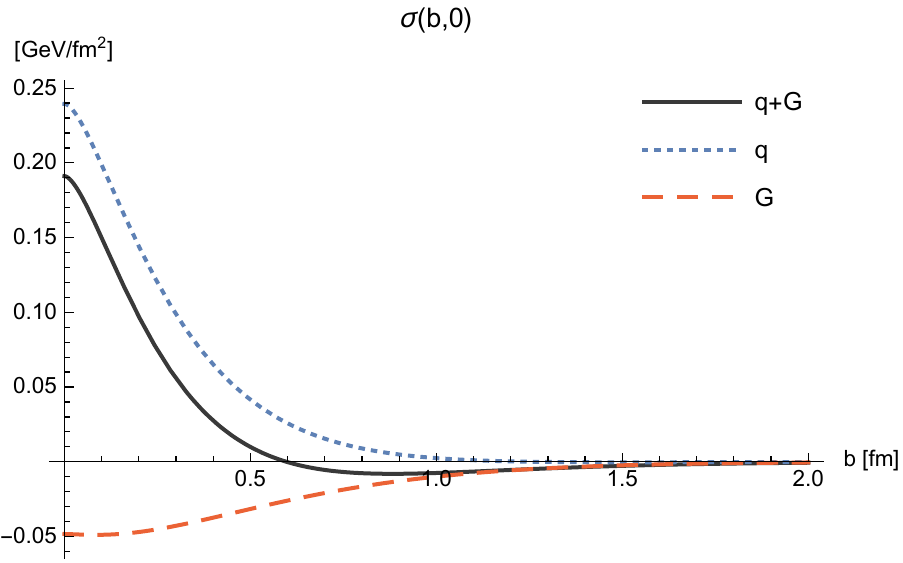}
    \hspace{0.5em}
    \includegraphics[width=0.49\textwidth]{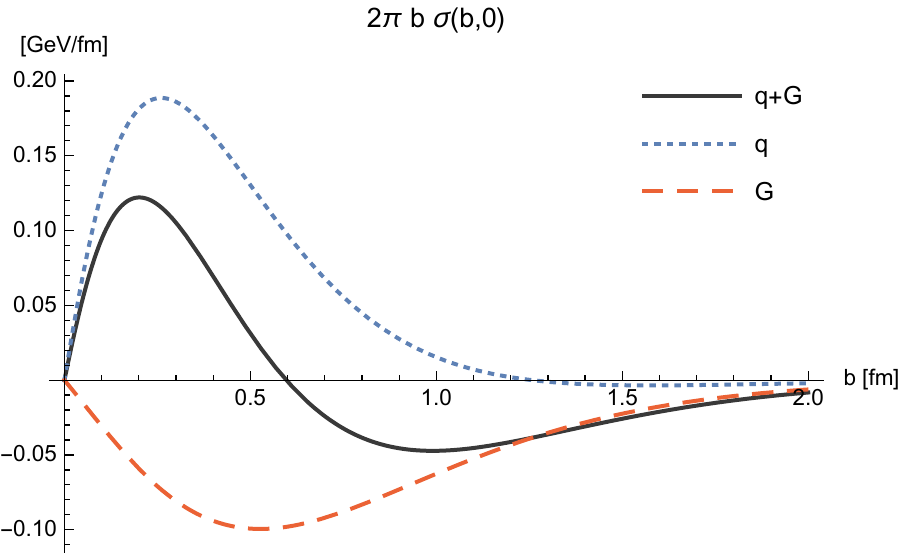}
    \\
	(a)\hspace*{0.45\linewidth}(b)
     \caption{\label{fig:sigma}
    Plots of the two-dimensional isotropic pressure, (a) $\sigma(b,0)$ and (b) $2\pi\,b\,\sigma(b,0)$, using the multipole model~\eqref{eq:model} with parameters given in Table~\ref{tab:model}, see Eq.~\eqref{eq:sigma} or~\eqref{eq:sigmabis} for the definition in terms of GFFs. 
    }
\end{figure}

\begin{figure}[t!]
    \centering
    \includegraphics[width=0.49\textwidth]{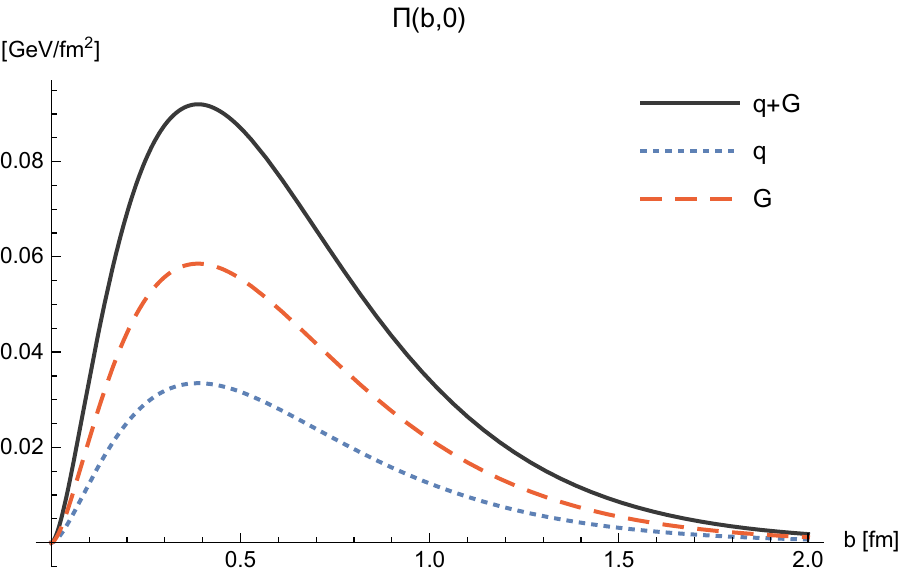}
    \hspace{0.5em}
    \includegraphics[width=0.49\textwidth]{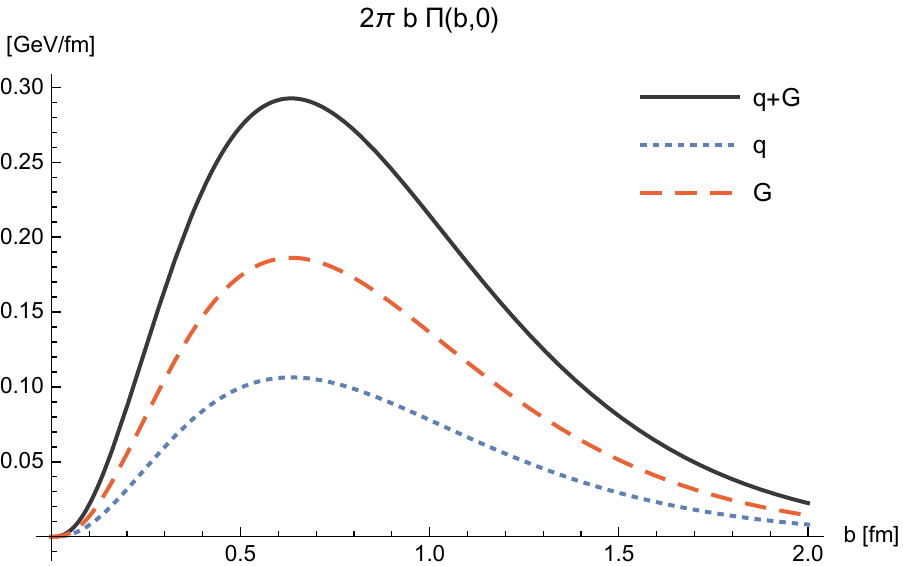}
    \\
	(a)\hspace*{0.45\linewidth}(b)
    \caption{\label{fig:Pi}
    Plots of the two-dimensional pressure anisotropy, (a) $\Pi(b,0)$ and (b) $2\pi\,b\,\Pi(b,0)$, using the multipole model~\eqref{eq:model} with parameters given in Table~\ref{tab:model}, see Eq.~\eqref{eq:Pi} or~\eqref{eq:Pibis} for the definition in terms of GFFs.
    }
\end{figure}

These distributions are illustrated in Figs.~\ref{fig:rho}--\ref{fig:Pi}
in units of GeV/fm$^2=178.27$ g/cm$^2$ using the multipole model~\eqref{eq:model} with parameters given in Table~\ref{tab:model}. Their behavior turns out to be similar to the corresponding three-dimensional distributions, see Figs.~\ref{fig:epsilon}-\ref{fig:s}, except for the gluon contribution to the radial pressure, compare Figs.~\ref{fig:pr} and~\ref{fig:sigmar}. This is an effect of the projection onto the transverse plane which mixes three-dimensional radial and tangential pressures together, as indicated by Eq.~\eqref{2Dsigmar}. While the latter have the same sign for quarks, they are of opposite sign for gluons. 


Similarly to the three-dimensional case, we can relate the value of the GFF $C_a(t)$ at $t=0$ to a weighted integral of the pressure anisotropy
\begin{equation}
\int\ud^2\uvec b_\perp\,b^2\,\Pi_a(b,0)=-\frac{8}{M}\,C_a(0)\,.
\end{equation}
Summing over the constituents, we also find the additional relations
\begin{equation}\label{eq:C0bis}
\int\ud^2\uvec b_\perp\,b^2\sigma_r(b,0)=-\frac{2}{M}\,C(0)\,,\qquad \int\ud^2\uvec b_\perp\,b^2\sigma_t(b,0)=\frac{6}{M}\,C(0)\,,\qquad \int\ud^2\uvec b_\perp\,b^2\sigma(b,0)=\frac{2}{M}\,C(0)\,,
\end{equation}
which are simply the two-dimensional version of those appearing in Eq.~\eqref{eq:C0}.
\newline

The last term in Eq.~\eqref{eq:EMT_EF} makes the EMT asymmetric. In particular, the density of longitudinal momentum $\mathcal T^{03}_a$ is not equal to the longitudinal flux of energy $\mathcal T^{30}_a$ when the GFF $D_a(t)$ is not identically zero. Using Poincar\'e invariance, the antisymmetric part of the EMT $T^{[\mu\nu]}$ can be expressed in terms of the intrinsic angular momentum current $S^{\lambda\mu\nu}$ as follows~\cite{Leader:2013jra,Lorce:2017wkb}
\begin{equation}
T^{[\mu\nu]}(x)=-\partial_\lambda S^{\lambda\mu\nu}(x)\,.
\end{equation}
In the case of QCD, we get for the corresponding off-forward matrix elements
\begin{equation}
\langle\!\langle T^{[\mu\nu]}_q(0)\rangle\!\rangle=-\frac{i}{2}\,\epsilon^{\mu\nu\Delta\lambda}\,\langle\!\langle \overline\psi(0)\gamma_\lambda\gamma_5\psi(0)\rangle\!\rangle\,,
\end{equation} 
where the matrix elements of the axial-vector current are parametrized as follows
\begin{equation}
\bra{p',\uvec s} \overline\psi(0)\gamma^\mu\gamma_5\psi(0)\ket{p,\uvec s}=
\bar u(p',\uvec s)\left[
\gamma^\mu\gamma_5\, G^q_A(t) + \frac{\Delta^\mu\gamma_5}{2M}\,G^q_P(t)
\right]u(p,\uvec s)
\end{equation}
in terms of the axial-vector and induced pseudoscalar form factors. According to Ref.~\cite{Lorce:2017isp}, the corresponding bilinears can be expressed in instant form as
\begin{align}
\bar u'\gamma_5u&=\mathcal N^{-1}\left[(P^0+M)\,(\Delta\cdot S)-\Delta^0(P\cdot S)\right],\\
\bar u'\gamma^\mu\gamma_5 u&=\mathcal N^{-1}\left\{2\left[S^\mu(P^2+P^0M)-(P^\mu+\eta^{\mu0}M)\,(P\cdot S)+\frac{\Delta^\mu(\Delta\cdot S)}{4}\right]+i\epsilon^{\mu P\Delta 0}\right\}\,.
\end{align}
For an unpolarized target $S^\mu=0$, we can write
\begin{equation}
\langle\!\langle \overline\psi(0)\gamma^\mu\gamma_5\psi(0)\rangle\!\rangle=\frac{i\epsilon^{\mu P\Delta 0}}{2P^0\mathcal N}\,G_A^q(t)\,,
\end{equation}
and hence we find in the EF
\begin{equation}
\langle\!\langle T^{[\mu\nu]}_q(0)\rangle\!\rangle\big|_\text{EF}=\frac{\uvec\Delta^2_\perp}{4P^0(P^0+M)}\,P^{[\mu}\eta^{\nu]0}\,G^q_A(t)\,.
\end{equation}
Using now the expression in Eq.~\eqref{eq:EMT_EF} for the left-hand side, we recover the relation $D_q(t)=-G^q_A(t)$.

\begin{figure}[t!]
    \centering
    \includegraphics[trim={0 4cm 0 0}, width=0.45\linewidth]{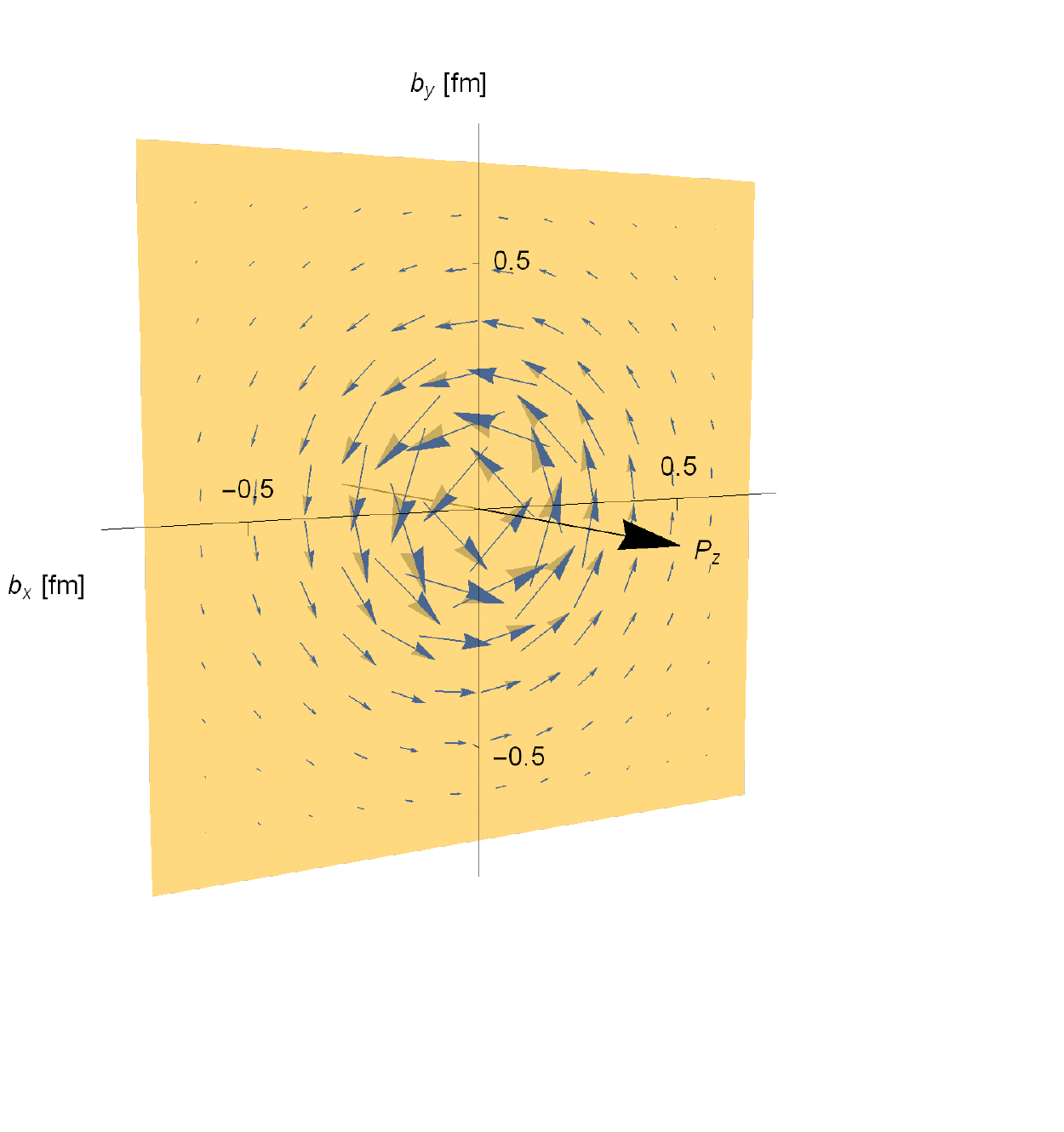}
    \includegraphics[width=0.45\linewidth]{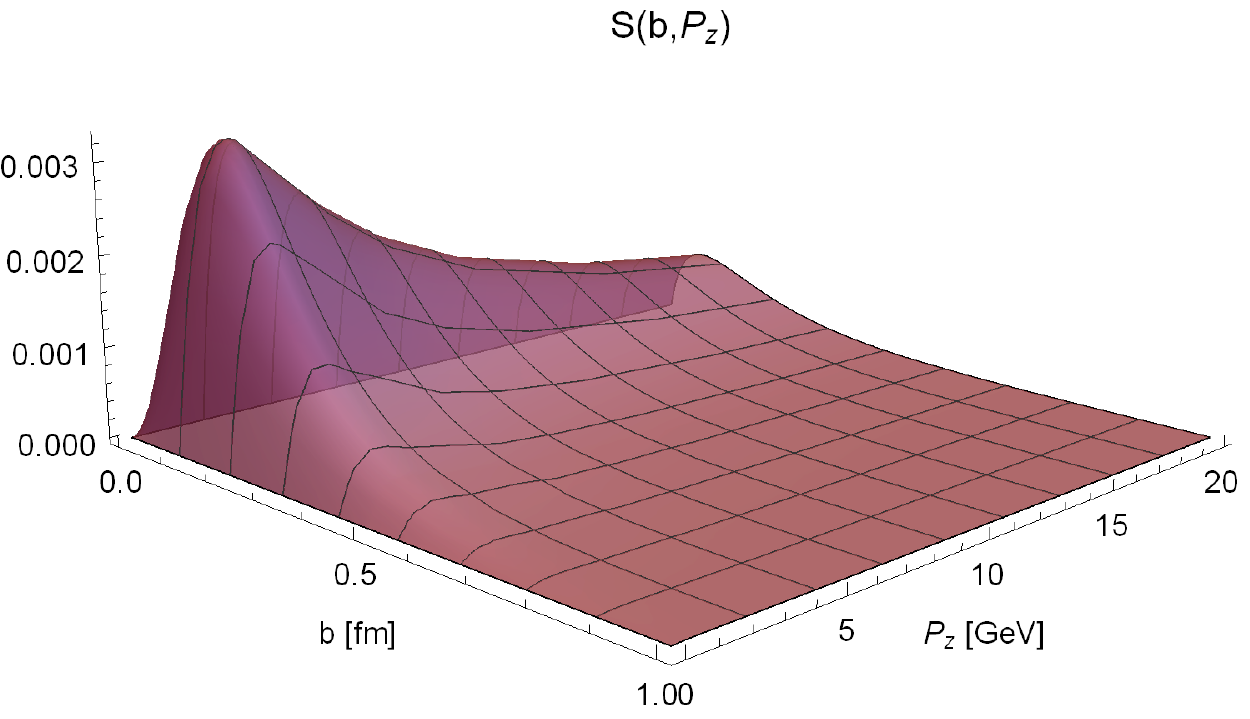}
    \vspace{1cm}
    \\
	(a)\hspace*{0.45\linewidth}(b)
    \caption{Quark spin distribution $\uvec {\mathsf S}(\uvec b_\perp;\uvec P)$ for $\uvec P = (0\,,0\,,P_z)$ using the multipole model~\eqref{eq:model} with parameters given in Table~\ref{tab:model}, see Eq.~\eqref{eq:Si}. (a) Distribution of the quark spin $\uvec {\mathsf S}$ in impact-parameter space for $P_z$ = 0\,.1 GeV, and (b) magnitude of the quark spin density $\mathsf S$ as a function of the impact-parameter $b$ and the target momentum $P_z$.}
	\label{fig:spin}
\end{figure}

We are now ready to discuss the distribution of spin inside an unpolarized target. The quark spin operator being given by $\frac{1}{2}\,\overline\psi(0)\gamma^i\gamma_5\psi(0)$, the corresponding distribution in the EF is given by
\begin{equation}
\label{eq:Si}
\mathsf S^i(\uvec b_\perp;\uvec P)=-(\uvec P\times\uvec\nabla)^i\int\frac{\ud^2\uvec\Delta_\perp}{(2\pi)^2}\,e^{-i\uvec\Delta_\perp\cdot\uvec b_\perp}\,\frac{G^q_A(t)}{4P^0(P^0+M)}.
\end{equation}
We see that even in an unpolarized target, a nonzero quark spin distribution appears when the target is moving, see Fig.~\ref{fig:spin}. The spin direction is orthogonal to both the target momentum and the impact parameter. This is reminiscent of transverse shifts observed for transversely polarized moving target, see~\cite{Lorce:2018zpf} and references therein. In the latter case, the transverse shifts appear because of a nonzero net orbital angular momentum in the system. In the present case, the appearance of transverse spin distribution is a result of spin-orbit coupling~\cite{Lorce:2011kd,Lorce:2014mxa,Bhoonah:2017olu}. 

\subsection{Infinite-momentum frame}

The infinite-momentum frame (IMF) is a special case of the EF obtained by considering the limit $P_z\to\infty$~\cite{Weinberg:1966jm}. We find that the (1+2)-dimensional unpolarized off-forward amplitude \eqref{eq:EMT_EF} reduces in the IMF to
\begin{align}
\label{eq:EMTIMF}
\langle\!\langle T^{\alpha\beta}_a(0)\rangle\!\rangle\big|_\text{IMF}&=\eta^{\alpha0}\eta^{\beta 0}\left(E_P\,A_a(t)-\frac{\uvec\Delta^2_\perp}{4M}\,B_a(t)-\frac{\uvec\Delta^2_\perp}{8E_P}\,A_a(t)\right)\\
&+\frac{M^2}{E_P}\left(\frac{\Delta^\alpha_\perp\Delta^\beta_\perp +\eta^{\alpha\beta}\uvec \Delta^2_\perp}{M^2}\, C_a(t)+\eta^{\alpha\beta}\bar C_a(t)\right)+\mathcal O(P^{-2}_z)\nn\,.
\end{align}
After two-dimensional Fourier transform, we obtain for $\uvec P_\perp=\uvec 0_\perp$
\begin{align}
\mathsf T^{\alpha\beta}_a(\uvec b_\perp;\uvec 0_\perp,P_z)=\frac{M^2}{E_P}&\left\{\eta^{\alpha0}\eta^{\beta0}\,\frac{E^2_P}{M^2}\left[\mathsf A_a(b)+\frac{1}{4E_P^2}\frac{1}{b}\frac{\ud}{\ud b}\!\left(b\,\frac{\ud}{\ud b}\!\left[\frac{E_P}{M}\,\mathsf B_a(b)+\frac{\mathsf A_a(b)}{2}\right]-4\mathsf C_a(b)\right)\right]\right.\\
&\left. +\eta^{\alpha\beta} \left[ \bar{\mathsf{C}}_a(b) - \frac{1}{M^2}\frac{\ud^2\mathsf C_a(b)}{\ud b^2}\right]-\frac{x^\alpha\, x^\beta}{b^2}\, \frac{1}{M^2}\,b\,\frac{\ud}{\ud b}\!\left(\frac{1}{b}\frac{\ud\mathsf C_a(b)}{\ud b}\right)\right\}+\mathcal O(P^{-2}_z)\nn\,,
\end{align}
where $x^\alpha=(0,\uvec b_\perp)$ and therefore, by comparison with Eqs.~\eqref{2DEF} and~\eqref{2DEFbis}, we can write
\begin{align}
\label{eq:rhoIMF}
\rho_a(b,P_z) &\approx M\,\mathsf A_a(b),\\
\label{eq:sigmarIMF}
\sigma_{r,a}(b,P_z) &\approx M\left\{-\bar{\mathsf{C}}_a(b)+ \frac{1}{M^2}\frac{1}{b}\frac{\ud\mathsf C_a(b)}{\ud b} \right\},
\\
\label{eq:sigmatIMF}
\sigma_{t,a}(b,P_z) &\approx M\left\{-\bar{\mathsf{C}}_a(b) + \frac{1}{M^2}\frac{\ud^2\mathsf C_a(b)}{\ud b^2} \right\},\\
\label{eq:sigmaIMF}
\sigma_a(b,P_z) &\approx M\left\{-\bar{\mathsf{C}}_a(b)+\frac{1}{2} \frac{1}{M^2}\frac{1}{b}\frac{\ud}{\ud b}\!\left(b\,\frac{\ud\mathsf C_a(b)}{\ud b}\right) \right\},
\\
\label{eq:PiIMF}
\Pi_a(b,P_z) &\approx M\left\{-\frac{1}{M^2}\,b\,\frac{\ud}{\ud b}\!\left(\frac{1}{b}\frac{\ud\mathsf C_a(b)}{\ud b}\right) \right\},
\end{align}
keeping only the leading terms in $P_z^2\gg P^2$ and taking the boost factor $\gamma=E_P/M$ into account in the identification $t^{\alpha\beta}\sim\gamma\,\mathsf T^{\alpha\beta}$. 

\begin{figure}[ht!]
    \centering
    \includegraphics[width=0.49\textwidth]{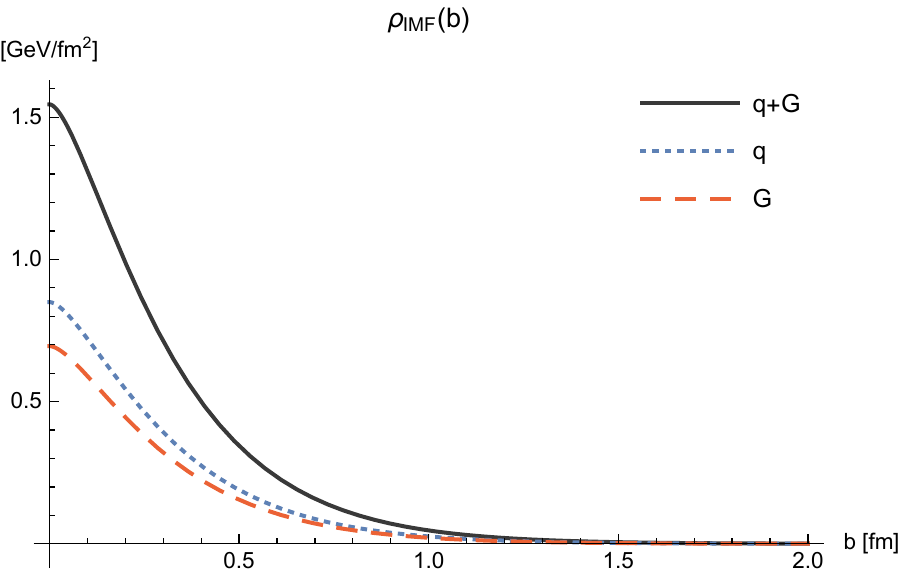}
    \hspace{0.5em}
    \includegraphics[width=0.49\textwidth]{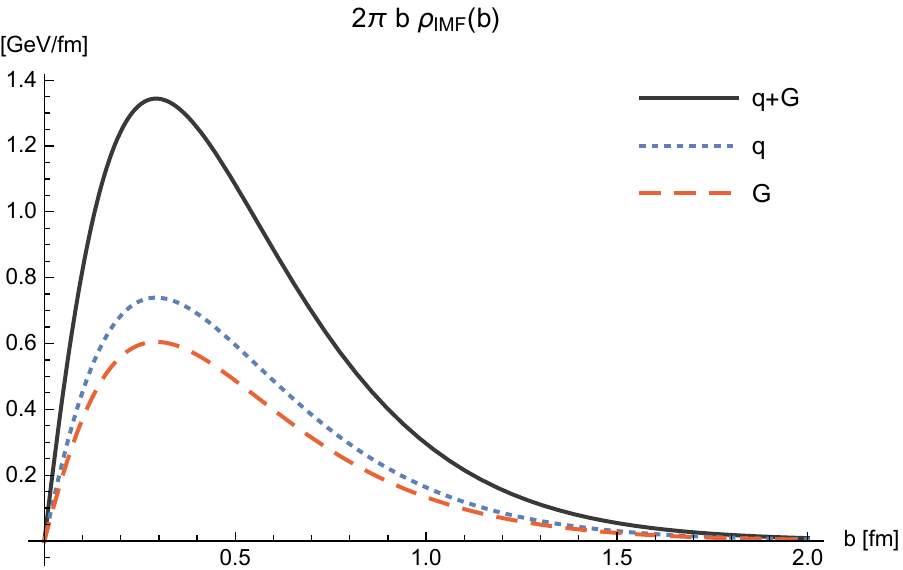}
    \\
	(a)\hspace*{0.45\linewidth}(b)
	\\[2em]
	\includegraphics[width=0.49\textwidth]{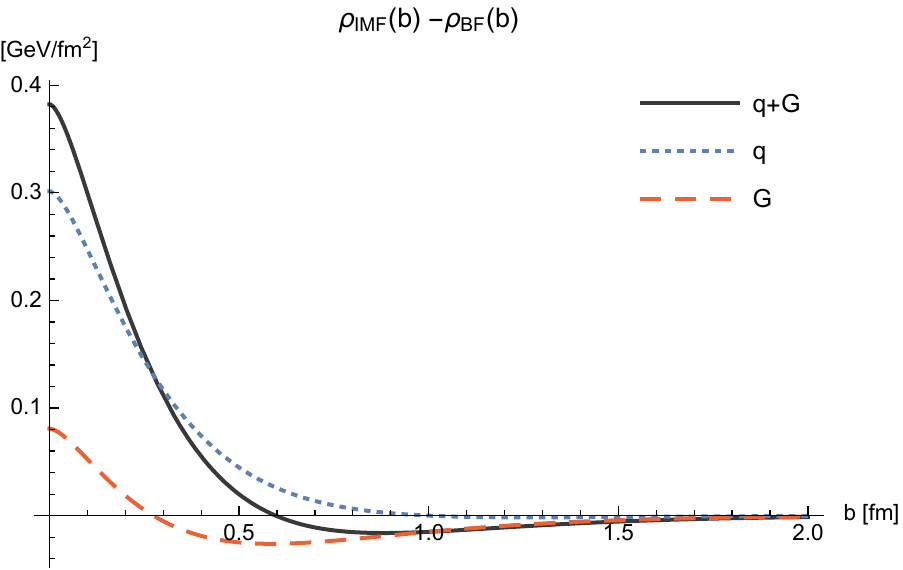}
    \hspace{0.5em}
    \includegraphics[width=0.49\textwidth]{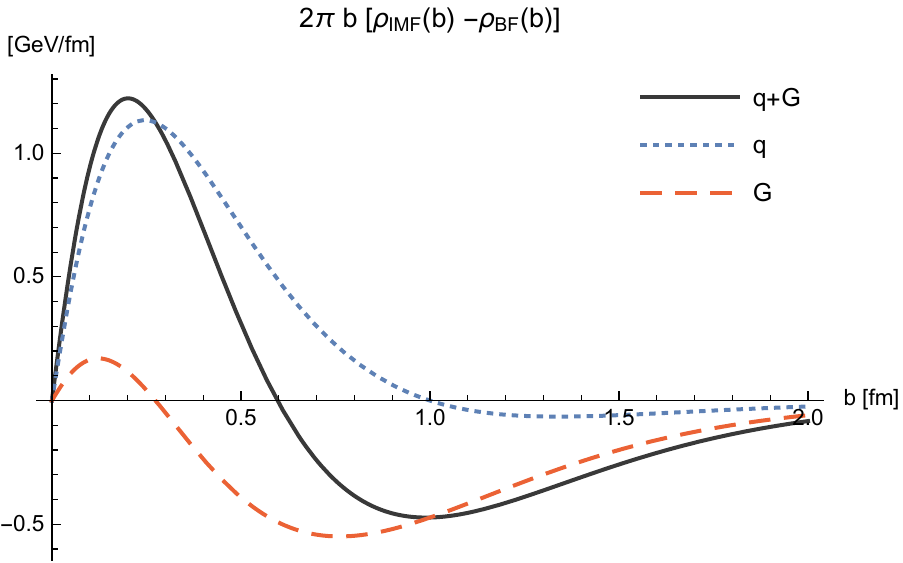}
    \\
	(c)\hspace*{0.45\linewidth}(d)
    \caption{\label{fig:rhoIMF}Plots of the two-dimensional energy density in the IMF, (a) $\rho_\text{IMF}(b)=\lim_{P_z\to\infty}\rho(b,P_z)$ and (b) $2\pi\,b\,\rho_\text{IMF}(b)$, and their difference with the corresponding densities in the BF (c) $\rho_\text{IMF}(b)-\rho_\text{BF}(b)$ with $\rho_\text{BF} = \rho(b,0)$ and (d) $2\pi\,b\,[\rho_\text{IMF}(b)-\rho_\text{BF}(b)]$, using the multipole model~\eqref{eq:model} with parameters given in Table~\ref{tab:model}, see Eqs.~\eqref{eq:rho}  or~\eqref{eq:rhobis}, and~\eqref{eq:rhoIMF} for the definition in terms of GFFs.
    }
\end{figure}

While the two-dimensional pressures are the same in both the BF ($P_z=0$) and IMF ($P_z\to\infty$), the energy densities~\eqref{eq:rho} and~\eqref{eq:rhoIMF} differ. This may be interpreted by the expectation that kinetic energy grows with $P_z$ whereas binding energy associated with pressure forces remains constant. In the IMF, kinetic energy becomes by far the dominant contribution and we recover the parton picture where quarks and gluons behave as almost free massless particles. Since the two-dimensional pressures in the IMF coincides with those shown in Figs.~\ref{fig:sigmar}-\ref{fig:Pi}, we simply illustrate in Fig.~\ref{fig:rhoIMF} the energy density in the IMF, using the multipole model~\eqref{eq:model} with parameters given in Table~\ref{tab:model}. The comparison with the energy density in the BF shows that the kinetic energy is more concentrated around the center of the nucleon than the binding energy. Naturally, once summed over all the constituents and integrated over the impact parameter space, the difference disappears.

\section{Distributions in front form}\label{sec:4}

The interpretation of form factors in the Breit frame are known to be plagued by relativistic corrections~\cite{Burkardt:2000za,Polyakov:2002yz,Belitsky:2005qn}. In particular, one could multiply the off-forward amplitude~\eqref{off-forward} by some function $f(P^0/E_P)$ normalized as $f(1)=1$ to correct for Lorentz contractions effects. While such correction factor does not change the integrated quantities, it does change the spatial distributions since it introduces an additional $\uvec\Delta$-dependence. Moreover, the correction factor cannot be determined in practice in a model-independent way because Lorentz boosts depend on the dynamics of the interacting system.

An interpretation free of such relativistic corrections can however be obtained using the light-front (LF) formalism~\cite{Dirac:1949cp,Brodsky:1997de}, which amounts to adopting the point of view of a massless observer~\cite{Lorce:2018zpf}. This remarkable feature is explained by the fact that, in the LF formalism, the subgroup of Lorentz transformations associated with the transverse plane is Galilean~\cite{Susskind:1967rg,Burkardt:2002hr}. Burkardt used this formalism and introduced the boost-invariant impact-parameter distributions (IPDs) of quarks and gluons~\cite{Burkardt:2000za,Burkardt:2002hr}. Including parton transverse momentum to the picture led then to the concept of relativistic phase-space (or Wigner) distributions~\cite{Lorce:2011kd,Lorce:2011ni,Lorce:2012ce,Lorce:2015sqe}. Longitudinal LF momentum IPDs~\cite{Abidin:2008sb,Selyugin:2009ic,Son:2014sna,Chakrabarti:2015lba,Mondal:2015fok,Mondal:2016xsm,SattaryNikkhoo:2018odd,Kumar:2017dbf,Kumano:2017lhr,Kaur:2018ewq} and longitudinal angular momentum IPDs~\cite{Adhikari:2016dir,Lorce:2017wkb,Kumar:2017dbf} have also recently been discussed in the literature. Our aim here is to introduce the IPDs associated with the EMT.

\subsection{Light-front components}

The LF components of a four-vector are given by
\begin{equation}
a^\mu=[a^+,a^-,\uvec a_\perp]\,,
\end{equation}
where $a^\pm=(a^0\pm a^3)/\sqrt{2}$. In terms of these, the scalar product of two four-vectors reads
\begin{equation}\label{eq:LFscalarproduct}
a\cdot b=a^+ b^-+a^-b^+-\uvec a_\perp\cdot\uvec b_\perp.
\end{equation}
One conventionally chooses $x^+$ to the represent the LF time coordinate. It then follows from Eq.~\eqref{eq:LFscalarproduct} that $p^-$ represents the LF energy. The other conjugate components $x^-$ and $p^+$ represent the longitudinal LF coordinate and momentum, respectively. Without loss of generality, we choose the $z$-axis along $\uvec P$, so that we simply have
\begin{equation}
\label{eq:P_perp}
P^\mu=[P^+,P^-,\uvec 0_\perp].
\end{equation}
We will denote the spatial LF three-vectors as $\tilde{\uvec a}$ and use a dotted notation to keep the tensor expressions compact. A LF three-vector with a dotted index will indicate a minus first component followed by the transverse ones, while an undotted index will indicate a plus first component followed by the transverse ones, i.e. $a^{\dot\alpha}=(a^-,\uvec a_\perp)$ and $a^{\alpha}=(a^+,\uvec a_\perp)$. Thus, in position space, the spatial LF coordinates read $x^{\dot\alpha}=(x^-,\uvec x_\perp)$, and in momentum space they read $p^{\alpha}=(p^+,\uvec p_\perp)$. We can then rewrite the scalar product of two four-vectors~\eqref{eq:LFscalarproduct} as
\begin{equation}
x\cdot p=x^+p^-+\tilde{\uvec x}\cdot\tilde{\uvec p}  
\end{equation}
with $\tilde{\uvec x}\cdot\tilde{\uvec p}=x_{\dot \alpha}\, p^\alpha=x^{\dot \alpha}\, p_\alpha=x^-p^+-\uvec x_\perp\cdot\uvec p_\perp$. Using LF components is convenient because they behave in a simple way under LF boosts~\cite{Dirac:1949cp,Brodsky:1997de}. In particular, performing a longitudinal LF boost amounts to a mere rescaling of the LF components $[a^+,a^-,\uvec a_\perp]\mapsto [e^{-\omega} a^+,e^{\omega}a^-,\uvec a_\perp]$, where $\gamma = \cosh \omega$.

Because of the Galilean symmetry in the transverse plane, it is interesting to organize the LF components of the EMT as follows~\cite{Soper:1976bb}
\begin{equation}\label{LFEMTcomp}
T^{\mu\nu}=\left(\begin{array}{ccc|c}
T^{+-}&T^{+1}&T^{+2}&T^{++}\\
T^{1-}&T^{11}&T^{12}&T^{1+}\\
T^{2-}&T^{21}&T^{22}&T^{2+}\\
\hline
T^{--}&T^{-1}&T^{-2}&T^{-+}
\end{array}\right).
\end{equation} 
The upper left corner corresponds to a (1+2)-dimensional Galilean EMT $T^{\alpha\dot\beta}$. The upper right corner corresponds to a ``mass'' current $J^\alpha= T^{\alpha +}$, where the role of ``mass'' in the transverse plane is played by the longitudinal LF momentum. Since we are considering only the static EMT, the above currents are conserved in the (1+2)-dimensional subspace
\begin{equation}
\partial_{\dot\alpha} T^{\alpha\dot\beta}=\partial_-T^{+\dot\beta}+\partial_iT^{i\dot\beta}=0\,,\qquad \partial_{\dot\alpha} J^\alpha=0\,.
\end{equation}
Together, they form the so-called covariant non-relativistic stress-energy tensor $T^{\alpha\mu}$ appearing in the context of Newton-Cartan geometries, which find important applications in condensed matter and in the study of non-relativistic holographic systems, see e.g.~\cite{Geracie:2016bkg} and references therein.  The last line in Eq.~\eqref{LFEMTcomp}, which describes the flux of energy and momentum along the spatial LF direction $x^-$, does not have any known simple interpretation within the Galilean picture.

\subsection{Light-front amplitudes}
We can repeat the same procedure as in Section~\ref{sec:3} in the LF formalism. Integrating the covariant phase-space density operator~\eqref{eq:RPcov} over $P^2$ and $\Delta^-$ leads to
\begin{equation}\label{eq:RPnoncov2}
\rho_{R,P}=\int\frac{\ud^3\tilde{\uvec\Delta}}{(2\pi)^3\,2P^+}\,e^{-i\Delta\cdot R}\,\ket{P-\tfrac{\Delta}{2}}\bra{P+\tfrac{\Delta}{2}}\,,
\end{equation}
with $\Delta^- = -\Delta^+\,P^-/P^+$, since $\uvec P_\perp=\uvec 0_\perp$~\eqref{eq:P_perp}, and
\begin{equation}\label{eq:Delta02}
p^-=\frac{\frac{\uvec\Delta_\perp^2}{4}+M^2}{2\left(P^+-\frac{\Delta^+}{2}\right)}\,,\qquad p'^-=\frac{\frac{\uvec\Delta_\perp^2}{4}+M^2}{2\left(P^++\frac{\Delta^+}{2}\right)}\,.
\end{equation}
The non-explicitly covariant form~\eqref{eq:RPnoncov2} coincides with that in Ref.~\cite{Lorce:2017wkb} if we replace the normalization factor $2P^+$ by $2\sqrt{p'^+p^+}$. Once again, the difference in the normalization comes from the fact that ``position'' states in Ref.~\cite{Lorce:2017wkb} were defined with the non-relativistic normalization $\langle x'|x\rangle=\delta^{(3)}(\tilde{\uvec x}' -\tilde{\uvec x})$ at equal LF times. This difference will however not concern us since we will essentially be interested in the case $\Delta^-=p'^--p^-=0$. 

Setting the origin at the average position of the system, the LF distribution of energy and momentum inside the system with canonical polarization $\uvec s$ and average LF momentum $\tilde{\uvec P}=(P^+,\uvec 0_\perp)$ is given by the Fourier transform
\begin{equation}\label{eq:TFT2}
\mathcal T^{\mu\nu}_a(\tilde{\uvec x};\tilde{\uvec P})=
\langle T^{\mu\nu}_a(\tilde{\uvec x})\rangle_{0,P}=\int\frac{\ud^3\tilde{\uvec\Delta}}{(2\pi)^3}\,e^{i\tilde{\uvec\Delta}\cdot\tilde{\uvec x}}\,\langle\!\langle T^{\mu\nu}_a(0)\rangle\!\rangle
\end{equation} 
of the LF off-forward amplitude~\cite{Lorce:2017wkb} 
\begin{equation}\label{off-forward2}
\langle\!\langle T^{\mu\nu}_a(0)\rangle\!\rangle=\frac{\bra{p',\uvec s}T^{\mu\nu}_a(0)\ket{p,\uvec s}}{2P^+}\,.
\end{equation}
Note that $x^+=0$ because the LF positions $\uvec X^-$ and $\uvec R^-$ are considered at the same LF time $X^+=R^+$. If we allow $X^+$ and $R^+$ to be different, the above static EMT $\mathcal T^{\mu\nu}_a(\tilde{\uvec x};\tilde{\uvec P})$ can be recovered by considering the LF time average $\int {\ud X^+}/[2\pi\,\delta(0)]$ in a frame where the LF energy transfer vanishes $\Delta^-=0$. 

The Dirac bilinears appearing in the parametrization of the EMT~\eqref{eq:TmunuUU} can be expressed in front form as~\cite{Lorce:2017isp}
\begin{align}
\bar u'u&=\mathcal N^{-1}\left[2MP^++i\epsilon^{+P\Delta S_\perp}\right],\\
\bar u'i\sigma^{\mu\Delta} u&=\mathcal N^{-1}\left\{M(\eta^{\mu +}\Delta^2-\Delta^\mu\Delta^+)+2\left[P^+\,i\epsilon^{\mu\Delta P S_\perp}-\frac{\Delta^2}{4}\,i\epsilon^{\mu\Delta S_\perp +}-\left[(P\cdot S_\perp)+Ms_z\right]i\epsilon^{\mu\Delta P+}\right]\right\},
\end{align}
where $S^\mu_\perp=[0\,,0\,,\uvec s_\perp]$ and $\mathcal N=\sqrt{p'^+}\sqrt{p^+}$. The unpolarized off-forward LF amplitude then reads
\begin{align}\label{Tred2}
\langle\!\langle T^{\mu\nu}_a(0)\rangle\!\rangle&=\frac{M}{\mathcal N}\left\{\frac{P^\mu P^\nu}{M}\,A_a(t)
+ \frac{\Delta^\mu\Delta^\nu - \eta^{\mu\nu}\Delta^2}{M}\, C_a(t)
+ M \eta^{\mu\nu}\bar C_a(t)\right\}\\
&+\frac{\Delta^2}{4P^+\mathcal N}\left\{P^{\{\mu}\eta^{\nu\}+}\,\frac{A_a(t)+B_a(t)}{2}+P^{[\mu}\eta^{\nu]+}\,\frac{D_a(t)}{2}\right\}\nn\\
&-\frac{\Delta^+}{4P^+\mathcal N}\left\{P^{\{\mu}\Delta^{\nu\}}\,\frac{A_a(t)+B_a(t)}{2}+P^{[\mu}\Delta^{\nu]}\,\frac{D_a(t)}{2}\right\}\nn\,.
\end{align}
Note that the structure is slightly simpler than the corresponding expression in instant form~\eqref{Tred}.

The static LF EMT can receive a (quasi-)probabilistic interpretation only when no LF energy is transferred to the system $\Delta^-=0$~\cite{Lorce:2017wkb}. Since the onshell conditions impose that $\Delta^-=-\Delta^+P^-/P^+$ and since $P^\pm>0$ for a massive target, we will consider the following cases:
\begin{enumerate}[label=(\alph*)]
\item $\Delta^+=0$ --- Drell-Yan frame (DYF);
\item $P^+\to\infty$ --- infinite-momentum frame (IMF).
\end{enumerate}
Note that in both cases the normalization factor appearing in Eq.~\eqref{Tred2} reduces to $\mathcal N\approx P^+$.

\subsection{Drell-Yan frame}
The Drell-Yan frame defined by $\Delta^+=0$ can be seen as the LF version of the elastic frame defined by $\tilde{\uvec P}\cdot\tilde{\uvec\Delta}=0$. Distributions in the DYF can be obtained by integrating the static LF EMT over the longitudinal LF coordinate $x^-$
\begin{equation}
\mathsf T^{\mu\nu}_a(\uvec b_\perp;\tilde{\uvec P})=\int\ud x^-\,\mathcal T^{\mu\nu}_a(\tilde{\uvec x};\tilde{\uvec P})=\int\frac{\ud^2\uvec\Delta_\perp}{(2\pi)^2}\,e^{-i\uvec\Delta_\perp\cdot\uvec b_\perp}\,\langle\!\langle T^{\mu\nu}_a(0)\rangle\!\rangle\big|_\text{DYF}\,,
\end{equation}
where $\uvec b_\perp=\uvec x_\perp$ is the same impact parameter as in instant form. The unpolarized off-forward LF amplitude~\eqref{off-forward2} in the DYF takes the form
\begin{align}
\label{eq:EMT_DYF}
\langle\!\langle T^{\mu\nu}_a(0)\rangle\!\rangle\big|_\text{DYF}&=\frac{M}{P^+}\left\{\frac{P^\mu P^\nu}{M}\,A_a(t)
+ \frac{\Delta^\mu_\perp\Delta^\nu_\perp +\eta^{\mu\nu}\uvec \Delta^2_\perp}{M}\, C_a(t)
+ M \eta^{\mu\nu}\bar C_a(t)\right\}\\
&-\frac{\uvec\Delta^2_\perp}{4(P^+)^2}\left\{P^{\{\mu}\eta^{\nu\}+}\,\frac{A_a(t)+B_a(t)}{2}+P^{[\mu}\eta^{\nu]+}\,\frac{D_a(t)}{2}\right\},\nn
\end{align}
where $\Delta^\mu_\perp=[0,0,\uvec\Delta_\perp]$ and $P^-=\frac{M^2}{2P^+}\left(1+\frac{\uvec\Delta^2_\perp}{4M^2}\right)$. If we further integrate the static LF EMT over the impact-parameter $\uvec b_\perp$, which amounts to setting $\uvec\Delta_\perp=\uvec 0_\perp$ in Eq.~\eqref{eq:EMT_DYF}, we recover the FL expression~\eqref{eq:TmunuFL} with $E_P$ replaced by $P^+$.

\begin{figure}[t!]
    \centering
    \includegraphics[width=0.49\textwidth]{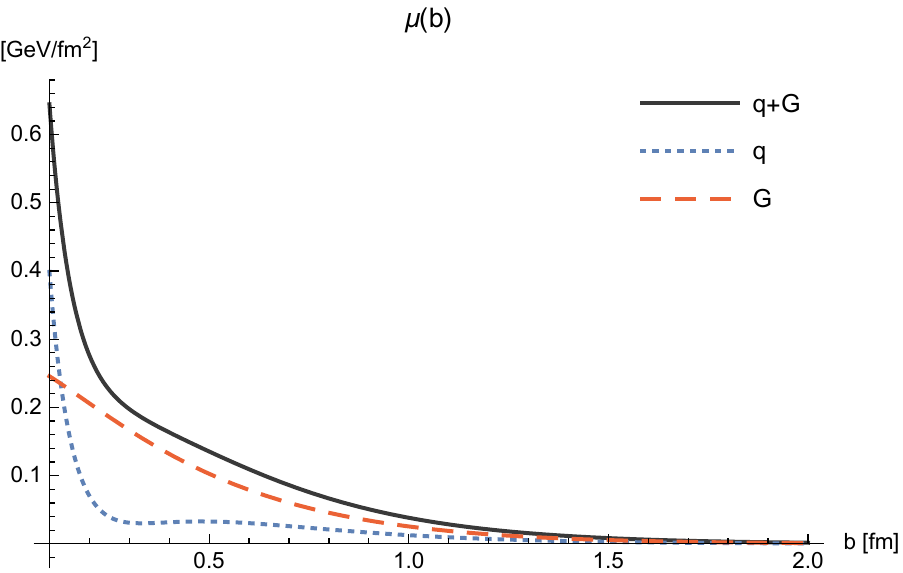}
    \hspace{0.5em}
    \includegraphics[width=0.49\textwidth]{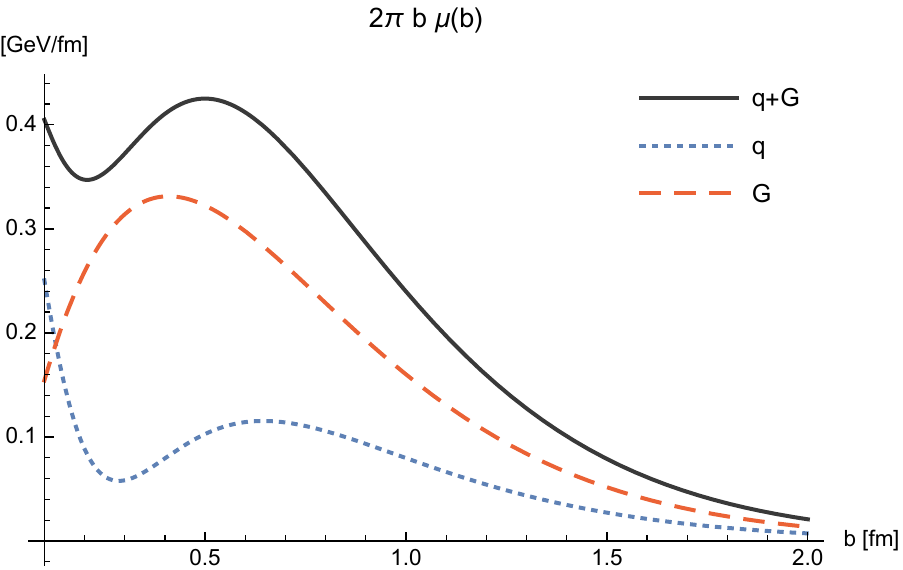}
    \\
	(a)\hspace*{0.45\linewidth}(b)
    \caption{\label{fig:mu}
    Plots of the two-dimensional Galilean energy density, (a) $\mu(b)$ and (b) $2\pi\,b\,\mu(b)$, using the multipole model~\eqref{eq:model} with parameters given in Table~\ref{tab:model}, see Eq.~\eqref{eq:mu_LF} for the definition in terms of GFFs.
    }
\end{figure}

However for $\uvec\Delta_\perp\neq\uvec 0_\perp$, unlike $P^0$ in instant form~\eqref{eq:EMT_EF}, $P^+$ in front form is an independent variable which does not depend on $\uvec\Delta_\perp$. One can therefore write a relatively simple expression for the static LF EMT in terms of Fourier transforms of GFFs\footnote{Keeping terms up to order $1/(P^+)^2$, we observe that $\mathsf T^{\mu\nu}_a(\uvec b_\perp)$ has the same structure as the asymmetric anisotropic version of the EMT for a type-II fluid used in General Relativity to describe gravitational collapse~\cite{Hawking:1973uf,Husain:1995bf,Wang:1998qx}.}. The (1+2)-dimensional Galilean ``mass'' current takes the simple form
\begin{equation}
J^\alpha_a(\uvec b_\perp;\tilde{\uvec P})=\mathsf T^{\alpha +}_a(\uvec b_\perp;\tilde{\uvec P})=P^+\,\eta^{\alpha -}\,\mathsf A_a(b)\,.
\end{equation}
Its LF time component $J^+_a=T^{++}_a$ can be interpreted as the density of longitudinal LF momentum carried by constituent type $a$ and has been discussed in~\cite{Abidin:2008sb,Selyugin:2009ic,Son:2014sna,Chakrabarti:2015lba,Mondal:2015fok,Mondal:2016xsm,SattaryNikkhoo:2018odd,Kumar:2017dbf,Kumano:2017lhr,Kaur:2018ewq}. As expected, the distribution of longitudinal LF momentum coincides with the properly rescaled instant-form energy density in IMF $\mu_a(b)=\rho_\text{IMF}(b)P^+/M$. For the (1+2)-dimensional Galilean EMT, we find
\begin{align}
\mathsf T^{\alpha \dot\beta}_a(\uvec b_\perp;\tilde{\uvec P})&=\frac{M^2}{P^+}\left\{\eta^{\alpha -}\eta^{\dot\beta +}\left[\frac{\mathsf A_a(b)}{2}+\frac{1}{4M^2}\frac{1}{b}\frac{\ud}{\ud b}\!\left(\frac{b}{2}\,\frac{\ud}{\ud b}\left[\mathsf B_a(b)+\mathsf D_a(b)\right]-4\mathsf C_a(b)\right)\right]\right.\\
&\left.\qquad\quad\,+\eta^{\alpha\dot\beta} \left[ \bar{\mathsf{C}}_a(b) - \frac{1}{M^2}\frac{\ud^2\mathsf C_a(b)}{\ud b^2}\right]-\frac{x^\alpha  x^{\dot\beta}}{b^2}\, \frac{1}{M^2}\,b\,\frac{\ud}{\ud b}\!\left(\frac{1}{b}\frac{\ud\mathsf C_a(b)}{\ud b}\right)\right\},\nn
\end{align}
where $x^\mu=[0,0,\uvec b_\perp]$. Note that the Galilean EMT $\mathsf T^{\alpha \dot\beta}_a$ has a global boost factor $M/P^+$. We can get rid of this factor by integrating $\mathcal T^{\mu\nu}_a(\tilde{\uvec x};\tilde{\uvec P})$ over the coordinate $r_L= r^-P^+/M$ invariant under longitudinal LF boosts instead of $r^-$. The structure of the Galilean EMT then looks like
\begin{equation}
t^{\alpha\dot\beta}(\uvec b_\perp)=[\mu(b)+\sigma_t(b)]\,n^\alpha \bar n^{\dot\beta} - \sigma_t(b)\eta^{\alpha\dot\beta}+[\sigma_r(b)-\sigma_t(b)]\,\chi^\alpha \chi^{\dot\beta}
\end{equation}
where $n^\alpha=\eta^{\alpha -}$, $\bar n^{\dot\beta}=\eta^{\dot\beta+}$, and $\chi^\mu=[0\,,0\,,\uvec b_\perp/b]$. This tensor can alternatively be written as
\begin{equation}
t^{\alpha\dot\beta}(\uvec b_\perp)=[\mu(b)+\sigma(b)]\,n^\alpha \bar n^{\dot\beta} - \sigma(b)\eta^{\alpha\dot\beta}+\Pi(b)\left(\chi^\alpha \chi^{\dot\beta}-\frac{1}{2}\,l^{\alpha\dot\beta}\right)\,,
\end{equation}
with $l^{\alpha\dot\beta}=n^\alpha \bar n^{\dot\beta}-\eta^{\alpha\dot\beta}$. We can therefore define $P^+$-independent two-dimensional Galilean energy density, radial pressure, tangential pressure, isotropic pressure, and pressure anisotropy associated with constituent type $a$ as
\begin{align}
\label{eq:mu_LF}
\mu_a(b) &= M\left\{\frac{\mathsf A_a(b)}{2}+\bar{\mathsf C}_a(b) + \frac{1}{4M^2}\frac{1}{b}\frac{\ud}{\ud b}\!\left(b\,\frac{\ud}{\ud b}\!\left[\frac{\mathsf B_a(b)+\mathsf D_a(b)}{2}-4\mathsf C_a(b)\right]\right)\right\},\\
\label{eq:sigmar_LF}
\sigma_{r,a}(b) &= M\left\{-\bar{\mathsf{C}}_a(b)+ \frac{1}{M^2}\frac{1}{b}\frac{\ud\mathsf C_a(b)}{\ud b} \right\},
\\
\label{eq:sigmat_LF}
\sigma_{t,a}(b) &= M\left\{-\bar{\mathsf{C}}_a(b) + \frac{1}{M^2}\frac{\ud^2\mathsf C_a(b)}{\ud b^2} \right\},\\
\label{eq:sigma_LF}
\sigma_a(b) &= M\left\{-\bar{\mathsf{C}}_a(b)+\frac{1}{2} \frac{1}{M^2}\frac{1}{b}\frac{\ud}{\ud b}\!\left(b\,\frac{\ud\mathsf C_a(b)}{\ud b}\right) \right\},
\\
\label{eq:Pi_LF}
\Pi_a(b) &= M\left\{-\frac{1}{M^2}\,b\,\frac{\ud}{\ud b}\!\left(\frac{1}{b}\frac{\ud\mathsf C_a(b)}{\ud b}\right) \right\}.
\end{align}
The two-dimensional pressures are the same as the ones obtained in the BF~\eqref{eq:sigmar}-\eqref{eq:Pi}. For this reason, we used the same notation as in instant form. It is also not surprising that the energy densities $\rho_a$ and $\mu_a$ defined, respectively, in instant form and front form differ because they are simply related to different components of the EMT. The latter is illustrated in Fig.~\ref{fig:mu} using the multipole model~\eqref{eq:model} with parameters given in Table~\ref{tab:model}.
The difference in the behavior around $b=0$ between the quark and gluon contributions comes from the fact $D_g(t)=0$ whereas $D_q(t)\neq 0$. One might also be at first sight puzzled by the fact that total LF energy is given by $\sum_a\int\ud^2b_\perp\,\mu_a(b)=M/2$ instead of $M$. This simply comes from our definition of the LF components which implies that $\int\ud^3\tilde x\,\mathcal T^{+-}(\tilde{\uvec x};\tilde{\uvec P})=M^2/2P^+$.

\subsection{Infinite-momentum frame}

In Ref.~\cite{Brodsky:2006in,Brodsky:2006ku} three-dimensional LF distributions for finite $P^+$ have been defined by means of a Fourier transform with respect to the longitudinal LF boost-invariant skewness variable $\xi= -\Delta^+/2P^+$ instead of $\Delta^+$. The problem with these distributions is that the (quasi-)probabilistic interpretation is lost owing to the non-vanishing LF energy transfer $\Delta^-\neq 0$. Moreover, the center of the target with respect to which the transverse coordinates are defined differs between the initial and final states when $\xi\neq 0$~\cite{Diehl:2002he}.

The above issues disappear when $\xi=0$. In the literature, one usually considers finite $P^+$ and hence $\Delta^+=0$, reducing the LF distributions to two spatial dimensions. This option was discussed in the previous section. The other possibility is to consider $P^+\gg\Delta^+,\sqrt{P^2}$, i.e. the IMF within the LF formalism. In this case, we can formally define another type of three-dimensional LF distributions free of the aforementioned problems. To the best of our knowledge, this is the first time that such an option is explored. The reason can likely be attributed to the fact that one usually has in mind reaching the IMF through an infinite longitudinal boost. In that case, both $P^+$ and $\Delta^+$ will get large with their ratio $\xi$ fixed. Actually, one should just consider $P^+\to\infty$ with $\Delta^+$ fixed. The information about the longitudinal spatial structure is then encoded around $\xi\approx 0$.

Expanding the unpolarized off-forward LF amplitude~\eqref{Tred2} in powers of $1/P^+$, we find
\begin{align}
\label{eq:Tmunu_LF_IMF}
\langle\!\langle T^{\mu\nu}_a(0)\rangle\!\rangle&=P^+\,\eta^{\mu-}\eta^{\nu-}\left\{A_a(t)
-\frac{(\Delta^+)^2}{4(P^+)^2}\left[\tfrac{1}{2}\,A_a(t)+B_a(t)\right]\right\}\\
&+\frac{M}{P^+}\left\{M\,\eta^{-\{\mu}\eta^{\nu\}+}\,\frac{A_a(t)}{2}+\frac{\Delta^\mu\Delta^\nu + \eta^{\mu\nu}\uvec\Delta^2_\perp}{M}\, C_a(t)
+ M \eta^{\mu\nu}\bar C_a(t)\right\}\nn\\
&-\frac{\uvec\Delta^2_\perp}{4P^+}\left\{\eta^{-\{\mu}\eta^{\nu\}+}\,\frac{B_a(t)}{2}+\eta^{-[\mu}\eta^{\nu]+}\,\frac{D_a(t)}{2}\right\}\nn\\
&-\frac{\Delta^+}{4P^+}\left\{\eta^{-\{\mu}\Delta^{\nu\}}_\perp\,\frac{A_a(t)+B_a(t)}{2}+\eta^{-[\mu}\Delta^{\nu]}_\perp\,\frac{D_a(t)}{2}\right\}+\mathcal O(1/(P^+)^2)\nn\,.
\end{align}
Its Fourier transform $\mathcal T^{\mu\nu}_a$ can be expressed in terms of 3-dimensional Fourier transforms of GFFs. The expression we obtain is however so complicated that we were not able to recognize the EMT structure of any known continuous medium discussed in the literature. Note also that since $\Delta^-\propto 1/(P^+)^2$, we can write $t\approx-\uvec\Delta^2_\perp$. The GFFs therefore do not contribute to the $\Delta^+$-dependence in the LF IMF and the longitudinal structure is essentially determined by Lorentz symmetry. As a final remark, we naturally recover the DYF results in the limit $\Delta^+\to 0$. This means that the projection of the distributions defined in the LF IMF onto the transverse plane coincides with the two-dimensional distributions defined in the DYF.

\section{Discussion}\label{sec:5}

Having defined the notions of energy density and pressure inside the nucleon, we can go on and discuss the questions of hydrostatic equilibrium and stability constraints. While the former are automatically satisfied once GFFs are determined, the latter provide new constraints particularly useful for the phenomenology of high-energy scatterings involving nucleons.

\subsection{Hydrostatic equilibrium}

\subsubsection{Three-dimensional case}

Conservation of the total EMT $\partial_\mu T^{\mu\nu}(x)=0$ implies that the static total EMT $\mathcal T^{\mu\nu}=\sum_a\mathcal T^{\mu\nu}_a$ satisfies in the BF
\begin{equation}\label{EMTconservation}
\nabla^i\mathcal T^{ij}(\uvec r;\uvec 0)=0\,,
\end{equation}
or equivalently
\begin{equation}\label{eq:conservationEMT}
\frac{\ud p_r(r)}{\ud r}=-\frac{2s(r)}{r},
\end{equation}
which is the equation of hydrostatic equilibrium in the presence of pressure anisotropy\footnote{For spherically symmetric compact stars, the line element in Schwarzschild coordinates is given by
\begin{equation*}
\ud s^2=e^{\nu(r,t)}\,\ud t^2-e^{-\lambda(r,t)}\,\ud r^2-r^2\left(\ud\theta^2+\sin^2\theta\,\ud\varphi^2\right),
\end{equation*}
and the equation of hydrostatic equilibrium, which directly derives from the vanishing of the covariant divergence of the EMT in the static limit, reads
\begin{equation}\label{balancestar}
\frac{\ud p_r(r)}{\ud r}=-\frac{1}{2}\frac{\ud \nu(r)}{\ud r}\big[\varepsilon(r)+p_r(r)\big]-\frac{2s(r)}{r}
\end{equation}
in the presence of anisotropic matter~\cite{Bowers:1974tgi}. If we switch off gravitational effects by sending Newton's constant to zero, both functions $\nu(r,t)$ and $\lambda(r,t)$ vanish and we recover Eq.~\eqref{eq:conservationEMT}.}. The RHS of Eq.~\eqref{eq:conservationEMT} represents the effects of a force arising from the anisotropic nature of the medium. When $p_r(r)<p_t(r)$, we get $s(r)<0$ and the force is repulsive\footnote{This seems to be the favored case for neutrons stars since it allows the construction of more compact and more stable objects than with ordinary isotropic matter~\cite{Gokhroo:1994fbj,Dev:2003qd}, relaxing therefore the tension between observations and theoretical bounds.}. When $p_r(r)>p_t(r)$, we get $s(r)>0$ and the force is attractive. This force is the mechanical origin of the surface tension between a liquid and its vapor~\cite{Marchand:2011}. In the bulk of ordinary fluids, the density is essentially constant and the pressure is isotropic. The density changes however drastically across the interface, and generates an asymmetry in the stress tensor which can be understood as arising from a difference of ranges between attractive and repulsive interactions. The interface being usually extremely thin for ordinary fluids, the asymmetry can usually be modelled by a simple surface tension. This picture is supported  by a molecular dynamics simulation of molecules interacting via a Lennard-Jones potential~\cite{Marchand:2011,Weijs:2011}, which shows that the transition in the time-averaged density profile from the high-density liquid to the low-density gas takes place in a very narrow region that is a few molecules wide. At the same time, the profile of the stress anisotropy indicates that there is a force localized in the same narrow region acting in the direction parallel to the interface. 

If $r$ is the coordinate normal to the interface, the surface tension $\gamma$ is obtained from the Bakker equation~\cite{bakker:1928} (also known as the Kirkwood and Buff method~\cite{Kirkwood:1949})
\begin{equation}
\gamma=\int_D\ud r\,s(r),
\end{equation}
where $D$ is the domain where $s(r)$ is significant. When the domain is very narrow, we can make the approximation $s(r)\approx\gamma\,\delta(r-R)$. Integrating then the equation of hydrostatic equilibrium~\eqref{eq:conservationEMT} over the radial coordinate yields the well-known Young-Laplace relation $p_r(0)=2\gamma/R$ for a spherical drop with radius $R$~\cite{Thomson:1958}. For compact stars and hadrons, the anisotropic stress spreads over a significant fraction of the volume of the system and cannot be realistically approximated by a delta function.
\newline

\begin{figure}[t!]
    \centering
    \includegraphics[width=0.49\textwidth]{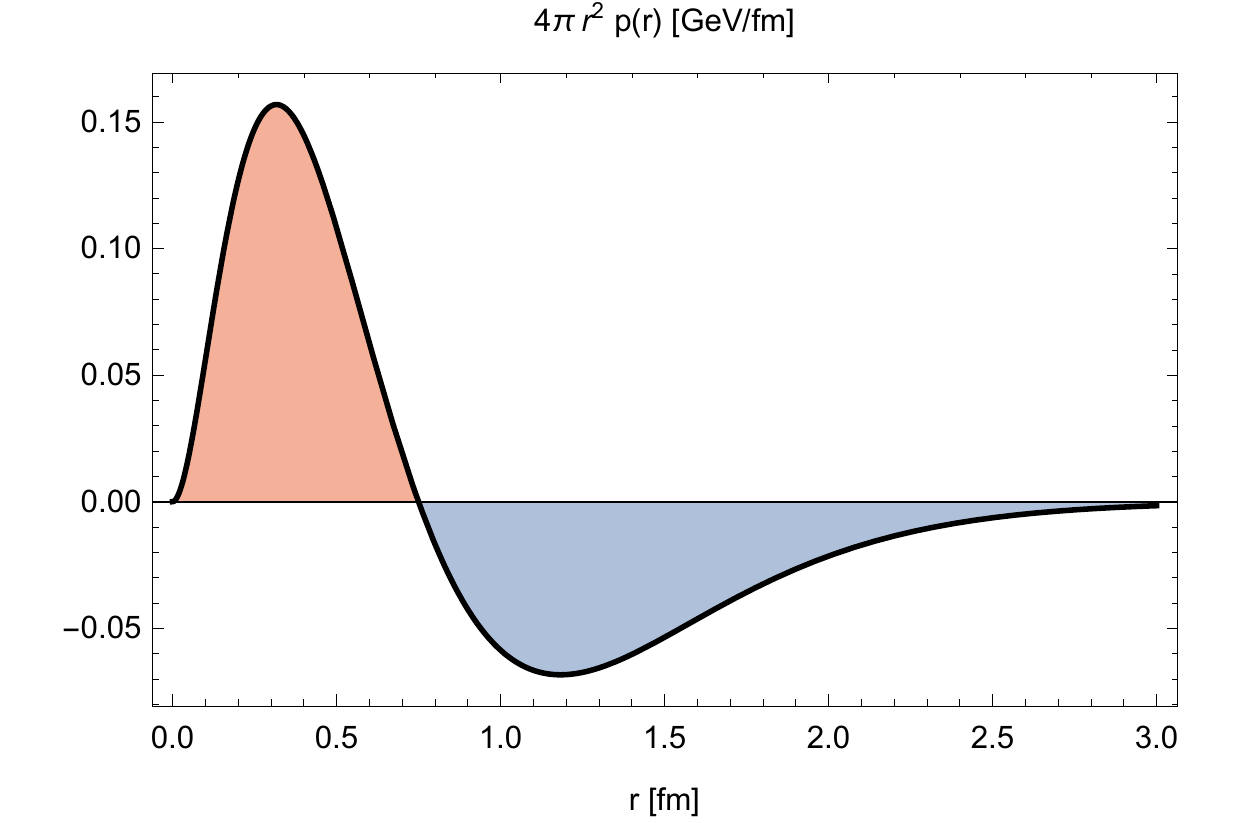}
    \includegraphics[width=0.49\textwidth]{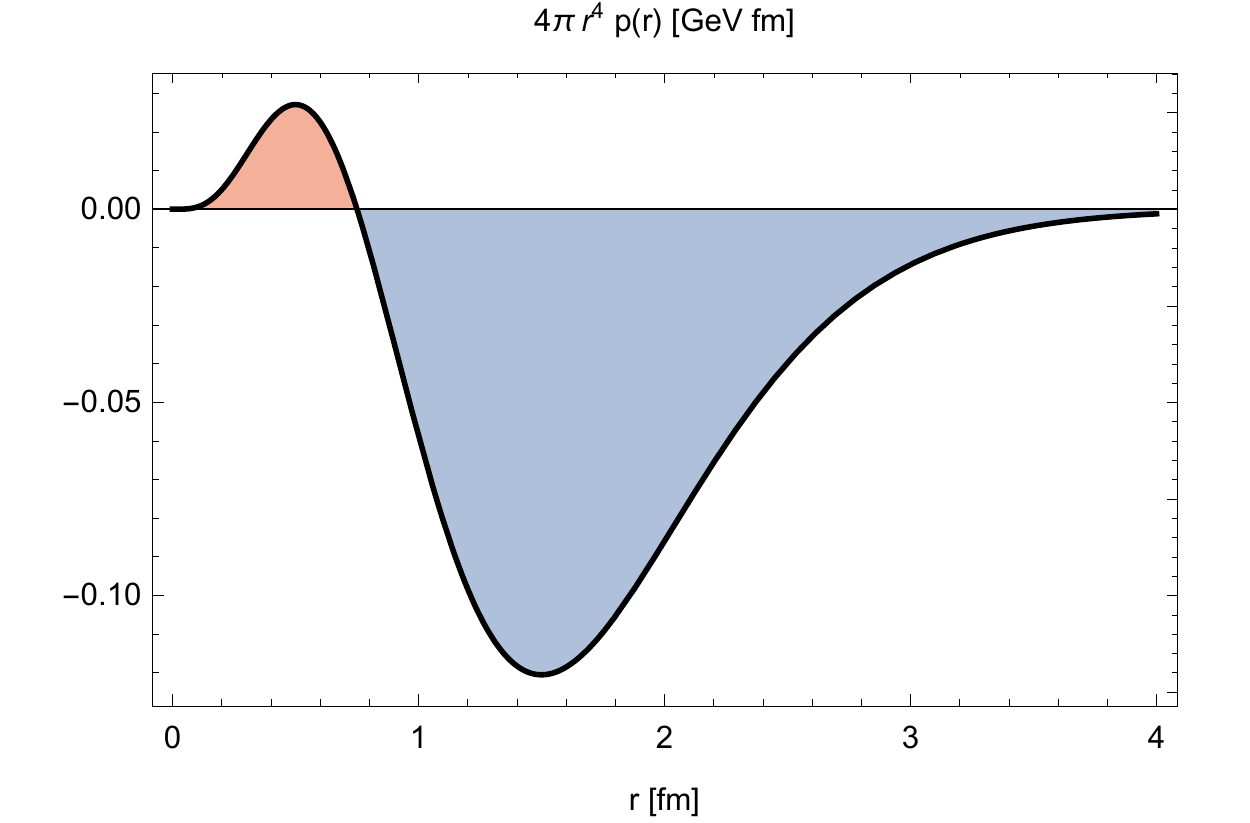}
    \\[1em]
	(a)\hspace*{0.45\linewidth}(b)
    \\[1em]
    \caption{Illustration of the hydrostatic equilibrium in the Breit frame. (a) integrand of the von Laue condition~\eqref{eq:vonLaue} and (b) the same integrand multiplied by $r^2$.
    }
    \label{fig:stability_3D}
\end{figure}

Many relations can be derived from Eq.~\eqref{eq:conservationEMT}, as discussed in~\cite{Goeke:2007fp,Polyakov:2018zvc}. Let $f(r)$ be some radial function, one finds using integration by parts
\begin{equation}
\left[f(r)\,p_r(r)\right]_0^\infty=\int_0^\infty\ud r\left[\frac{\ud f(r)}{\ud r}\,p_r(r)-2\,\frac{f(r)}{r}\,s(r)\right].
\end{equation}
For $f(r)=1$, one obtains the generalized Young-Laplace relation
\begin{equation}
p_r(0)=2\int_0^\infty\ud r\,\frac{s(r)}{r}.
\end{equation}
If $p_r(r)$ is finite at $r=0$ and decays faster than $1/r^N$ with $N>0$ for $r\to \infty$, the choice $f(r)=r^N$ leads to
\begin{equation}
\label{eq:vonLaue-generalization}
\int_0^\infty\ud r\,r^{N-1}\left[Np_r(r)-2s(r)\right]=0\,.
\end{equation}
The case $N=3$ is known as the von Laue condition~\cite{Laue:1911lrk}
\begin{equation}
\label{eq:vonLaue}
\int_0^\infty\ud r\,r^2\,p(r)=0\,,
\end{equation}
and indicates that the isotropic pressure has to change sign. We illustrate this in Fig.~\ref{fig:stability_3D} using the multipole model~\eqref{eq:model} with parameters given in Table~\ref{tab:model}.
As one can see from panel (a) showing the integrand of the von Laue condition~\eqref{eq:vonLaue}, the net positive pressure (i.e. repulsive force) of the inner region is exactly balanced by the net negative pressure (i.e. attractive force) of the outer region. Multiplying this integrand by an additional factor $r^2$ as in Eq.~\eqref{eq:C0} explains why the gravitational charge $C(0)$ turns out to be negative.

Similarly, the case $N=2$
\begin{equation}
\label{eq:stability_3D_pt}
\int_0^\infty\ud r\,r\,p_t(r)=0
\end{equation}
indicates that the tangential pressure also changes sign. The net tangential force in the inner region is repulsive whereas it is negative in the outer region, leading once more to the conclusion that $C(0)<0$ based on Eq.~\eqref{eq:C0}. The other remarkable values are $N=1,\frac{6}{5},\frac{12}{5},4,6$
\begin{align}
\label{eq:stability_3D_pt-s}
\int_0^\infty\ud r\,p_t(r)&=\int_0^\infty\ud r\,s(r)\,,\\
\label{eq:stability_3D_p-s}
\int_0^\infty\ud r\,r^{1/5}\,p(r)&=\int_0^\infty\ud r\,r^{1/5}\,s(r)\,,\\
\label{eq:stability_3D_p+pt}
\int_0^\infty\ud r\,r^{7/5}\,p(r)&=-\int_0^\infty\ud r\,r^{7/5}\,p_t(r)\,,\\
\int_0^\infty\ud r\,r^3\,p_r(r)&=-\int_0^\infty\ud r\,r^3\,p_t(r)\,,\\
\label{eq:stability_3D_p+pr}
\int_0^\infty\ud r\,r^5\,p(r)&=-\int_0^\infty\ud r\,r^5\,p_r(r)\,.
\end{align}

All the above relations are automatically satisfied by our expressions~\eqref{eq:pr} and~\eqref{eq:s} when summed over the constituents. The reason for this is that the parametrization~\eqref{eq:TmunuUU} together with the sum rules~\eqref{eq:A} already encode the conservation of the total EMT, implying that the equation of hydrostatic equilibrium~\eqref{eq:conservationEMT} is identically satisfied. These relations can however be particularly useful to test model predictions where full Lorentz covariance are often absent.

\subsubsection{Two-dimensional case}

The discussion in the EF proceeds analogously to the BF. The main difference is that the spatial distributions are now two-dimensional instead of three-dimensional. Note also that since the two-dimensional pressure distributions are the same in both EF (with $P_z=0$ or $P_z\to\infty$) and DYF, the following results apply to both instant and front forms.

\begin{figure}[t!]
    \centering
    \includegraphics[width=0.49\textwidth]{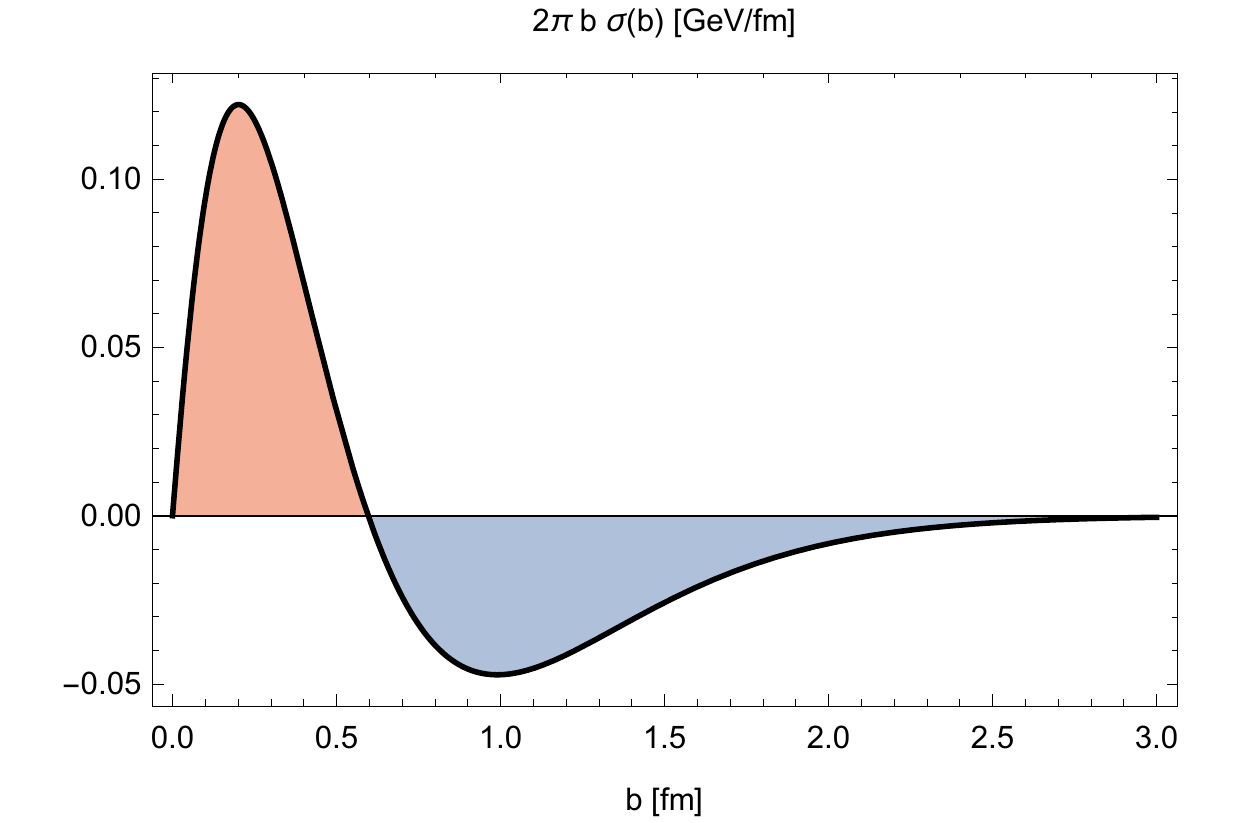}
    \includegraphics[width=0.49\textwidth]{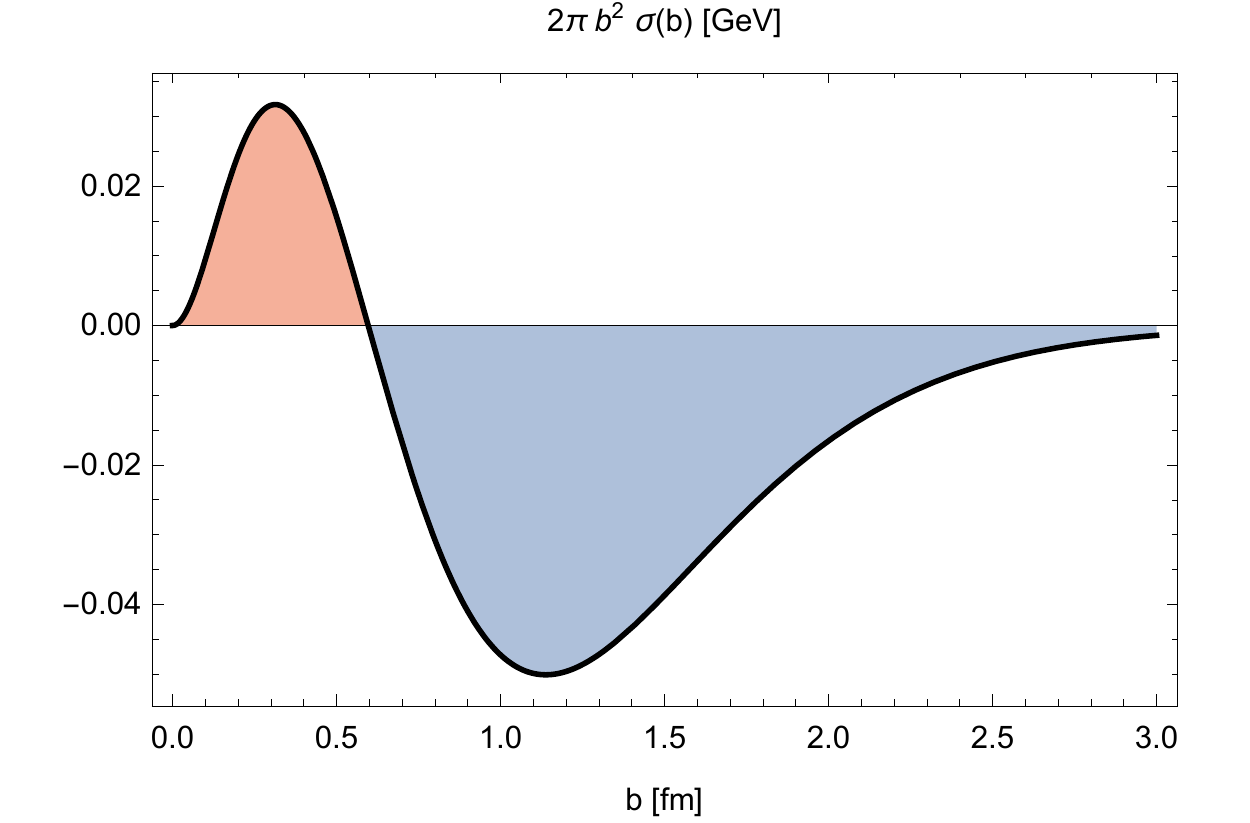}
    \\[1em]
	(a)\hspace*{0.45\linewidth}(b)
    \\[1em]
    \caption{Illustration of the hydrostatic equilibrium in the elastic frame. (a) integrand of the two-dimensional von Laue condition~\eqref{eq:stability_2D_p} and (b) the same integrand multiplied by $b$.
    }
    \label{fig:stability_2D}
\end{figure}

Using the conservation of the static total EMT in two dimensions
\begin{equation}
\nabla^i_\perp\mathsf T^{ij}(\uvec b_\perp;\uvec 0)=0\,,
\end{equation}
we find that the equation of hydrodynamic equilibrium in the EF takes the form
\begin{equation}
\frac{\ud \sigma_r(b)}{\ud b}=-\frac{\Pi(b)}{b}\,,
\end{equation}
where we omitted the $P_z$-dependence for convenience. If $f(b)$ is some radial function, we obtain using integration by parts
\begin{equation}
\left[f(b)\,\sigma_r(b)\right]_0^\infty=\int_0^\infty\ud b\left[\frac{\ud f(b)}{\ud b}\,\sigma_r(b)-\frac{f(b)}{b}\,\Pi(b)\right].
\end{equation}
For $f(b)=1$, we get
\begin{equation}\label{2DYoung}
\sigma_r(0)=\int_0^\infty\ud b\,\frac{\Pi(b)}{b}\,.
\end{equation}
When the pressure anisotropy is concentrated within a thin region, it can be approximated by $\Pi(b)\approx \tau\,\delta(b-R)$. Equation~\eqref{2DYoung} reduces then to the two-dimensional version of the Young-Laplace relation $\sigma_r(0)=\tau/R$, where $\tau$ can be thought of as some sort of effective string tension. More generally, the effective string tension can be defined from the two-dimensional version of the Bakker equation
\begin{equation}
\tau=\int_D\ud b\,\Pi(b).
\end{equation}
If $\sigma_r(b)$ is finite at $b=0$ and decays faster than $1/b^N$ with $N>0$ for $b\to \infty$, the choice $f(b)=b^N$ leads to
\begin{equation}
\int_0^\infty\ud b\,b^{N-1}\left[N\sigma_r(b)-\Pi(b)\right]=0\,.
\end{equation}
The case $N=2$ corresponds to the two-dimensional version of the von Laue condition
\begin{equation}
\label{eq:stability_2D_p}
\int_0^\infty\ud b\,b\,\sigma(b)=0\,,
\end{equation}
and indicates that the isotropic pressure has to change sign. We illustrate this in Fig.~\ref{fig:stability_2D} using the multipole model~\eqref{eq:model} with parameters given in Table~\ref{tab:model}.
As one can see from panel (a) showing the integrand of the two-dimensional version of the von Laue condition~\eqref{eq:stability_2D_p}, the picture is similar to the one in the BF. Namely, the net positive pressure of the inner region is exactly balanced by the net negative pressure of the outer region. Multiplying this integrand by an additional factor $b$ as in Eq.~\eqref{eq:C0bis} explains why the gravitational charge $C(0)$ turns out to be negative.

Similarly, the case $N=1$
\begin{equation}
\label{eq:stability_2D_pt}
\int_0^\infty\ud b\,\sigma_t(b)=0
\end{equation}
indicates that the tangential pressure also changes sign. Once again, the net tangential force in the inner region is repulsive whereas it is negative in the outer region, in agreement with $C(0)<0$ according to Eq.~\eqref{eq:C0bis}. The other remarkable values are $N=\frac{1}{2},\frac{2}{3},\frac{4}{3},4$
\begin{align}
\label{eq:stability_2D_pt-s}
\int_0^\infty\ud b\,b^{-1/2}\,\sigma_t(b)&=\int_0^\infty\ud b\,b^{-1/2}\,\Pi(b)\,,\\
\label{eq:stability_2D_p-s}
\int_0^\infty\ud b\,b^{-1/3}\,\sigma(b)&=\int_0^\infty\ud b\,b^{-1/3}\,\Pi(b)\,,\\
\label{eq:stability_2D_p+pt}
\int_0^\infty\ud b\,b^{1/3}\,\sigma(b)&=-\int_0^\infty\ud b\,b^{1/3}\,\sigma_t(b)\,,\\
\label{eq:stability_2D_p+pr}
\int_0^\infty\ud b\,b^3\,\sigma(b)&=-\int_0^\infty\ud b\,b^3\,\sigma_r(b)\,.
\end{align}

For the same reason as in the three-dimensional case, all the above relations are automatically satisfied by our expressions~\eqref{eq:sigma} and~\eqref{eq:Pi} when summed over the constituents.

\subsection{Stability}

We have seen earlier that the study of the nucleon EMT might provide some clues about the EoS for the matter lying in the heart of compact stars. The stability of compact stars made of anisotropic matter has been extensively discussed in~\cite{Herrera:1992lwz,Herrera:1997plx,Abreu:2007ew}. We suggest that applying these results to the case of the nucleon can in turn provide new constraints on the nucleon EMT, and hence on the GPDs.
\newline

For a stable system, it is expected that\footnote{Note that for compact stars it is also expected that $p_t(r)>0$, and hence $p(r)>0$. This does not contradict Eqs.~\eqref{eq:vonLaue} and~\eqref{eq:stability_3D_pt} since the gravitational force is attractive and long range, leading to a significant stress anisotropy. In some sense, we could interpret the gravitational contribution in Eq.~\eqref{balancestar} as $\frac{1}{2}\frac{\ud\nu(r)}{\ud r}\left[\varepsilon_r(r)+p_r(r)\right]=\frac{\ud p^\text{grav}_r(r)}{\ud r}+\frac{2s^\text{grav}(r)}{r}$, with the expectation that $p_r(r)>p^\text{grav}_r(r)$ and $\int_0^\infty\ud r\,r\left[p_t(r)+p^\text{grav}_t(r)\right]=0$.}
\begin{flalign}
\text{(i)}\ & \varepsilon(0)<\infty,\ p(0)<\infty\ \text{and}\ s(0)=0\,;
\\
\text{(ii)}\ & \varepsilon(r)>0\ \text{and}\ p_r(r)>0\,;
\\
\text{(iii)}\ & \frac{\ud\varepsilon(r)}{\ud r}<0\ \text{and}\ \frac{\ud p_r(r)}{\ud r}<0\,.
\end{flalign}
All these constraints are satisfied by our simple multipole model~\eqref{eq:model} with parameters given in Table~\ref{tab:model}. We observe in particular that the constraint (i) rules out the dipole Ansatz for the GFFs $B_a(t)$ and $C_a(t)$ sometimes used in the literature, because it generates a $1/r$ pole in $\varepsilon(r)$ and $p(r)$, and leads to $s(0)\neq 0$. We therefore used in our simple multipole model the tripole Ansatz which does not have the same problem and which agrees with the asymptotic behaviour expected from the quark counting rules~\cite{Matveev:1973ra,Brodsky:1973kr,Brodsky:1974vy,Lepage:1980fj}.

It is also expected that the (squared) radial and tangential speeds of sound defined as $v^2_{sr}(r)=\frac{\ud p_r(r)/\ud r}{\ud\varepsilon(r)/\ud r}$ and $v^2_{st}(r)=\frac{\ud p_t(r)/\ud r}{\ud\varepsilon(r)/\ud r}$ satisfy 
\begin{flalign}
\text{(iv)}\ & 0\leq v^2_{sr}(r)\leq 1\ \text{and}\ 0\leq v^2_{st}(r)\leq 1\,;
\\
\text{(v)}\ & |v^2_{st}(r)-v^2_{sr}(r)|\leq 1\,;
\\
\text{(vi)}\ & \Gamma(r)=\frac{\varepsilon(r)+p_r(r)}{p_r(r)}\,v^2_{sr}>\frac{4}{3}\,.
\end{flalign}
Violations of these additional constraints are observed over some range in $r$ within our simple multipole model.

Finally, there exist also energy conditions which reflect the principles of relativity and play an important role in General Relativity. They constitute an essential ingredient for establishing general results like the no-hair theorem, the laws of black hole thermodynamics or the singularity theorems of Penrose and Hawking~\cite{Hawking:1973uf,Wald:1984rg,Poisson:2009pwt,Senovilla:2014gza,Curiel:2014zba,Martin-Moruno:2017exc,Maeda:2018hqu}. The most popular ones are the Null Energy Condition (NEC), Weak Energy Condition (WEC), Strong Energy Condition (SEC), and Dominant Energy Condition (DEC)
\begin{align}
&\text{NEC}&\varepsilon(r)+p_i(r)&\geq 0\,,\\
&\text{WEC}&\varepsilon(r)+p_i(r)&\geq 0\qquad \text{and}\qquad \varepsilon(r)\geq 0\,,\\
&\text{SEC}&\varepsilon(r)+p_i(r)&\geq 0\qquad \text{and}\qquad \varepsilon(r)+3\,p(r)\geq 0\,,\\
&\text{DEC}&\varepsilon(r)&\geq |p_i(r)|,
\end{align}
where $i=r,t$. Some of these energy conditions have a simple physical interpretation. Namely, the weak energy condition arises from the requirement that the energy density is non-negative for any observer, and the dominant energy condition ensures that the energy flow cannot exceed the speed of light for any observer. All known forms of matter so far satisfy these energy conditions. Our simple multipole model~\eqref{eq:model} with parameters given in Table~\ref{tab:model} also satisfy these energy conditions except the SEC $\varepsilon(r)+3\,p(r)\geq 0$ for some range in $r$. 

All the above conditions on the distributions of energy density and pressure are extremely interesting since, when transposed to the nucleon case using our expressions~\eqref{eq:epsilon}-\eqref{eq:s}, they can provide new phenomenological constraints on the GFFs and hence on the GPDs of the nucleon\footnote{Note however that some of the bounds may be violated by quantum effects, see~\cite{Ford:1978qya,Fewster:2012yh,Fewster:2017wmn,Fewster:2018pey}.}. Recently, criterion (ii) has been considered in~\cite{Perevalova:2016dln} and led to the conclusion that $C(0)$ should be negative owing to Eq.~\eqref{eq:C0}. We note that this agrees with criterion (iii) which implies that $s(r)>0$ (as assumed in~\cite{Goeke:2007fp}) using the equation of hydrostatic equilibrium~\eqref{eq:conservationEMT}, and in turn $C(0)<0$ according to Eq.~\eqref{eq:sCrel}. Note also that the inequalities can in principle be easily transposed to the two-dimensional case, which in conjunction with our expressions~\eqref{eq:rho}-\eqref{eq:Pi}, may lead to yet further new constraints on the GPDs.

Satisfying all the constraints at the same time is not at all a trivial task. Some may even perhaps be unapplicable to the nucleon case. For these reasons, we refrain from developing here a more realistic model since its main purpose in the present study was only to illustrate the various distributions. A detailed analysis of the stability constraints in the hadronic context goes beyond the scope of the present paper and is left for future investigations.

\section{Conclusions}\label{sec:6}

We revisited the interpretation of spin-1/2 gravitational form factors in terms of the mechanical properties of hadrons. We rederived and significantly extended the existing literature, which so far has been exploiting only the Breit frame and the instant form of dynamics. In particular we discussed here both the instant and front forms of dynamics, and the separate quark and gluon contributions to the energy and pressure distributions in the Breit, elastic, infinite-momentum and Drell-Yan frames. Our key results are contained in Eqs.~(\ref{eq:epsilon})--(\ref{eq:s}) for the Breit frame, Eqs.~(\ref{eq:rho})--(\ref{eq:Pi}) for the elastic frame, and Eqs.~(\ref{eq:mu_LF})--(\ref{eq:Pi_LF}) for the Drell-Yan frame. We illustrated our argument with a simple phenomenological model, and highlighted the constraints coming from mechanical properties that should generically be satisfied in model-building. 

We put a special emphasis on the pressure anisotropy, which offers an innovative perspective on the nucleon structure from our rapidly growing knowledge on compact stellar objects. The direct observation of neutron star mergers in terrestrial gravitational wave observatories indeed already brought constraints on the equation of state of nuclear matter at high density and low temperature. Orders of magnitude and model studies suggest that the nucleon itself may be described with the same concepts and pictures. In particular, the stability and hydrostatic equilibrium conditions are presumably the best theoretical ingredients to elaborate on this picture since most of the gravitational form factors are within contemporary experimental reach through hard exclusive experiments and generalized parton distributions. However, determining how much we can learn about the physics of compact stars from the nucleon structure (or conversely) is still an exciting open question.

\section*{Acknowledgement}

We are grateful to Maxim Polyakov, Peter Schweitzer and Oleg Teryaev for fruitful discussions. This work has been supported by the Agence Nationale de la Recherche under the project ANR-16-CE31-0019, the P2IO LabEx (ANR-10-LABX-0038) in the framework "Investissements d'Avenir" (ANR-11-IDEX-0003-01) managed by the Agence Nationale de la Recherche, and the CEA-Enhanced Eurotalents Program co-funded by FP7 Marie Skłodowska-Curie COFUND Program (Grant agreement no 600382).

\bibliographystyle{apsrev4-1long}
\bibliography{bibliography}

\begin{thebibliography}{100}%
\makeatletter
\providecommand \@ifxundefined [1]{%
 \ifx #1\undefined \expandafter \@firstoftwo
 \else \expandafter \@secondoftwo
\fi
}%
\providecommand \@ifnum [1]{%
 \ifnum #1\expandafter \@firstoftwo
 \else \expandafter \@secondoftwo
\fi
}%
\providecommand \enquote [1]{``#1''}%
\providecommand \bibnamefont  [1]{#1}%
\providecommand \bibfnamefont [1]{#1}%
\providecommand \citenamefont [1]{#1}%
\providecommand\href[0]{\@sanitize\@href}%
\providecommand\@href[1]{\endgroup\@@startlink{#1}\endgroup\@@href}%
\providecommand\@@href[1]{#1\@@endlink}%
\providecommand \@sanitize [0]{\begingroup\catcode`\&12\catcode`\#12\relax}%
\@ifxundefined \pdfoutput {\@firstoftwo}{%
 \@ifnum{\z@=\pdfoutput}{\@firstoftwo}{\@secondoftwo}%
}{%
 \providecommand\@@startlink[1]{\leavevmode\special{html:<a href="#1">}}%
 \providecommand\@@endlink[0]{\special{html:</a>}}%
}{%
 \providecommand\@@startlink[1]{%
  \leavevmode
  \pdfstartlink
   attr{/Border[0 0 1 ]/H/I/C[0 1 1]}%
   user{/Subtype/Link/A<</Type/Action/S/URI/URI(#1)>>}%
  \relax
 }%
 \providecommand\@@endlink[0]{\pdfendlink}%
}%
\providecommand \url  [0]{\begingroup\@sanitize \@url }%
\providecommand \@url [1]{\endgroup\@href {#1}{\urlprefix}}%
\providecommand \urlprefix [0]{URL }%
\providecommand \Eprint[0]{\href }%
\@ifxundefined \urlstyle {%
  \providecommand \doi [1]{doi:\discretionary{}{}{}#1}%
}{%
  \providecommand \doi [0]{doi:\discretionary{}{}{}\begingroup
  \urlstyle{rm}\Url }%
}%
\providecommand \doibase [0]{http://dx.doi.org/}%
\providecommand \Doi[1]{\href{\doibase#1}}%
\providecommand \bibAnnote [3]{%
  \BibitemShut{#1}%
  \begin{quotation}\noindent
    \textsc{Key:}\ #2\\\textsc{Annotation:}\ #3%
  \end{quotation}%
}%
\providecommand \bibAnnoteFile [2]{%
  \IfFileExists{#2}{\bibAnnote {#1} {#2} {\input{#2}}}{}%
}%
\providecommand \typeout [0]{\immediate \write \m@ne }%
\providecommand \selectlanguage [0]{\@gobble}%
\providecommand \bibinfo [0]{\@secondoftwo}%
\providecommand \bibfield [0]{\@secondoftwo}%
\providecommand \translation [1]{[#1]}%
\providecommand \BibitemOpen[0]{}%
\providecommand \bibitemStop [0]{}%
\providecommand \bibitemNoStop [0]{.\EOS\space}%
\providecommand \EOS [0]{\spacefactor3000\relax}%
\providecommand \BibitemShut [1]{\csname bibitem#1\endcsname}%
\bibitem{Ji:1996ek}%
  \BibitemOpen
  \bibfield{author}{%
  \bibinfo {author} {\bibfnamefont{Xiang-Dong}\ \bibnamefont{Ji}},\ }%
  \bibfield{title}{%
  \enquote{\bibinfo {title} {{Gauge-Invariant Decomposition of Nucleon
  Spin}},}\ }%
  \bibfield{journal}{%
  \Doi{10.1103/PhysRevLett.78.610}{\bibinfo {journal} {Phys. Rev. Lett.}}\ }%
  \textbf{\bibinfo {volume} {78}},\ \bibinfo {pages} {610--613} (\bibinfo
  {year} {1997}),\
  \Eprint{http://arxiv.org/abs/hep-ph/9603249}{arXiv:hep-ph/9603249 [hep-ph]}%
  \bibAnnoteFile{NoStop}{Ji:1996ek}%
\bibitem{Ji:1996nm}%
  \BibitemOpen
  \bibfield{author}{%
  \bibinfo {author} {\bibfnamefont{Xiang-Dong}\ \bibnamefont{Ji}},\ }%
  \bibfield{title}{%
  \enquote{\bibinfo {title} {{Deeply virtual Compton scattering}},}\ }%
  \bibfield{journal}{%
  \Doi{10.1103/PhysRevD.55.7114}{\bibinfo {journal} {Phys. Rev.}}\ }%
  \textbf{\bibinfo {volume} {D55}},\ \bibinfo {pages} {7114--7125} (\bibinfo
  {year} {1997}),\
  \Eprint{http://arxiv.org/abs/hep-ph/9609381}{arXiv:hep-ph/9609381 [hep-ph]}%
  \bibAnnoteFile{NoStop}{Ji:1996nm}%
\bibitem{Accardi:2012qut}%
  \BibitemOpen
  \bibfield{author}{%
  \bibinfo {author} {\bibfnamefont{A.}~\bibnamefont{Accardi}} \emph{et~al.},\
  }%
  \bibfield{title}{%
  \enquote{\bibinfo {title} {{Electron Ion Collider: The Next QCD Frontier}},}\
  }%
  \bibfield{journal}{%
  \Doi{10.1140/epja/i2016-16268-9}{\bibinfo {journal} {Eur. Phys. J.}}\ }%
  \textbf{\bibinfo {volume} {A52}},\ \bibinfo {pages} {268} (\bibinfo {year}
  {2016}),\ \Eprint{http://arxiv.org/abs/1212.1701}{arXiv:1212.1701 [nucl-ex]}%
  \bibAnnoteFile{NoStop}{Accardi:2012qut}%
\bibitem{Yang:2018bft}%
  \BibitemOpen
  \bibfield{author}{%
  \bibinfo {author} {\bibfnamefont{Yi-Bo}\ \bibnamefont{Yang}}, \bibinfo
  {author} {\bibfnamefont{Ming}\ \bibnamefont{Gong}}, \bibinfo {author}
  {\bibfnamefont{Jian}\ \bibnamefont{Liang}}, \bibinfo {author}
  {\bibfnamefont{Huey-Wen}\ \bibnamefont{Lin}}, \bibinfo {author}
  {\bibfnamefont{Keh-Fei}\ \bibnamefont{Liu}}, \bibinfo {author}
  {\bibfnamefont{Dimitra}\ \bibnamefont{Pefkou}},\ and\ \bibinfo {author}
  {\bibfnamefont{Phiala}\ \bibnamefont{Shanahan}},\ }%
  \bibfield{title}{%
  \enquote{\bibinfo {title} {{Nonperturbatively renormalized glue momentum
  fraction at the physical pion mass from lattice QCD}},}\ }%
  \bibfield{journal}{%
  \Doi{10.1103/PhysRevD.98.074506}{\bibinfo {journal} {Phys. Rev.}}\ }%
  \textbf{\bibinfo {volume} {D98}},\ \bibinfo {pages} {074506} (\bibinfo {year}
  {2018}),\ \Eprint{http://arxiv.org/abs/1805.00531}{arXiv:1805.00531
  [hep-lat]}%
  \bibAnnoteFile{NoStop}{Yang:2018bft}%
\bibitem{Alexandrou:2018xnp}%
  \BibitemOpen
  \bibfield{author}{%
  \bibinfo {author} {\bibfnamefont{Constantia}\ \bibnamefont{Alexandrou}},
  \bibinfo {author} {\bibfnamefont{Martha}\ \bibnamefont{Constantinou}},
  \bibinfo {author} {\bibfnamefont{Kyriakos}\ \bibnamefont{Hadjiyiannakou}},
  \bibinfo {author} {\bibfnamefont{Karl}\ \bibnamefont{Jansen}}, \bibinfo
  {author} {\bibfnamefont{Christos}\ \bibnamefont{Kallidonis}}, \bibinfo
  {author} {\bibfnamefont{Giannis}\ \bibnamefont{Koutsou}},\ and\ \bibinfo
  {author} {\bibfnamefont{Alejandro}\ \bibnamefont{Vaquero Avilés-Casco}},\ }%
  \bibfield{title}{%
  \enquote{\bibinfo {title} {{Nucleon spin structure from lattice QCD}},}\ }%
  \bibfield{booktitle}{%
  \emph{\bibinfo {booktitle} {{Proceedings, 26th International Workshop on Deep
  Inelastic Scattering and Related Subjects (DIS 2018): Port Island, Kobe,
  Japan, April 16-20, 2018}}},\ }%
  \bibfield{journal}{%
  \Doi{10.22323/1.316.0148}{\bibinfo {journal} {PoS}}\ }%
  \textbf{\bibinfo {volume} {DIS2018}},\ \bibinfo {pages} {148} (\bibinfo
  {year} {2018}),\ \Eprint{http://arxiv.org/abs/1807.11214}{arXiv:1807.11214
  [hep-lat]}%
  \bibAnnoteFile{NoStop}{Alexandrou:2018xnp}%
\bibitem{Yang:2018nqn}%
  \BibitemOpen
  \bibfield{author}{%
  \bibinfo {author} {\bibfnamefont{Yi-Bo}\ \bibnamefont{Yang}}, \bibinfo
  {author} {\bibfnamefont{Jian}\ \bibnamefont{Liang}}, \bibinfo {author}
  {\bibfnamefont{Yu-Jiang}\ \bibnamefont{Bi}}, \bibinfo {author}
  {\bibfnamefont{Ying}\ \bibnamefont{Chen}}, \bibinfo {author}
  {\bibfnamefont{Terrence}\ \bibnamefont{Draper}}, \bibinfo {author}
  {\bibfnamefont{Keh-Fei}\ \bibnamefont{Liu}},\ and\ \bibinfo {author}
  {\bibfnamefont{Zhaofeng}\ \bibnamefont{Liu}},\ }%
  \bibfield{title}{%
  \enquote{\bibinfo {title} {{Proton Mass Decomposition from the QCD Energy
  Momentum Tensor}},}\ }%
  \bibfield{journal}{%
  \Doi{10.1103/PhysRevLett.121.212001}{\bibinfo {journal} {Phys. Rev. Lett.}}\
  }%
  \textbf{\bibinfo {volume} {121}},\ \bibinfo {pages} {212001} (\bibinfo {year}
  {2018}),\ \Eprint{http://arxiv.org/abs/1808.08677}{arXiv:1808.08677
  [hep-lat]}%
  \bibAnnoteFile{NoStop}{Yang:2018nqn}%
\bibitem{Shanahan:2018pib}%
  \BibitemOpen
  \bibfield{author}{%
  \bibinfo {author} {\bibfnamefont{P.~E.}\ \bibnamefont{Shanahan}}\ and\
  \bibinfo {author} {\bibfnamefont{W.}~\bibnamefont{Detmold}},\ }%
  \bibfield{title}{%
  \enquote{\bibinfo {title} {{Gluon gravitational form factors of the nucleon
  and the pion from lattice QCD}},}\ }%
   (\bibinfo {year} {2018}),\
  \Eprint{http://arxiv.org/abs/1810.04626}{arXiv:1810.04626 [hep-lat]}%
  \bibAnnoteFile{NoStop}{Shanahan:2018pib}%
\bibitem{Shanahan:2018nnv}%
  \BibitemOpen
  \bibfield{author}{%
  \bibinfo {author} {\bibfnamefont{P.~E.}\ \bibnamefont{Shanahan}}\ and\
  \bibinfo {author} {\bibfnamefont{W.}~\bibnamefont{Detmold}},\ }%
  \bibfield{title}{%
  \enquote{\bibinfo {title} {{The pressure distribution and shear forces inside
  the proton}},}\ }%
   (\bibinfo {year} {2018}),\
  \Eprint{http://arxiv.org/abs/1810.07589}{arXiv:1810.07589 [nucl-th]}%
  \bibAnnoteFile{NoStop}{Shanahan:2018nnv}%
\bibitem{Pais:1949vdk}%
  \BibitemOpen
  \bibfield{author}{%
  \bibinfo {author} {\bibfnamefont{A.}~\bibnamefont{Pais}}\ and\ \bibinfo
  {author} {\bibfnamefont{S.~T.}\ \bibnamefont{Epstein}},\ }%
  \bibfield{title}{%
  \enquote{\bibinfo {title} {{Note on Relativistic Properties of
  Self-Energies}},}\ }%
  \bibfield{journal}{%
  \Doi{10.1103/RevModPhys.21.445}{\bibinfo {journal} {Rev. Mod. Phys.}}\ }%
  \textbf{\bibinfo {volume} {21}},\ \bibinfo {pages} {445--446} (\bibinfo
  {year} {1949})%
  \bibAnnoteFile{NoStop}{Pais:1949vdk}%
\bibitem{Polyakov:2002yz}%
  \BibitemOpen
  \bibfield{author}{%
  \bibinfo {author} {\bibfnamefont{M.~V.}\ \bibnamefont{Polyakov}},\ }%
  \bibfield{title}{%
  \enquote{\bibinfo {title} {{Generalized parton distributions and strong
  forces inside nucleons and nuclei}},}\ }%
  \bibfield{journal}{%
  \Doi{10.1016/S0370-2693(03)00036-4}{\bibinfo {journal} {Phys. Lett.}}\ }%
  \textbf{\bibinfo {volume} {B555}},\ \bibinfo {pages} {57--62} (\bibinfo
  {year} {2003}),\
  \Eprint{http://arxiv.org/abs/hep-ph/0210165}{arXiv:hep-ph/0210165 [hep-ph]}%
  \bibAnnoteFile{NoStop}{Polyakov:2002yz}%
\bibitem{Polyakov:2018zvc}%
  \BibitemOpen
  \bibfield{author}{%
  \bibinfo {author} {\bibfnamefont{Maxim~V.}\ \bibnamefont{Polyakov}}\ and\
  \bibinfo {author} {\bibfnamefont{Peter}\ \bibnamefont{Schweitzer}},\ }%
  \bibfield{title}{%
  \enquote{\bibinfo {title} {{Forces inside hadrons: pressure, surface tension,
  mechanical radius, and all that}},}\ }%
  \bibfield{journal}{%
  \Doi{10.1142/S0217751X18300259}{\bibinfo {journal} {Int. J. Mod. Phys.}}\ }%
  \textbf{\bibinfo {volume} {A33}},\ \bibinfo {pages} {1830025} (\bibinfo
  {year} {2018}),\ \Eprint{http://arxiv.org/abs/1805.06596}{arXiv:1805.06596
  [hep-ph]}%
  \bibAnnoteFile{NoStop}{Polyakov:2018zvc}%
\bibitem{Ruderman:1972aj}%
  \BibitemOpen
  \bibfield{author}{%
  \bibinfo {author} {\bibfnamefont{M.}~\bibnamefont{Ruderman}},\ }%
  \bibfield{title}{%
  \enquote{\bibinfo {title} {{Pulsars: structure and dynamics}},}\ }%
  \bibfield{journal}{%
  \Doi{10.1146/annurev.aa.10.090172.002235}{\bibinfo {journal} {Ann. Rev.
  Astron. Astrophys.}}\ }%
  \textbf{\bibinfo {volume} {10}},\ \bibinfo {pages} {427--476} (\bibinfo
  {year} {1972})%
  \bibAnnoteFile{NoStop}{Ruderman:1972aj}%
\bibitem{Canuto:1974ft}%
  \BibitemOpen
  \bibfield{author}{%
  \bibinfo {author} {\bibfnamefont{V.}~\bibnamefont{Canuto}},\ }%
  \bibfield{title}{%
  \enquote{\bibinfo {title} {{Equation of State at Ultrahigh Densities. 1.}}.}\
  }%
  \bibfield{journal}{%
  \Doi{10.1146/annurev.aa.12.090174.001123}{\bibinfo {journal} {Ann. Rev.
  Astron. Astrophys.}}\ }%
  \textbf{\bibinfo {volume} {12}},\ \bibinfo {pages} {167--214} (\bibinfo
  {year} {1974})%
  \bibAnnoteFile{NoStop}{Canuto:1974ft}%
\bibitem{Alcock:1986hz}%
  \BibitemOpen
  \bibfield{author}{%
  \bibinfo {author} {\bibfnamefont{Charles}\ \bibnamefont{Alcock}}, \bibinfo
  {author} {\bibfnamefont{Edward}\ \bibnamefont{Farhi}},\ and\ \bibinfo
  {author} {\bibfnamefont{Angela}\ \bibnamefont{Olinto}},\ }%
  \bibfield{title}{%
  \enquote{\bibinfo {title} {{Strange stars}},}\ }%
  \bibfield{journal}{%
  \Doi{10.1086/164679}{\bibinfo {journal} {Astrophys. J.}}\ }%
  \textbf{\bibinfo {volume} {310}},\ \bibinfo {pages} {261--272} (\bibinfo
  {year} {1986})%
  \bibAnnoteFile{NoStop}{Alcock:1986hz}%
\bibitem{Haensel:1986qb}%
  \BibitemOpen
  \bibfield{author}{%
  \bibinfo {author} {\bibfnamefont{P.}~\bibnamefont{Haensel}}, \bibinfo
  {author} {\bibfnamefont{J.~L.}\ \bibnamefont{Zdunik}},\ and\ \bibinfo
  {author} {\bibfnamefont{R.}~\bibnamefont{Schaeffer}},\ }%
  \bibfield{title}{%
  \enquote{\bibinfo {title} {{Strange quark stars}},}\ }%
  \bibfield{journal}{%
  \bibinfo {journal} {Astron. Astrophys.}\ }%
  \textbf{\bibinfo {volume} {160}},\ \bibinfo {pages} {121--128} (\bibinfo
  {year} {1986})%
  \bibAnnoteFile{NoStop}{Haensel:1986qb}%
\bibitem{Weber:2004kj}%
  \BibitemOpen
  \bibfield{author}{%
  \bibinfo {author} {\bibfnamefont{Fridolin}\ \bibnamefont{Weber}},\ }%
  \bibfield{title}{%
  \enquote{\bibinfo {title} {{Strange quark matter and compact stars}},}\ }%
  \bibfield{journal}{%
  \Doi{10.1016/j.ppnp.2004.07.001}{\bibinfo {journal} {Prog. Part. Nucl.
  Phys.}}\ }%
  \textbf{\bibinfo {volume} {54}},\ \bibinfo {pages} {193--288} (\bibinfo
  {year} {2005}),\
  \Eprint{http://arxiv.org/abs/astro-ph/0407155}{arXiv:astro-ph/0407155
  [astro-ph]}%
  \bibAnnoteFile{NoStop}{Weber:2004kj}%
\bibitem{PerezGarcia:2010ap}%
  \BibitemOpen
  \bibfield{author}{%
  \bibinfo {author} {\bibfnamefont{M.~Angeles}\ \bibnamefont{Perez-Garcia}},
  \bibinfo {author} {\bibfnamefont{Joseph}\ \bibnamefont{Silk}},\ and\ \bibinfo
  {author} {\bibfnamefont{Jirina~R.}\ \bibnamefont{Stone}},\ }%
  \bibfield{title}{%
  \enquote{\bibinfo {title} {{Dark matter, neutron stars and strange quark
  matter}},}\ }%
  \bibfield{journal}{%
  \Doi{10.1103/PhysRevLett.105.141101}{\bibinfo {journal} {Phys. Rev. Lett.}}\
  }%
  \textbf{\bibinfo {volume} {105}},\ \bibinfo {pages} {141101} (\bibinfo {year}
  {2010}),\ \Eprint{http://arxiv.org/abs/1007.1421}{arXiv:1007.1421
  [astro-ph.CO]}%
  \bibAnnoteFile{NoStop}{PerezGarcia:2010ap}%
\bibitem{Rodrigues:2011zza}%
  \BibitemOpen
  \bibfield{author}{%
  \bibinfo {author} {\bibfnamefont{Hilario}\ \bibnamefont{Rodrigues}}, \bibinfo
  {author} {\bibfnamefont{Sergio~Barbosa}\ \bibnamefont{Duarte}},\ and\
  \bibinfo {author} {\bibfnamefont{Jose Carlos~T.}\
  \bibnamefont{De~Oliveira}},\ }%
  \bibfield{title}{%
  \enquote{\bibinfo {title} {{Massive compact stars as quark stars}},}\ }%
  \bibfield{journal}{%
  \Doi{10.1088/0004-637X/730/1/31}{\bibinfo {journal} {Astrophys. J.}}\ }%
  \textbf{\bibinfo {volume} {730}},\ \bibinfo {pages} {31} (\bibinfo {year}
  {2011}),\ \Eprint{http://arxiv.org/abs/1407.4704}{arXiv:1407.4704
  [astro-ph.HE]}%
  \bibAnnoteFile{NoStop}{Rodrigues:2011zza}%
\bibitem{Bowers:1974tgi}%
  \BibitemOpen
  \bibfield{author}{%
  \bibinfo {author} {\bibfnamefont{Richard~L.}\ \bibnamefont{Bowers}}\ and\
  \bibinfo {author} {\bibfnamefont{E.~P.~T.}\ \bibnamefont{Liang}},\ }%
  \bibfield{title}{%
  \enquote{\bibinfo {title} {{Anisotropic Spheres in General Relativity}},}\ }%
  \bibfield{journal}{%
  \Doi{10.1086/152760}{\bibinfo {journal} {Astrophys. J.}}\ }%
  \textbf{\bibinfo {volume} {188}},\ \bibinfo {pages} {657--665} (\bibinfo
  {year} {1974})%
  \bibAnnoteFile{NoStop}{Bowers:1974tgi}%
\bibitem{Herrera:1997plx}%
  \BibitemOpen
  \bibfield{author}{%
  \bibinfo {author} {\bibfnamefont{L.}~\bibnamefont{Herrera}}\ and\ \bibinfo
  {author} {\bibfnamefont{N.~O.}\ \bibnamefont{Santos}},\ }%
  \bibfield{title}{%
  \enquote{\bibinfo {title} {{Local anisotropy in self-gravitating systems}},}\
  }%
  \bibfield{journal}{%
  \Doi{10.1016/S0370-1573(96)00042-7}{\bibinfo {journal} {Phys. Rept.}}\ }%
  \textbf{\bibinfo {volume} {286}},\ \bibinfo {pages} {53--130} (\bibinfo
  {year} {1997})%
  \bibAnnoteFile{NoStop}{Herrera:1997plx}%
\bibitem{Mak:2001eb}%
  \BibitemOpen
  \bibfield{author}{%
  \bibinfo {author} {\bibfnamefont{M.~K.}\ \bibnamefont{Mak}}\ and\ \bibinfo
  {author} {\bibfnamefont{T.}~\bibnamefont{Harko}},\ }%
  \bibfield{title}{%
  \enquote{\bibinfo {title} {{Anisotropic stars in general relativity}},}\ }%
  \bibfield{journal}{%
  \Doi{10.1098/rspa.2002.1014}{\bibinfo {journal} {Proc. Roy. Soc. Lond.}}\ }%
  \textbf{\bibinfo {volume} {A459}},\ \bibinfo {pages} {393--408} (\bibinfo
  {year} {2003}),\
  \Eprint{http://arxiv.org/abs/gr-qc/0110103}{arXiv:gr-qc/0110103 [gr-qc]}%
  \bibAnnoteFile{NoStop}{Mak:2001eb}%
\bibitem{Dev:2000gt}%
  \BibitemOpen
  \bibfield{author}{%
  \bibinfo {author} {\bibfnamefont{Krsna}\ \bibnamefont{Dev}}\ and\ \bibinfo
  {author} {\bibfnamefont{Marcelo}\ \bibnamefont{Gleiser}},\ }%
  \bibfield{title}{%
  \enquote{\bibinfo {title} {{Anisotropic stars: Exact solutions}},}\ }%
  \bibfield{journal}{%
  \Doi{10.1023/A:1020707906543}{\bibinfo {journal} {Gen. Rel. Grav.}}\ }%
  \textbf{\bibinfo {volume} {34}},\ \bibinfo {pages} {1793--1818} (\bibinfo
  {year} {2002}),\
  \Eprint{http://arxiv.org/abs/astro-ph/0012265}{arXiv:astro-ph/0012265
  [astro-ph]}%
  \bibAnnoteFile{NoStop}{Dev:2000gt}%
\bibitem{Dev:2003qd}%
  \BibitemOpen
  \bibfield{author}{%
  \bibinfo {author} {\bibfnamefont{Krsna}\ \bibnamefont{Dev}}\ and\ \bibinfo
  {author} {\bibfnamefont{Marcelo}\ \bibnamefont{Gleiser}},\ }%
  \bibfield{title}{%
  \enquote{\bibinfo {title} {{Anisotropic stars. 2. Stability}},}\ }%
  \bibfield{journal}{%
  \Doi{10.1023/A:1024534702166}{\bibinfo {journal} {Gen. Rel. Grav.}}\ }%
  \textbf{\bibinfo {volume} {35}},\ \bibinfo {pages} {1435--1457} (\bibinfo
  {year} {2003}),\
  \Eprint{http://arxiv.org/abs/gr-qc/0303077}{arXiv:gr-qc/0303077 [gr-qc]}%
  \bibAnnoteFile{NoStop}{Dev:2003qd}%
\bibitem{Silva:2014fca}%
  \BibitemOpen
  \bibfield{author}{%
  \bibinfo {author} {\bibfnamefont{Hector~O.}\ \bibnamefont{Silva}}, \bibinfo
  {author} {\bibfnamefont{Caio F.~B.}\ \bibnamefont{Macedo}}, \bibinfo {author}
  {\bibfnamefont{Emanuele}\ \bibnamefont{Berti}},\ and\ \bibinfo {author}
  {\bibfnamefont{Luís C.~B.}\ \bibnamefont{Crispino}},\ }%
  \bibfield{title}{%
  \enquote{\bibinfo {title} {{Slowly rotating anisotropic neutron stars in
  general relativity and scalar–tensor theory}},}\ }%
  \bibfield{journal}{%
  \Doi{10.1088/0264-9381/32/14/145008}{\bibinfo {journal} {Class. Quant.
  Grav.}}\ }%
  \textbf{\bibinfo {volume} {32}},\ \bibinfo {pages} {145008} (\bibinfo {year}
  {2015}),\ \Eprint{http://arxiv.org/abs/1411.6286}{arXiv:1411.6286 [gr-qc]}%
  \bibAnnoteFile{NoStop}{Silva:2014fca}%
\bibitem{Boehmer:2006ye}%
  \BibitemOpen
  \bibfield{author}{%
  \bibinfo {author} {\bibfnamefont{C.~G.}\ \bibnamefont{Boehmer}}\ and\
  \bibinfo {author} {\bibfnamefont{T.}~\bibnamefont{Harko}},\ }%
  \bibfield{title}{%
  \enquote{\bibinfo {title} {{Bounds on the basic physical parameters for
  anisotropic compact general relativistic objects}},}\ }%
  \bibfield{journal}{%
  \Doi{10.1088/0264-9381/23/22/023}{\bibinfo {journal} {Class. Quant. Grav.}}\
  }%
  \textbf{\bibinfo {volume} {23}},\ \bibinfo {pages} {6479--6491} (\bibinfo
  {year} {2006}),\
  \Eprint{http://arxiv.org/abs/gr-qc/0609061}{arXiv:gr-qc/0609061 [gr-qc]}%
  \bibAnnoteFile{NoStop}{Boehmer:2006ye}%
\bibitem{Goeke:2007fp}%
  \BibitemOpen
  \bibfield{author}{%
  \bibinfo {author} {\bibfnamefont{K.}~\bibnamefont{Goeke}}, \bibinfo {author}
  {\bibfnamefont{J.}~\bibnamefont{Grabis}}, \bibinfo {author}
  {\bibfnamefont{J.}~\bibnamefont{Ossmann}}, \bibinfo {author}
  {\bibfnamefont{M.~V.}\ \bibnamefont{Polyakov}}, \bibinfo {author}
  {\bibfnamefont{P.}~\bibnamefont{Schweitzer}}, \bibinfo {author}
  {\bibfnamefont{A.}~\bibnamefont{Silva}},\ and\ \bibinfo {author}
  {\bibfnamefont{D.}~\bibnamefont{Urbano}},\ }%
  \bibfield{title}{%
  \enquote{\bibinfo {title} {{Nucleon form-factors of the energy momentum
  tensor in the chiral quark-soliton model}},}\ }%
  \bibfield{journal}{%
  \Doi{10.1103/PhysRevD.75.094021}{\bibinfo {journal} {Phys. Rev.}}\ }%
  \textbf{\bibinfo {volume} {D75}},\ \bibinfo {pages} {094021} (\bibinfo {year}
  {2007}),\ \Eprint{http://arxiv.org/abs/hep-ph/0702030}{arXiv:hep-ph/0702030
  [hep-ph]}%
  \bibAnnoteFile{NoStop}{Goeke:2007fp}%
\bibitem{Burkardt:2002hr}%
  \BibitemOpen
  \bibfield{author}{%
  \bibinfo {author} {\bibfnamefont{Matthias}\ \bibnamefont{Burkardt}},\ }%
  \bibfield{title}{%
  \enquote{\bibinfo {title} {{Impact parameter space interpretation for
  generalized parton distributions}},}\ }%
  \bibfield{journal}{%
  \Doi{10.1142/S0217751X03012370}{\bibinfo {journal} {Int. J. Mod. Phys.}}\ }%
  \textbf{\bibinfo {volume} {A18}},\ \bibinfo {pages} {173--208} (\bibinfo
  {year} {2003}),\
  \Eprint{http://arxiv.org/abs/hep-ph/0207047}{arXiv:hep-ph/0207047 [hep-ph]}%
  \bibAnnoteFile{NoStop}{Burkardt:2002hr}%
\bibitem{Leader:2013jra}%
  \BibitemOpen
  \bibfield{author}{%
  \bibinfo {author} {\bibfnamefont{E.}~\bibnamefont{Leader}}\ and\ \bibinfo
  {author} {\bibfnamefont{C.}~\bibnamefont{Lorcé}},\ }%
  \bibfield{title}{%
  \enquote{\bibinfo {title} {{The angular momentum controversy: What’s it all
  about and does it matter?}}.}\ }%
  \bibfield{journal}{%
  \Doi{10.1016/j.physrep.2014.02.010}{\bibinfo {journal} {Phys. Rept.}}\ }%
  \textbf{\bibinfo {volume} {541}},\ \bibinfo {pages} {163--248} (\bibinfo
  {year} {2014}),\ \Eprint{http://arxiv.org/abs/1309.4235}{arXiv:1309.4235
  [hep-ph]}%
  \bibAnnoteFile{NoStop}{Leader:2013jra}%
\bibitem{Lorce:2017wkb}%
  \BibitemOpen
  \bibfield{author}{%
  \bibinfo {author} {\bibfnamefont{Cédric}\ \bibnamefont{Lorcé}}, \bibinfo
  {author} {\bibfnamefont{Luca}\ \bibnamefont{Mantovani}},\ and\ \bibinfo
  {author} {\bibfnamefont{Barbara}\ \bibnamefont{Pasquini}},\ }%
  \bibfield{title}{%
  \enquote{\bibinfo {title} {{Spatial distribution of angular momentum inside
  the nucleon}},}\ }%
  \bibfield{journal}{%
  \Doi{10.1016/j.physletb.2017.11.018}{\bibinfo {journal} {Phys. Lett.}}\ }%
  \textbf{\bibinfo {volume} {B776}},\ \bibinfo {pages} {38--47} (\bibinfo
  {year} {2018}),\ \Eprint{http://arxiv.org/abs/1704.08557}{arXiv:1704.08557
  [hep-ph]}%
  \bibAnnoteFile{NoStop}{Lorce:2017wkb}%
\bibitem{Hillery:1983ms}%
  \BibitemOpen
  \bibfield{author}{%
  \bibinfo {author} {\bibfnamefont{M.}~\bibnamefont{Hillery}}, \bibinfo
  {author} {\bibfnamefont{R.~F.}\ \bibnamefont{O'Connell}}, \bibinfo {author}
  {\bibfnamefont{M.~O.}\ \bibnamefont{Scully}},\ and\ \bibinfo {author}
  {\bibfnamefont{Eugene~P.}\ \bibnamefont{Wigner}},\ }%
  \bibfield{title}{%
  \enquote{\bibinfo {title} {{Distribution functions in physics:
  Fundamentals}},}\ }%
  \bibfield{journal}{%
  \Doi{10.1016/0370-1573(84)90160-1}{\bibinfo {journal} {Phys. Rept.}}\ }%
  \textbf{\bibinfo {volume} {106}},\ \bibinfo {pages} {121--167} (\bibinfo
  {year} {1984})%
  \bibAnnoteFile{NoStop}{Hillery:1983ms}%
\bibitem{Villars:1950pkp}%
  \BibitemOpen
  \bibfield{author}{%
  \bibinfo {author} {\bibfnamefont{F.}~\bibnamefont{Villars}},\ }%
  \bibfield{title}{%
  \enquote{\bibinfo {title} {{On the Energy-Momentum Tensor of the
  Electron}},}\ }%
  \bibfield{journal}{%
  \Doi{10.1103/PhysRev.79.122}{\bibinfo {journal} {Phys. Rev.}}\ }%
  \textbf{\bibinfo {volume} {79}},\ \bibinfo {pages} {122--128} (\bibinfo
  {year} {1950})%
  \bibAnnoteFile{NoStop}{Villars:1950pkp}%
\bibitem{Kobzarev:1962wt}%
  \BibitemOpen
  \bibfield{author}{%
  \bibinfo {author} {\bibfnamefont{I.~{\relax Yu}.}\ \bibnamefont{Kobzarev}}\
  and\ \bibinfo {author} {\bibfnamefont{L.~B.}\ \bibnamefont{Okun}},\ }%
  \bibfield{title}{%
  \enquote{\bibinfo {title} {{GRAVITATIONAL INTERACTION OF FERMIONS}},}\ }%
  \bibfield{journal}{%
  \bibinfo {journal} {Zh. Eksp. Teor. Fiz.}\ }%
  \textbf{\bibinfo {volume} {43}},\ \bibinfo {pages} {1904--1909} (\bibinfo
  {year} {1962}),\ \bibinfo {note} {[Sov. Phys. JETP16,1343 (1963)]}%
  \bibAnnoteFile{NoStop}{Kobzarev:1962wt}%
\bibitem{Kobzarev:1962wt2}%
  \BibitemOpen
  \bibfield{author}{%
  \bibinfo {author} {\bibfnamefont{I.~{\relax Yu}.}\ \bibnamefont{Kobzarev}}\
  and\ \bibinfo {author} {\bibfnamefont{L.~B.}\ \bibnamefont{Okun}},\ }%
  \bibfield{title}{%
  \enquote{\bibinfo {title} {{GRAVITATIONAL INTERACTION OF FERMIONS}},}\ }%
  \bibfield{journal}{%
  \bibinfo {journal} {Sov. Phys. JETP}\ }%
  \textbf{\bibinfo {volume} {16}},\ \bibinfo {pages} {1343} (\bibinfo {year}
  {1963}),\ \bibinfo {note} {[Zh. Eksp. Teor. Fiz.43, 1904 (1962)]}%
  \bibAnnoteFile{NoStop}{Kobzarev:1962wt2}%
\bibitem{Pagels:1966zza}%
  \BibitemOpen
  \bibfield{author}{%
  \bibinfo {author} {\bibfnamefont{Heinz}\ \bibnamefont{Pagels}},\ }%
  \bibfield{title}{%
  \enquote{\bibinfo {title} {{Energy-Momentum Structure Form Factors of
  Particles}},}\ }%
  \bibfield{journal}{%
  \Doi{10.1103/PhysRev.144.1250}{\bibinfo {journal} {Phys. Rev.}}\ }%
  \textbf{\bibinfo {volume} {144}},\ \bibinfo {pages} {1250--1260} (\bibinfo
  {year} {1966})%
  \bibAnnoteFile{NoStop}{Pagels:1966zza}%
\bibitem{Bakker:2004ib}%
  \BibitemOpen
  \bibfield{author}{%
  \bibinfo {author} {\bibfnamefont{B.~L.~G.}\ \bibnamefont{Bakker}}, \bibinfo
  {author} {\bibfnamefont{E.}~\bibnamefont{Leader}},\ and\ \bibinfo {author}
  {\bibfnamefont{T.~L.}\ \bibnamefont{Trueman}},\ }%
  \bibfield{title}{%
  \enquote{\bibinfo {title} {{A Critique of the angular momentum sum rules and
  a new angular momentum sum rule}},}\ }%
  \bibfield{journal}{%
  \Doi{10.1103/PhysRevD.70.114001}{\bibinfo {journal} {Phys. Rev.}}\ }%
  \textbf{\bibinfo {volume} {D70}},\ \bibinfo {pages} {114001} (\bibinfo {year}
  {2004}),\ \Eprint{http://arxiv.org/abs/hep-ph/0406139}{arXiv:hep-ph/0406139
  [hep-ph]}%
  \bibAnnoteFile{NoStop}{Bakker:2004ib}%
\bibitem{Lorce:2015lna}%
  \BibitemOpen
  \bibfield{author}{%
  \bibinfo {author} {\bibfnamefont{Cédric}\ \bibnamefont{Lorcé}},\ }%
  \bibfield{title}{%
  \enquote{\bibinfo {title} {{The light-front gauge-invariant energy-momentum
  tensor}},}\ }%
  \bibfield{journal}{%
  \Doi{10.1007/JHEP08(2015)045}{\bibinfo {journal} {JHEP}}\ }%
  \textbf{\bibinfo {volume} {08}},\ \bibinfo {pages} {045} (\bibinfo {year}
  {2015}),\ \Eprint{http://arxiv.org/abs/1502.06656}{arXiv:1502.06656
  [hep-ph]}%
  \bibAnnoteFile{NoStop}{Lorce:2015lna}%
\bibitem{Teryaev:1999su}%
  \BibitemOpen
  \bibfield{author}{%
  \bibinfo {author} {\bibfnamefont{O.~V.}\ \bibnamefont{Teryaev}},\ }%
  \bibfield{title}{%
  \enquote{\bibinfo {title} {{Spin structure of nucleon and equivalence
  principle}},}\ }%
   (\bibinfo {year} {1999}),\
  \Eprint{http://arxiv.org/abs/hep-ph/9904376}{arXiv:hep-ph/9904376 [hep-ph]}%
  \bibAnnoteFile{NoStop}{Teryaev:1999su}%
\bibitem{Lowdon:2017idv}%
  \BibitemOpen
  \bibfield{author}{%
  \bibinfo {author} {\bibfnamefont{Peter}\ \bibnamefont{Lowdon}}, \bibinfo
  {author} {\bibfnamefont{Kelly Yu-Ju}\ \bibnamefont{Chiu}},\ and\ \bibinfo
  {author} {\bibfnamefont{Stanley~J.}\ \bibnamefont{Brodsky}},\ }%
  \bibfield{title}{%
  \enquote{\bibinfo {title} {{Rigorous constraints on the matrix elements of
  the energy-momentum tensor}},}\ }%
  \bibfield{journal}{%
  \Doi{10.1016/j.physletb.2017.09.050}{\bibinfo {journal} {Phys. Lett.}}\ }%
  \textbf{\bibinfo {volume} {B774}},\ \bibinfo {pages} {1--6} (\bibinfo {year}
  {2017}),\ \Eprint{http://arxiv.org/abs/1707.06313}{arXiv:1707.06313
  [hep-th]}%
  \bibAnnoteFile{NoStop}{Lowdon:2017idv}%
\bibitem{Teryaev:2016edw}%
  \BibitemOpen
  \bibfield{author}{%
  \bibinfo {author} {\bibfnamefont{O.~V.}\ \bibnamefont{Teryaev}},\ }%
  \bibfield{title}{%
  \enquote{\bibinfo {title} {{Gravitational form factors and nucleon spin
  structure}},}\ }%
  \bibfield{journal}{%
  \Doi{10.1007/s11467-016-0573-6}{\bibinfo {journal} {Front. Phys.(Beijing)}}\
  }%
  \textbf{\bibinfo {volume} {11}},\ \bibinfo {pages} {111207} (\bibinfo {year}
  {2016})%
  \bibAnnoteFile{NoStop}{Teryaev:2016edw}%
\bibitem{Brodsky:2000ii}%
  \BibitemOpen
  \bibfield{author}{%
  \bibinfo {author} {\bibfnamefont{Stanley~J.}\ \bibnamefont{Brodsky}},
  \bibinfo {author} {\bibfnamefont{Dae~Sung}\ \bibnamefont{Hwang}}, \bibinfo
  {author} {\bibfnamefont{Bo-Qiang}\ \bibnamefont{Ma}},\ and\ \bibinfo {author}
  {\bibfnamefont{Ivan}\ \bibnamefont{Schmidt}},\ }%
  \bibfield{title}{%
  \enquote{\bibinfo {title} {{Light cone representation of the spin and orbital
  angular momentum of relativistic composite systems}},}\ }%
  \bibfield{journal}{%
  \Doi{10.1016/S0550-3213(00)00626-X}{\bibinfo {journal} {Nucl. Phys.}}\ }%
  \textbf{\bibinfo {volume} {B593}},\ \bibinfo {pages} {311--335} (\bibinfo
  {year} {2001}),\
  \Eprint{http://arxiv.org/abs/hep-th/0003082}{arXiv:hep-th/0003082 [hep-th]}%
  \bibAnnoteFile{NoStop}{Brodsky:2000ii}%
\bibitem{Kim:2012ts}%
  \BibitemOpen
  \bibfield{author}{%
  \bibinfo {author} {\bibfnamefont{Hyun-Chul}\ \bibnamefont{Kim}}, \bibinfo
  {author} {\bibfnamefont{Peter}\ \bibnamefont{Schweitzer}},\ and\ \bibinfo
  {author} {\bibfnamefont{Ulugbek}\ \bibnamefont{Yakhshiev}},\ }%
  \bibfield{title}{%
  \enquote{\bibinfo {title} {{Energy-momentum tensor form factors of the
  nucleon in nuclear matter}},}\ }%
  \bibfield{journal}{%
  \Doi{10.1016/j.physletb.2012.10.055}{\bibinfo {journal} {Phys. Lett.}}\ }%
  \textbf{\bibinfo {volume} {B718}},\ \bibinfo {pages} {625--631} (\bibinfo
  {year} {2012}),\ \Eprint{http://arxiv.org/abs/1205.5228}{arXiv:1205.5228
  [hep-ph]}%
  \bibAnnoteFile{NoStop}{Kim:2012ts}%
\bibitem{Hagler:2007xi}%
  \BibitemOpen
  \bibfield{author}{%
  \bibinfo {author} {\bibfnamefont{Ph.}\ \bibnamefont{Hagler}} \emph{et~al.}
  (\bibinfo {collaboration} {LHPC}),\ }%
  \bibfield{title}{%
  \enquote{\bibinfo {title} {{Nucleon Generalized Parton Distributions from
  Full Lattice QCD}},}\ }%
  \bibfield{journal}{%
  \Doi{10.1103/PhysRevD.77.094502}{\bibinfo {journal} {Phys. Rev.}}\ }%
  \textbf{\bibinfo {volume} {D77}},\ \bibinfo {pages} {094502} (\bibinfo {year}
  {2008}),\ \Eprint{http://arxiv.org/abs/0705.4295}{arXiv:0705.4295 [hep-lat]}%
  \bibAnnoteFile{NoStop}{Hagler:2007xi}%
\bibitem{Bali:2016wqg}%
  \BibitemOpen
  \bibfield{author}{%
  \bibinfo {author} {\bibfnamefont{Gunnar}\ \bibnamefont{Bali}}, \bibinfo
  {author} {\bibfnamefont{Sara}\ \bibnamefont{Collins}}, \bibinfo {author}
  {\bibfnamefont{Meinulf}\ \bibnamefont{Göckeler}}, \bibinfo {author}
  {\bibfnamefont{Rudolf}\ \bibnamefont{Rödl}}, \bibinfo {author}
  {\bibfnamefont{Andreas}\ \bibnamefont{Schäfer}},\ and\ \bibinfo {author}
  {\bibfnamefont{Andre}\ \bibnamefont{Sternbeck}},\ }%
  \bibfield{title}{%
  \enquote{\bibinfo {title} {{Nucleon generalized form factors from lattice QCD
  with nearly physical quark masses}},}\ }%
  \bibfield{booktitle}{%
  \emph{\bibinfo {booktitle} {{Proceedings, 33rd International Symposium on
  Lattice Field Theory (Lattice 2015): Kobe, Japan, July 14-18, 2015}}},\ }%
  \bibfield{journal}{%
  \bibinfo {journal} {PoS}\ }%
  \textbf{\bibinfo {volume} {LATTICE2015}},\ \bibinfo {pages} {118} (\bibinfo
  {year} {2016}),\ \Eprint{http://arxiv.org/abs/1601.04818}{arXiv:1601.04818
  [hep-lat]}%
  \bibAnnoteFile{NoStop}{Bali:2016wqg}%
\bibitem{Alexandrou:2017oeh}%
  \BibitemOpen
  \bibfield{author}{%
  \bibinfo {author} {\bibfnamefont{C.}~\bibnamefont{Alexandrou}}, \bibinfo
  {author} {\bibfnamefont{M.}~\bibnamefont{Constantinou}}, \bibinfo {author}
  {\bibfnamefont{K.}~\bibnamefont{Hadjiyiannakou}}, \bibinfo {author}
  {\bibfnamefont{K.}~\bibnamefont{Jansen}}, \bibinfo {author}
  {\bibfnamefont{C.}~\bibnamefont{Kallidonis}}, \bibinfo {author}
  {\bibfnamefont{G.}~\bibnamefont{Koutsou}}, \bibinfo {author}
  {\bibfnamefont{A.}~\bibnamefont{Vaquero Avilés-Casco}},\ and\ \bibinfo
  {author} {\bibfnamefont{C.}~\bibnamefont{Wiese}},\ }%
  \bibfield{title}{%
  \enquote{\bibinfo {title} {{Nucleon Spin and Momentum Decomposition Using
  Lattice QCD Simulations}},}\ }%
  \bibfield{journal}{%
  \Doi{10.1103/PhysRevLett.119.142002}{\bibinfo {journal} {Phys. Rev. Lett.}}\
  }%
  \textbf{\bibinfo {volume} {119}},\ \bibinfo {pages} {142002} (\bibinfo {year}
  {2017}),\ \Eprint{http://arxiv.org/abs/1706.02973}{arXiv:1706.02973
  [hep-lat]}%
  \bibAnnoteFile{NoStop}{Alexandrou:2017oeh}%
\bibitem{Ji:1998pc}%
  \BibitemOpen
  \bibfield{author}{%
  \bibinfo {author} {\bibfnamefont{Xiang-Dong}\ \bibnamefont{Ji}},\ }%
  \bibfield{title}{%
  \enquote{\bibinfo {title} {{Off forward parton distributions}},}\ }%
  \bibfield{journal}{%
  \Doi{10.1088/0954-3899/24/7/002}{\bibinfo {journal} {J. Phys.}}\ }%
  \textbf{\bibinfo {volume} {G24}},\ \bibinfo {pages} {1181--1205} (\bibinfo
  {year} {1998}),\
  \Eprint{http://arxiv.org/abs/hep-ph/9807358}{arXiv:hep-ph/9807358 [hep-ph]}%
  \bibAnnoteFile{NoStop}{Ji:1998pc}%
\bibitem{Mueller:1998fv}%
  \BibitemOpen
  \bibfield{author}{%
  \bibinfo {author} {\bibfnamefont{Dieter}\ \bibnamefont{Müller}}, \bibinfo
  {author} {\bibfnamefont{D.}~\bibnamefont{Robaschik}}, \bibinfo {author}
  {\bibfnamefont{B.}~\bibnamefont{Geyer}}, \bibinfo {author}
  {\bibfnamefont{F.~M.}\ \bibnamefont{Dittes}},\ and\ \bibinfo {author}
  {\bibfnamefont{J.}~\bibnamefont{Ho$\check{\text{r}}$ej$\check{\text{s}}$i}},\
  }%
  \bibfield{title}{%
  \enquote{\bibinfo {title} {{Wave functions, evolution equations and evolution
  kernels from light ray operators of QCD}},}\ }%
  \bibfield{journal}{%
  \Doi{10.1002/prop.2190420202}{\bibinfo {journal} {Fortsch. Phys.}}\ }%
  \textbf{\bibinfo {volume} {42}},\ \bibinfo {pages} {101--141} (\bibinfo
  {year} {1994}),\
  \Eprint{http://arxiv.org/abs/hep-ph/9812448}{arXiv:hep-ph/9812448 [hep-ph]}%
  \bibAnnoteFile{NoStop}{Mueller:1998fv}%
\bibitem{Radyushkin:1996nd}%
  \BibitemOpen
  \bibfield{author}{%
  \bibinfo {author} {\bibfnamefont{A.V.}\ \bibnamefont{Radyushkin}},\ }%
  \bibfield{title}{%
  \enquote{\bibinfo {title} {{Scaling limit of deeply virtual Compton
  scattering}},}\ }%
  \bibfield{journal}{%
  \Doi{10.1016/0370-2693(96)00528-X}{\bibinfo {journal} {Phys.Lett.}}\ }%
  \textbf{\bibinfo {volume} {B380}},\ \bibinfo {pages} {417--425} (\bibinfo
  {year} {1996}),\
  \Eprint{http://arxiv.org/abs/hep-ph/9604317}{arXiv:hep-ph/9604317 [hep-ph]}%
  \bibAnnoteFile{NoStop}{Radyushkin:1996nd}%
\bibitem{Kumericki:2016ehc}%
  \BibitemOpen
  \bibfield{author}{%
  \bibinfo {author} {\bibfnamefont{Kresimir}\ \bibnamefont{Kumericki}},
  \bibinfo {author} {\bibfnamefont{Simonetta}\ \bibnamefont{Liuti}},\ and\
  \bibinfo {author} {\bibfnamefont{Herve}\ \bibnamefont{Moutarde}},\ }%
  \bibfield{title}{%
  \enquote{\bibinfo {title} {{GPD phenomenology and DVCS fitting}},}\ }%
  \bibfield{journal}{%
  \Doi{10.1140/epja/i2016-16157-3}{\bibinfo {journal} {Eur. Phys. J.}}\ }%
  \textbf{\bibinfo {volume} {A52}},\ \bibinfo {pages} {157} (\bibinfo {year}
  {2016}),\ \Eprint{http://arxiv.org/abs/1602.02763}{arXiv:1602.02763
  [hep-ph]}%
  \bibAnnoteFile{NoStop}{Kumericki:2016ehc}%
\bibitem{Favart:2015umi}%
  \BibitemOpen
  \bibfield{author}{%
  \bibinfo {author} {\bibfnamefont{L.}~\bibnamefont{Favart}}, \bibinfo {author}
  {\bibfnamefont{M.}~\bibnamefont{Guidal}}, \bibinfo {author}
  {\bibfnamefont{T.}~\bibnamefont{Horn}},\ and\ \bibinfo {author}
  {\bibfnamefont{P.}~\bibnamefont{Kroll}},\ }%
  \bibfield{title}{%
  \enquote{\bibinfo {title} {{Deeply Virtual Meson Production on the
  nucleon}},}\ }%
  \bibfield{journal}{%
  \Doi{10.1140/epja/i2016-16158-2}{\bibinfo {journal} {Eur. Phys. J.}}\ }%
  \textbf{\bibinfo {volume} {A52}},\ \bibinfo {pages} {158} (\bibinfo {year}
  {2016}),\ \Eprint{http://arxiv.org/abs/1511.04535}{arXiv:1511.04535
  [hep-ph]}%
  \bibAnnoteFile{NoStop}{Favart:2015umi}%
\bibitem{Kumano:2017lhr}%
  \BibitemOpen
  \bibfield{author}{%
  \bibinfo {author} {\bibfnamefont{S.}~\bibnamefont{Kumano}}, \bibinfo {author}
  {\bibfnamefont{Qin-Tao}\ \bibnamefont{Song}},\ and\ \bibinfo {author}
  {\bibfnamefont{O.~V.}\ \bibnamefont{Teryaev}},\ }%
  \bibfield{title}{%
  \enquote{\bibinfo {title} {{Hadron tomography by generalized distribution
  amplitudes in pion-pair production process $\gamma^* \gamma \rightarrow \pi^0
  \pi^0 $ and gravitational form factors for pion}},}\ }%
  \bibfield{journal}{%
  \Doi{10.1103/PhysRevD.97.014020}{\bibinfo {journal} {Phys. Rev.}}\ }%
  \textbf{\bibinfo {volume} {D97}},\ \bibinfo {pages} {014020} (\bibinfo {year}
  {2018}),\ \Eprint{http://arxiv.org/abs/1711.08088}{arXiv:1711.08088
  [hep-ph]}%
  \bibAnnoteFile{NoStop}{Kumano:2017lhr}%
\bibitem{Leader:2012ar}%
  \BibitemOpen
  \bibfield{author}{%
  \bibinfo {author} {\bibfnamefont{Elliot}\ \bibnamefont{Leader}},\ }%
  \bibfield{title}{%
  \enquote{\bibinfo {title} {{A critical assessment of the angular momentum sum
  rules}},}\ }%
  \bibfield{journal}{%
  \Doi{10.1016/j.physletb.2013.01.050, 10.1016/j.physletb.2013.09.029}{\bibinfo
  {journal} {Phys. Lett.}}\ }%
  \textbf{\bibinfo {volume} {B720}},\ \bibinfo {pages} {120--124} (\bibinfo
  {year} {2013}),\ \bibinfo {note} {[Erratum: Phys. Lett.B726,927(2013)]},\
  \Eprint{http://arxiv.org/abs/1211.3957}{arXiv:1211.3957 [hep-ph]}%
  \bibAnnoteFile{NoStop}{Leader:2012ar}%
\bibitem{Tanaka:2018wea}%
  \BibitemOpen
  \bibfield{author}{%
  \bibinfo {author} {\bibfnamefont{Kazuhiro}\ \bibnamefont{Tanaka}},\ }%
  \bibfield{title}{%
  \enquote{\bibinfo {title} {{Operator relations for gravitational form factors
  of a spin-0 hadron}},}\ }%
  \bibfield{journal}{%
  \Doi{10.1103/PhysRevD.98.034009}{\bibinfo {journal} {Phys. Rev.}}\ }%
  \textbf{\bibinfo {volume} {D98}},\ \bibinfo {pages} {034009} (\bibinfo {year}
  {2018}),\ \Eprint{http://arxiv.org/abs/1806.10591}{arXiv:1806.10591
  [hep-ph]}%
  \bibAnnoteFile{NoStop}{Tanaka:2018wea}%
\bibitem{Alarcon:2011zs}%
  \BibitemOpen
  \bibfield{author}{%
  \bibinfo {author} {\bibfnamefont{J.~M.}\ \bibnamefont{Alarcon}}, \bibinfo
  {author} {\bibfnamefont{J.}~\bibnamefont{Martin~Camalich}},\ and\ \bibinfo
  {author} {\bibfnamefont{J.~A.}\ \bibnamefont{Oller}},\ }%
  \bibfield{title}{%
  \enquote{\bibinfo {title} {{The chiral representation of the $\pi N$
  scattering amplitude and the pion-nucleon sigma term}},}\ }%
  \bibfield{journal}{%
  \Doi{10.1103/PhysRevD.85.051503}{\bibinfo {journal} {Phys. Rev.}}\ }%
  \textbf{\bibinfo {volume} {D85}},\ \bibinfo {pages} {051503} (\bibinfo {year}
  {2012}),\ \Eprint{http://arxiv.org/abs/1110.3797}{arXiv:1110.3797 [hep-ph]}%
  \bibAnnoteFile{NoStop}{Alarcon:2011zs}%
\bibitem{Hoferichter:2015dsa}%
  \BibitemOpen
  \bibfield{author}{%
  \bibinfo {author} {\bibfnamefont{Martin}\ \bibnamefont{Hoferichter}},
  \bibinfo {author} {\bibfnamefont{J.}~\bibnamefont{Ruiz~de Elvira}}, \bibinfo
  {author} {\bibfnamefont{Bastian}\ \bibnamefont{Kubis}},\ and\ \bibinfo
  {author} {\bibfnamefont{Ulf-G.}\ \bibnamefont{Meißner}},\ }%
  \bibfield{title}{%
  \enquote{\bibinfo {title} {{High-Precision Determination of the Pion-Nucleon
  $\sigma$ Term from Roy-Steiner Equations}},}\ }%
  \bibfield{journal}{%
  \Doi{10.1103/PhysRevLett.115.092301}{\bibinfo {journal} {Phys. Rev. Lett.}}\
  }%
  \textbf{\bibinfo {volume} {115}},\ \bibinfo {pages} {092301} (\bibinfo {year}
  {2015}),\ \Eprint{http://arxiv.org/abs/1506.04142}{arXiv:1506.04142
  [hep-ph]}%
  \bibAnnoteFile{NoStop}{Hoferichter:2015dsa}%
\bibitem{Kharzeev:2002fm}%
  \BibitemOpen
  \bibfield{author}{%
  \bibinfo {author} {\bibfnamefont{D.~E.}\ \bibnamefont{Kharzeev}}\ and\
  \bibinfo {author} {\bibfnamefont{J.}~\bibnamefont{Raufeisen}},\ }%
  \bibfield{title}{%
  \enquote{\bibinfo {title} {{High-energy nuclear interactions and QCD: An
  Introduction}},}\ }%
  \bibfield{booktitle}{%
  \emph{\bibinfo {booktitle} {{New states of matter in hadronic interactions.
  Proceedings, Pan-American Advanced Study Institute, PASI 2002, Campos do
  Jordao, Sao Paulo, Brazil, January 7-18, 2002}}},\ }%
  \bibfield{journal}{%
  \Doi{10.1063/1.1513676}{\bibinfo {journal} {AIP Conf. Proc.}}\ }%
  \textbf{\bibinfo {volume} {631}},\ \bibinfo {pages} {27--69} (\bibinfo {year}
  {2002}),\ \Eprint{http://arxiv.org/abs/nucl-th/0206073}{arXiv:nucl-th/0206073
  [nucl-th]}%
  \bibAnnoteFile{NoStop}{Kharzeev:2002fm}%
\bibitem{Krein:2017usp}%
  \BibitemOpen
  \bibfield{author}{%
  \bibinfo {author} {\bibfnamefont{G.}~\bibnamefont{Krein}}, \bibinfo {author}
  {\bibfnamefont{A.~W.}\ \bibnamefont{Thomas}},\ and\ \bibinfo {author}
  {\bibfnamefont{K.}~\bibnamefont{Tsushima}},\ }%
  \bibfield{title}{%
  \enquote{\bibinfo {title} {{Nuclear-bound quarkonia and heavy-flavor
  hadrons}},}\ }%
  \bibfield{journal}{%
  \Doi{10.1016/j.ppnp.2018.02.001}{\bibinfo {journal} {Prog. Part. Nucl.
  Phys.}}\ }%
  \textbf{\bibinfo {volume} {100}},\ \bibinfo {pages} {161--210} (\bibinfo
  {year} {2018}),\ \Eprint{http://arxiv.org/abs/1706.02688}{arXiv:1706.02688
  [hep-ph]}%
  \bibAnnoteFile{NoStop}{Krein:2017usp}%
\bibitem{Joosten:2018gyo}%
  \BibitemOpen
  \bibfield{author}{%
  \bibinfo {author} {\bibfnamefont{S.}~\bibnamefont{Joosten}}\ and\ \bibinfo
  {author} {\bibfnamefont{Z.~E.}\ \bibnamefont{Meziani}},\ }%
  \bibfield{title}{%
  \enquote{\bibinfo {title} {{Heavy Quarkonium Production at Threshold: from
  JLab to EIC}},}\ }%
  \bibfield{booktitle}{%
  \emph{\bibinfo {booktitle} {{Proceedings, QCD Evolution Workshop (QCD 2017):
  Newport News, VA, USA, May 22-26, 2017}}},\ }%
  \bibfield{journal}{%
  \bibinfo {journal} {PoS}\ }%
  \textbf{\bibinfo {volume} {QCDEV2017}},\ \bibinfo {pages} {017} (\bibinfo
  {year} {2018}),\ \Eprint{http://arxiv.org/abs/1802.02616}{arXiv:1802.02616
  [hep-ex]}%
  \bibAnnoteFile{NoStop}{Joosten:2018gyo}%
\bibitem{Joosten:2018fql}%
  \BibitemOpen
  \bibfield{author}{%
  \bibinfo {author} {\bibfnamefont{Sylvester}\ \bibnamefont{Joosten}},\ }%
  \bibfield{title}{%
  \enquote{\bibinfo {title} {{Probing the gluonic structure of the nucleon
  through quarkonium production}},}\ }%
   (\bibinfo {year} {2018}),\
  \Eprint{http://arxiv.org/abs/1803.08615}{arXiv:1803.08615 [hep-ph]}%
  \bibAnnoteFile{NoStop}{Joosten:2018fql}%
\bibitem{Hatta:2018ina}%
  \BibitemOpen
  \bibfield{author}{%
  \bibinfo {author} {\bibfnamefont{Yoshitaka}\ \bibnamefont{Hatta}}\ and\
  \bibinfo {author} {\bibfnamefont{Di-Lun}\ \bibnamefont{Yang}},\ }%
  \bibfield{title}{%
  \enquote{\bibinfo {title} {{Holographic $J/\psi$ production near threshold
  and the proton mass problem}},}\ }%
  \bibfield{journal}{%
  \Doi{10.1103/PhysRevD.98.074003}{\bibinfo {journal} {Phys. Rev.}}\ }%
  \textbf{\bibinfo {volume} {D98}},\ \bibinfo {pages} {074003} (\bibinfo {year}
  {2018}),\ \Eprint{http://arxiv.org/abs/1808.02163}{arXiv:1808.02163
  [hep-ph]}%
  \bibAnnoteFile{NoStop}{Hatta:2018ina}%
\bibitem{Bernard:2001rs}%
  \BibitemOpen
  \bibfield{author}{%
  \bibinfo {author} {\bibfnamefont{Veronique}\ \bibnamefont{Bernard}}, \bibinfo
  {author} {\bibfnamefont{Latifa}\ \bibnamefont{Elouadrhiri}},\ and\ \bibinfo
  {author} {\bibfnamefont{Ulf-G.}\ \bibnamefont{Meissner}},\ }%
  \bibfield{title}{%
  \enquote{\bibinfo {title} {{Axial structure of the nucleon: Topical
  Review}},}\ }%
  \bibfield{journal}{%
  \Doi{10.1088/0954-3899/28/1/201}{\bibinfo {journal} {J. Phys.}}\ }%
  \textbf{\bibinfo {volume} {G28}},\ \bibinfo {pages} {R1--R35} (\bibinfo
  {year} {2002}),\
  \Eprint{http://arxiv.org/abs/hep-ph/0107088}{arXiv:hep-ph/0107088 [hep-ph]}%
  \bibAnnoteFile{NoStop}{Bernard:2001rs}%
\bibitem{Harland-Lang:2014zoa}%
  \BibitemOpen
  \bibfield{author}{%
  \bibinfo {author} {\bibfnamefont{L.~A.}\ \bibnamefont{Harland-Lang}},
  \bibinfo {author} {\bibfnamefont{A.~D.}\ \bibnamefont{Martin}}, \bibinfo
  {author} {\bibfnamefont{P.}~\bibnamefont{Motylinski}},\ and\ \bibinfo
  {author} {\bibfnamefont{R.~S.}\ \bibnamefont{Thorne}},\ }%
  \bibfield{title}{%
  \enquote{\bibinfo {title} {{Parton distributions in the LHC era: MMHT 2014
  PDFs}},}\ }%
  \bibfield{journal}{%
  \Doi{10.1140/epjc/s10052-015-3397-6}{\bibinfo {journal} {Eur. Phys. J.}}\ }%
  \textbf{\bibinfo {volume} {C75}},\ \bibinfo {pages} {204} (\bibinfo {year}
  {2015}),\ \Eprint{http://arxiv.org/abs/1412.3989}{arXiv:1412.3989 [hep-ph]}%
  \bibAnnoteFile{NoStop}{Harland-Lang:2014zoa}%
\bibitem{Mondal:2015fok}%
  \BibitemOpen
  \bibfield{author}{%
  \bibinfo {author} {\bibfnamefont{Chandan}\ \bibnamefont{Mondal}},\ }%
  \bibfield{title}{%
  \enquote{\bibinfo {title} {{Longitudinal momentum densities in transverse
  plane for nucleons}},}\ }%
  \bibfield{journal}{%
  \Doi{10.1140/epjc/s10052-016-3922-2}{\bibinfo {journal} {Eur. Phys. J.}}\ }%
  \textbf{\bibinfo {volume} {C76}},\ \bibinfo {pages} {74} (\bibinfo {year}
  {2016}),\ \Eprint{http://arxiv.org/abs/1511.01736}{arXiv:1511.01736
  [hep-ph]}%
  \bibAnnoteFile{NoStop}{Mondal:2015fok}%
\bibitem{Kumar:2017dbf}%
  \BibitemOpen
  \bibfield{author}{%
  \bibinfo {author} {\bibfnamefont{Narinder}\ \bibnamefont{Kumar}}, \bibinfo
  {author} {\bibfnamefont{Chandan}\ \bibnamefont{Mondal}},\ and\ \bibinfo
  {author} {\bibfnamefont{Neetika}\ \bibnamefont{Sharma}},\ }%
  \bibfield{title}{%
  \enquote{\bibinfo {title} {{Gravitational form factors and angular momentum
  densities in light-front quark-diquark model}},}\ }%
  \bibfield{journal}{%
  \Doi{10.1140/epja/i2017-12433-0}{\bibinfo {journal} {Eur. Phys. J.}}\ }%
  \textbf{\bibinfo {volume} {A53}},\ \bibinfo {pages} {237} (\bibinfo {year}
  {2017}),\ \Eprint{http://arxiv.org/abs/1712.02110}{arXiv:1712.02110
  [hep-ph]}%
  \bibAnnoteFile{NoStop}{Kumar:2017dbf}%
\bibitem{Deka:2013zha}%
  \BibitemOpen
  \bibfield{author}{%
  \bibinfo {author} {\bibfnamefont{M.}~\bibnamefont{Deka}} \emph{et~al.},\ }%
  \bibfield{title}{%
  \enquote{\bibinfo {title} {{Lattice study of quark and glue momenta and
  angular momenta in the nucleon}},}\ }%
  \bibfield{journal}{%
  \Doi{10.1103/PhysRevD.91.014505}{\bibinfo {journal} {Phys. Rev.}}\ }%
  \textbf{\bibinfo {volume} {D91}},\ \bibinfo {pages} {014505} (\bibinfo {year}
  {2015}),\ \Eprint{http://arxiv.org/abs/1312.4816}{arXiv:1312.4816 [hep-lat]}%
  \bibAnnoteFile{NoStop}{Deka:2013zha}%
\bibitem{Pasquini:2014vua}%
  \BibitemOpen
  \bibfield{author}{%
  \bibinfo {author} {\bibfnamefont{B.}~\bibnamefont{Pasquini}}, \bibinfo
  {author} {\bibfnamefont{M.~V.}\ \bibnamefont{Polyakov}},\ and\ \bibinfo
  {author} {\bibfnamefont{M.}~\bibnamefont{Vanderhaeghen}},\ }%
  \bibfield{title}{%
  \enquote{\bibinfo {title} {{Dispersive evaluation of the D-term form factor
  in deeply virtual Compton scattering}},}\ }%
  \bibfield{journal}{%
  \Doi{10.1016/j.physletb.2014.10.047}{\bibinfo {journal} {Phys. Lett.}}\ }%
  \textbf{\bibinfo {volume} {B739}},\ \bibinfo {pages} {133--138} (\bibinfo
  {year} {2014}),\ \Eprint{http://arxiv.org/abs/1407.5960}{arXiv:1407.5960
  [hep-ph]}%
  \bibAnnoteFile{NoStop}{Pasquini:2014vua}%
\bibitem{Burkert:2018bqq}%
  \BibitemOpen
  \bibfield{author}{%
  \bibinfo {author} {\bibfnamefont{V.~D.}\ \bibnamefont{Burkert}}, \bibinfo
  {author} {\bibfnamefont{L.}~\bibnamefont{Elouadrhiri}},\ and\ \bibinfo
  {author} {\bibfnamefont{F.~X.}\ \bibnamefont{Girod}},\ }%
  \bibfield{title}{%
  \enquote{\bibinfo {title} {{The pressure distribution inside the proton}},}\
  }%
  \bibfield{journal}{%
  \Doi{10.1038/s41586-018-0060-z}{\bibinfo {journal} {Nature}}\ }%
  \textbf{\bibinfo {volume} {557}},\ \bibinfo {pages} {396--399} (\bibinfo
  {year} {2018})%
  \bibAnnoteFile{NoStop}{Burkert:2018bqq}%
\bibitem{Lorce:2017xzd}%
  \BibitemOpen
  \bibfield{author}{%
  \bibinfo {author} {\bibfnamefont{Cédric}\ \bibnamefont{Lorcé}},\ }%
  \bibfield{title}{%
  \enquote{\bibinfo {title} {{On the hadron mass decomposition}},}\ }%
  \bibfield{journal}{%
  \Doi{10.1140/epjc/s10052-018-5561-2}{\bibinfo {journal} {Eur. Phys. J.}}\ }%
  \textbf{\bibinfo {volume} {C78}},\ \bibinfo {pages} {120} (\bibinfo {year}
  {2018}),\ \Eprint{http://arxiv.org/abs/1706.05853}{arXiv:1706.05853
  [hep-ph]}%
  \bibAnnoteFile{NoStop}{Lorce:2017xzd}%
\bibitem{Airapetian:2006vy}%
  \BibitemOpen
  \bibfield{author}{%
  \bibinfo {author} {\bibfnamefont{A.}~\bibnamefont{Airapetian}} \emph{et~al.}
  (\bibinfo {collaboration} {HERMES}),\ }%
  \bibfield{title}{%
  \enquote{\bibinfo {title} {{Precise determination of the spin structure
  function g(1) of the proton, deuteron and neutron}},}\ }%
  \bibfield{journal}{%
  \Doi{10.1103/PhysRevD.75.012007}{\bibinfo {journal} {Phys. Rev.}}\ }%
  \textbf{\bibinfo {volume} {D75}},\ \bibinfo {pages} {012007} (\bibinfo {year}
  {2007}),\ \Eprint{http://arxiv.org/abs/hep-ex/0609039}{arXiv:hep-ex/0609039
  [hep-ex]}%
  \bibAnnoteFile{NoStop}{Airapetian:2006vy}%
\bibitem{Polyakov:2018exb}%
  \BibitemOpen
  \bibfield{author}{%
  \bibinfo {author} {\bibfnamefont{Maxim~V.}\ \bibnamefont{Polyakov}}\ and\
  \bibinfo {author} {\bibfnamefont{Hyeon-Dong}\ \bibnamefont{Son}},\ }%
  \bibfield{title}{%
  \enquote{\bibinfo {title} {{Nucleon gravitational form factors from
  instantons: forces between quark and gluon subsystems}},}\ }%
  \bibfield{journal}{%
  \bibinfo {journal} {JHEP}\ }%
  \textbf{\bibinfo {volume} {09}},\ \bibinfo {pages} {156} (\bibinfo {year}
  {2018}),\ \Eprint{http://arxiv.org/abs/1808.00155}{arXiv:1808.00155
  [hep-ph]}%
  \bibAnnoteFile{NoStop}{Polyakov:2018exb}%
\bibitem{Alexandrou:2017hac}%
  \BibitemOpen
  \bibfield{author}{%
  \bibinfo {author} {\bibfnamefont{Constantia}\ \bibnamefont{Alexandrou}},
  \bibinfo {author} {\bibfnamefont{Martha}\ \bibnamefont{Constantinou}},
  \bibinfo {author} {\bibfnamefont{Kyriakos}\ \bibnamefont{Hadjiyiannakou}},
  \bibinfo {author} {\bibfnamefont{Karl}\ \bibnamefont{Jansen}}, \bibinfo
  {author} {\bibfnamefont{Christos}\ \bibnamefont{Kallidonis}}, \bibinfo
  {author} {\bibfnamefont{Giannis}\ \bibnamefont{Koutsou}},\ and\ \bibinfo
  {author} {\bibfnamefont{Alejandro}\ \bibnamefont{Vaquero Aviles-Casco}},\ }%
  \bibfield{title}{%
  \enquote{\bibinfo {title} {{Nucleon axial form factors using $N_f$ = 2
  twisted mass fermions with a physical value of the pion mass}},}\ }%
  \bibfield{journal}{%
  \Doi{10.1103/PhysRevD.96.054507}{\bibinfo {journal} {Phys. Rev.}}\ }%
  \textbf{\bibinfo {volume} {D96}},\ \bibinfo {pages} {054507} (\bibinfo {year}
  {2017}),\ \Eprint{http://arxiv.org/abs/1705.03399}{arXiv:1705.03399
  [hep-lat]}%
  \bibAnnoteFile{NoStop}{Alexandrou:2017hac}%
\bibitem{Lorce:2018zpf}%
  \BibitemOpen
  \bibfield{author}{%
  \bibinfo {author} {\bibfnamefont{Cédric}\ \bibnamefont{Lorcé}},\ }%
  \bibfield{title}{%
  \enquote{\bibinfo {title} {{The relativistic center of mass in field theory
  with spin}},}\ }%
  \bibfield{journal}{%
  \Doi{10.1140/epjc/s10052-018-6249-3}{\bibinfo {journal} {Eur. Phys. J.}}\ }%
  \textbf{\bibinfo {volume} {C78}},\ \bibinfo {pages} {785} (\bibinfo {year}
  {2018}),\ \Eprint{http://arxiv.org/abs/1805.05284}{arXiv:1805.05284
  [hep-ph]}%
  \bibAnnoteFile{NoStop}{Lorce:2018zpf}%
\bibitem{Lorce:2017isp}%
  \BibitemOpen
  \bibfield{author}{%
  \bibinfo {author} {\bibfnamefont{Cédric}\ \bibnamefont{Lorcé}},\ }%
  \bibfield{title}{%
  \enquote{\bibinfo {title} {{New explicit expressions for Dirac bilinears}},}\
  }%
  \bibfield{journal}{%
  \Doi{10.1103/PhysRevD.97.016005}{\bibinfo {journal} {Phys. Rev.}}\ }%
  \textbf{\bibinfo {volume} {D97}},\ \bibinfo {pages} {016005} (\bibinfo {year}
  {2018}),\ \Eprint{http://arxiv.org/abs/1705.08370}{arXiv:1705.08370
  [hep-ph]}%
  \bibAnnoteFile{NoStop}{Lorce:2017isp}%
\bibitem{Ji:1994av}%
  \BibitemOpen
  \bibfield{author}{%
  \bibinfo {author} {\bibfnamefont{Xiang-Dong}\ \bibnamefont{Ji}},\ }%
  \bibfield{title}{%
  \enquote{\bibinfo {title} {{A QCD analysis of the mass structure of the
  nucleon}},}\ }%
  \bibfield{journal}{%
  \Doi{10.1103/PhysRevLett.74.1071}{\bibinfo {journal} {Phys. Rev. Lett.}}\ }%
  \textbf{\bibinfo {volume} {74}},\ \bibinfo {pages} {1071--1074} (\bibinfo
  {year} {1995}),\
  \Eprint{http://arxiv.org/abs/hep-ph/9410274}{arXiv:hep-ph/9410274 [hep-ph]}%
  \bibAnnoteFile{NoStop}{Ji:1994av}%
\bibitem{Ji:1995sv}%
  \BibitemOpen
  \bibfield{author}{%
  \bibinfo {author} {\bibfnamefont{Xiang-Dong}\ \bibnamefont{Ji}},\ }%
  \bibfield{title}{%
  \enquote{\bibinfo {title} {{Breakup of hadron masses and energy - momentum
  tensor of QCD}},}\ }%
  \bibfield{journal}{%
  \Doi{10.1103/PhysRevD.52.271}{\bibinfo {journal} {Phys. Rev.}}\ }%
  \textbf{\bibinfo {volume} {D52}},\ \bibinfo {pages} {271--281} (\bibinfo
  {year} {1995}),\
  \Eprint{http://arxiv.org/abs/hep-ph/9502213}{arXiv:hep-ph/9502213 [hep-ph]}%
  \bibAnnoteFile{NoStop}{Ji:1995sv}%
\bibitem{Eckart:1940te}%
  \BibitemOpen
  \bibfield{author}{%
  \bibinfo {author} {\bibfnamefont{Carl}\ \bibnamefont{Eckart}},\ }%
  \bibfield{title}{%
  \enquote{\bibinfo {title} {{The Thermodynamics of irreversible processes. 3..
  Relativistic theory of the simple fluid}},}\ }%
  \bibfield{journal}{%
  \Doi{10.1103/PhysRev.58.919}{\bibinfo {journal} {Phys. Rev.}}\ }%
  \textbf{\bibinfo {volume} {58}},\ \bibinfo {pages} {919--924} (\bibinfo
  {year} {1940})%
  \bibAnnoteFile{NoStop}{Eckart:1940te}%
\bibitem{Gao:2015aax}%
  \BibitemOpen
  \bibfield{author}{%
  \bibinfo {author} {\bibfnamefont{Haiyan}\ \bibnamefont{Gao}}, \bibinfo
  {author} {\bibfnamefont{Tianbo}\ \bibnamefont{Liu}}, \bibinfo {author}
  {\bibfnamefont{Chao}\ \bibnamefont{Peng}}, \bibinfo {author}
  {\bibfnamefont{Zhihong}\ \bibnamefont{Ye}},\ and\ \bibinfo {author}
  {\bibfnamefont{Zhiwen}\ \bibnamefont{Zhao}},\ }%
  \bibfield{title}{%
  \enquote{\bibinfo {title} {{Proton Remains Puzzling}},}\ }%
  \bibfield{journal}{%
  \bibinfo {journal} {The Universe}\ }%
  \textbf{\bibinfo {volume} {3}},\ \bibinfo {pages} {18--30} (\bibinfo {year}
  {2015})%
  \bibAnnoteFile{NoStop}{Gao:2015aax}%
\bibitem{Hatta:2018sqd}%
  \BibitemOpen
  \bibfield{author}{%
  \bibinfo {author} {\bibfnamefont{Yoshitaka}\ \bibnamefont{Hatta}}, \bibinfo
  {author} {\bibfnamefont{Abha}\ \bibnamefont{Rajan}},\ and\ \bibinfo {author}
  {\bibfnamefont{Kazuhiro}\ \bibnamefont{Tanaka}},\ }%
  \bibfield{title}{%
  \enquote{\bibinfo {title} {{Quark and gluon contributions to the QCD trace
  anomaly}},}\ }%
  \bibfield{journal}{%
  \Doi{10.1007/JHEP12(2018)008}{\bibinfo {journal} {JHEP}}\ }%
  \textbf{\bibinfo {volume} {12}},\ \bibinfo {pages} {008} (\bibinfo {year}
  {2018}),\ \Eprint{http://arxiv.org/abs/1810.05116}{arXiv:1810.05116
  [hep-ph]}%
  \bibAnnoteFile{NoStop}{Hatta:2018sqd}%
\bibitem{Ji:1997gm}%
  \BibitemOpen
  \bibfield{author}{%
  \bibinfo {author} {\bibfnamefont{Xiang-Dong}\ \bibnamefont{Ji}}, \bibinfo
  {author} {\bibfnamefont{W.}~\bibnamefont{Melnitchouk}},\ and\ \bibinfo
  {author} {\bibfnamefont{X.}~\bibnamefont{Song}},\ }%
  \bibfield{title}{%
  \enquote{\bibinfo {title} {{A Study of off forward parton distributions}},}\
  }%
  \bibfield{journal}{%
  \Doi{10.1103/PhysRevD.56.5511}{\bibinfo {journal} {Phys. Rev.}}\ }%
  \textbf{\bibinfo {volume} {D56}},\ \bibinfo {pages} {5511--5523} (\bibinfo
  {year} {1997}),\
  \Eprint{http://arxiv.org/abs/hep-ph/9702379}{arXiv:hep-ph/9702379 [hep-ph]}%
  \bibAnnoteFile{NoStop}{Ji:1997gm}%
\bibitem{Sachs:1962zzc}%
  \BibitemOpen
  \bibfield{author}{%
  \bibinfo {author} {\bibfnamefont{R.~G.}\ \bibnamefont{Sachs}},\ }%
  \bibfield{title}{%
  \enquote{\bibinfo {title} {{High-Energy Behavior of Nucleon Electromagnetic
  Form Factors}},}\ }%
  \bibfield{journal}{%
  \Doi{10.1103/PhysRev.126.2256}{\bibinfo {journal} {Phys. Rev.}}\ }%
  \textbf{\bibinfo {volume} {126}},\ \bibinfo {pages} {2256--2260} (\bibinfo
  {year} {1962})%
  \bibAnnoteFile{NoStop}{Sachs:1962zzc}%
\bibitem{Bayin:1985cd}%
  \BibitemOpen
  \bibfield{author}{%
  \bibinfo {author} {\bibfnamefont{Selcuk~S.}\ \bibnamefont{Bayin}},\ }%
  \bibfield{title}{%
  \enquote{\bibinfo {title} {{Anisotropic fluids and cosmology}},}\ }%
  \bibfield{journal}{%
  \Doi{10.1086/164056}{\bibinfo {journal} {Astrophys. J.}}\ }%
  \textbf{\bibinfo {volume} {303}},\ \bibinfo {pages} {101--110} (\bibinfo
  {year} {1986})%
  \bibAnnoteFile{NoStop}{Bayin:1985cd}%
\bibitem{Lemaitre:1933gd2}%
  \BibitemOpen
  \bibfield{author}{%
  \bibinfo {author} {\bibfnamefont{G.}~\bibnamefont{Lemaitre}},\ }%
  \bibfield{title}{%
  \enquote{\bibinfo {title} {{The expanding universe}},}\ }%
  \bibfield{journal}{%
  \bibinfo {journal} {Annales Soc. Sci. Bruxelles}\ }%
  \textbf{\bibinfo {volume} {A53}},\ \bibinfo {pages} {51} (\bibinfo {year}
  {1933}),\ \bibinfo {note} {[Gen. Rel. Grav. 29, 641 (1997)]}%
  \bibAnnoteFile{NoStop}{Lemaitre:1933gd2}%
\bibitem{Lemaitre:1933gd}%
  \BibitemOpen
  \bibfield{author}{%
  \bibinfo {author} {\bibfnamefont{G.}~\bibnamefont{Lemaitre}},\ }%
  \bibfield{title}{%
  \enquote{\bibinfo {title} {{The expanding universe}},}\ }%
  \bibfield{journal}{%
  \Doi{10.1023/A:1018855621348}{\bibinfo {journal} {Gen. Rel. Grav.}}\ }%
  \textbf{\bibinfo {volume} {29}},\ \bibinfo {pages} {641--680} (\bibinfo
  {year} {1997}),\ \bibinfo {note} {[Annales Soc. Sci. Bruxelles
  A53,51(1933)]}%
  \bibAnnoteFile{NoStop}{Lemaitre:1933gd}%
\bibitem{Pohl:2010zza}%
  \BibitemOpen
  \bibfield{author}{%
  \bibinfo {author} {\bibfnamefont{Randolf}\ \bibnamefont{Pohl}}
  \emph{et~al.},\ }%
  \bibfield{title}{%
  \enquote{\bibinfo {title} {{The size of the proton}},}\ }%
  \bibfield{journal}{%
  \Doi{10.1038/nature09250}{\bibinfo {journal} {Nature}}\ }%
  \textbf{\bibinfo {volume} {466}},\ \bibinfo {pages} {213--216} (\bibinfo
  {year} {2010})%
  \bibAnnoteFile{NoStop}{Pohl:2010zza}%
\bibitem{Pohl1:2016xoo}%
  \BibitemOpen
  \bibfield{author}{%
  \bibinfo {author} {\bibfnamefont{Randolf}\ \bibnamefont{Pohl}} \emph{et~al.}
  (\bibinfo {collaboration} {CREMA}),\ }%
  \bibfield{title}{%
  \enquote{\bibinfo {title} {{Laser spectroscopy of muonic deuterium}},}\ }%
  \bibfield{journal}{%
  \Doi{10.1126/science.aaf2468}{\bibinfo {journal} {Science}}\ }%
  \textbf{\bibinfo {volume} {353}},\ \bibinfo {pages} {669--673} (\bibinfo
  {year} {2016})%
  \bibAnnoteFile{NoStop}{Pohl1:2016xoo}%
\bibitem{Arrington:2015yxa}%
  \BibitemOpen
  \bibfield{author}{%
  \bibinfo {author} {\bibfnamefont{John}\ \bibnamefont{Arrington}},\ }%
  \bibfield{title}{%
  \enquote{\bibinfo {title} {{An examination of proton charge radius
  extractions from e-p scattering data}},}\ }%
  \bibfield{journal}{%
  \Doi{10.1063/1.4922414}{\bibinfo {journal} {J. Phys. Chem. Ref. Data}}\ }%
  \textbf{\bibinfo {volume} {44}},\ \bibinfo {pages} {031203} (\bibinfo {year}
  {2015}),\ \Eprint{http://arxiv.org/abs/1506.00873}{arXiv:1506.00873
  [nucl-ex]}%
  \bibAnnoteFile{NoStop}{Arrington:2015yxa}%
\bibitem{Perevalova:2016dln}%
  \BibitemOpen
  \bibfield{author}{%
  \bibinfo {author} {\bibfnamefont{I.~A.}\ \bibnamefont{Perevalova}}, \bibinfo
  {author} {\bibfnamefont{M.~V.}\ \bibnamefont{Polyakov}},\ and\ \bibinfo
  {author} {\bibfnamefont{P.}~\bibnamefont{Schweitzer}},\ }%
  \bibfield{title}{%
  \enquote{\bibinfo {title} {{On LHCb pentaquarks as a baryon-$\psi$(2S) bound
  state: prediction of isospin-$\frac3{2}$ pentaquarks with hidden charm}},}\
  }%
  \bibfield{journal}{%
  \Doi{10.1103/PhysRevD.94.054024}{\bibinfo {journal} {Phys. Rev.}}\ }%
  \textbf{\bibinfo {volume} {D94}},\ \bibinfo {pages} {054024} (\bibinfo {year}
  {2016}),\ \Eprint{http://arxiv.org/abs/1607.07008}{arXiv:1607.07008
  [hep-ph]}%
  \bibAnnoteFile{NoStop}{Perevalova:2016dln}%
\bibitem{Hudson:2017xug}%
  \BibitemOpen
  \bibfield{author}{%
  \bibinfo {author} {\bibfnamefont{Jonathan}\ \bibnamefont{Hudson}}\ and\
  \bibinfo {author} {\bibfnamefont{Peter}\ \bibnamefont{Schweitzer}},\ }%
  \bibfield{title}{%
  \enquote{\bibinfo {title} {{D term and the structure of pointlike and
  composed spin-0 particles}},}\ }%
  \bibfield{journal}{%
  \Doi{10.1103/PhysRevD.96.114013}{\bibinfo {journal} {Phys. Rev.}}\ }%
  \textbf{\bibinfo {volume} {D96}},\ \bibinfo {pages} {114013} (\bibinfo {year}
  {2017}),\ \Eprint{http://arxiv.org/abs/1712.05316}{arXiv:1712.05316
  [hep-ph]}%
  \bibAnnoteFile{NoStop}{Hudson:2017xug}%
\bibitem{Hudson:2017oul}%
  \BibitemOpen
  \bibfield{author}{%
  \bibinfo {author} {\bibfnamefont{Jonathan}\ \bibnamefont{Hudson}}\ and\
  \bibinfo {author} {\bibfnamefont{Peter}\ \bibnamefont{Schweitzer}},\ }%
  \bibfield{title}{%
  \enquote{\bibinfo {title} {{Dynamic origins of fermionic D-terms}},}\ }%
  \bibfield{journal}{%
  \Doi{10.1103/PhysRevD.97.056003}{\bibinfo {journal} {Phys. Rev.}}\ }%
  \textbf{\bibinfo {volume} {D97}},\ \bibinfo {pages} {056003} (\bibinfo {year}
  {2018}),\ \Eprint{http://arxiv.org/abs/1712.05317}{arXiv:1712.05317
  [hep-ph]}%
  \bibAnnoteFile{NoStop}{Hudson:2017oul}%
\bibitem{Ozel:2016oaf}%
  \BibitemOpen
  \bibfield{author}{%
  \bibinfo {author} {\bibfnamefont{Feryal}\ \bibnamefont{Özel}}\ and\ \bibinfo
  {author} {\bibfnamefont{Paulo}\ \bibnamefont{Freire}},\ }%
  \bibfield{title}{%
  \enquote{\bibinfo {title} {{Masses, Radii, and the Equation of State of
  Neutron Stars}},}\ }%
  \bibfield{journal}{%
  \Doi{10.1146/annurev-astro-081915-023322}{\bibinfo {journal} {Ann. Rev.
  Astron. Astrophys.}}\ }%
  \textbf{\bibinfo {volume} {54}},\ \bibinfo {pages} {401--440} (\bibinfo
  {year} {2016}),\ \Eprint{http://arxiv.org/abs/1603.02698}{arXiv:1603.02698
  [astro-ph.HE]}%
  \bibAnnoteFile{NoStop}{Ozel:2016oaf}%
\bibitem{Danielewicz:2002pu}%
  \BibitemOpen
  \bibfield{author}{%
  \bibinfo {author} {\bibfnamefont{Pawel}\ \bibnamefont{Danielewicz}}, \bibinfo
  {author} {\bibfnamefont{Roy}\ \bibnamefont{Lacey}},\ and\ \bibinfo {author}
  {\bibfnamefont{William~G.}\ \bibnamefont{Lynch}},\ }%
  \bibfield{title}{%
  \enquote{\bibinfo {title} {{Determination of the equation of state of dense
  matter}},}\ }%
  \bibfield{journal}{%
  \Doi{10.1126/science.1078070}{\bibinfo {journal} {Science}}\ }%
  \textbf{\bibinfo {volume} {298}},\ \bibinfo {pages} {1592--1596} (\bibinfo
  {year} {2002}),\
  \Eprint{http://arxiv.org/abs/nucl-th/0208016}{arXiv:nucl-th/0208016
  [nucl-th]}%
  \bibAnnoteFile{NoStop}{Danielewicz:2002pu}%
\bibitem{Rezzolla:2016nxn}%
  \BibitemOpen
  \bibfield{author}{%
  \bibinfo {author} {\bibfnamefont{Luciano}\ \bibnamefont{Rezzolla}}\ and\
  \bibinfo {author} {\bibfnamefont{Kentaro}\ \bibnamefont{Takami}},\ }%
  \bibfield{title}{%
  \enquote{\bibinfo {title} {{Gravitational-wave signal from binary neutron
  stars: a systematic analysis of the spectral properties}},}\ }%
  \bibfield{journal}{%
  \Doi{10.1103/PhysRevD.93.124051}{\bibinfo {journal} {Phys. Rev.}}\ }%
  \textbf{\bibinfo {volume} {D93}},\ \bibinfo {pages} {124051} (\bibinfo {year}
  {2016}),\ \Eprint{http://arxiv.org/abs/1604.00246}{arXiv:1604.00246 [gr-qc]}%
  \bibAnnoteFile{NoStop}{Rezzolla:2016nxn}%
\bibitem{Annala:2017llu}%
  \BibitemOpen
  \bibfield{author}{%
  \bibinfo {author} {\bibfnamefont{Eemeli}\ \bibnamefont{Annala}}, \bibinfo
  {author} {\bibfnamefont{Tyler}\ \bibnamefont{Gorda}}, \bibinfo {author}
  {\bibfnamefont{Aleksi}\ \bibnamefont{Kurkela}},\ and\ \bibinfo {author}
  {\bibfnamefont{Aleksi}\ \bibnamefont{Vuorinen}},\ }%
  \bibfield{title}{%
  \enquote{\bibinfo {title} {{Gravitational-wave constraints on the
  neutron-star-matter Equation of State}},}\ }%
  \bibfield{journal}{%
  \Doi{10.1103/PhysRevLett.120.172703}{\bibinfo {journal} {Phys. Rev. Lett.}}\
  }%
  \textbf{\bibinfo {volume} {120}},\ \bibinfo {pages} {172703} (\bibinfo {year}
  {2018}),\ \Eprint{http://arxiv.org/abs/1711.02644}{arXiv:1711.02644
  [astro-ph.HE]}%
  \bibAnnoteFile{NoStop}{Annala:2017llu}%
\bibitem{Paschalidis:2017qmb}%
  \BibitemOpen
  \bibfield{author}{%
  \bibinfo {author} {\bibfnamefont{Vasileios}\ \bibnamefont{Paschalidis}},
  \bibinfo {author} {\bibfnamefont{Kent}\ \bibnamefont{Yagi}}, \bibinfo
  {author} {\bibfnamefont{David}\ \bibnamefont{Alvarez-Castillo}}, \bibinfo
  {author} {\bibfnamefont{David~B.}\ \bibnamefont{Blaschke}},\ and\ \bibinfo
  {author} {\bibfnamefont{Armen}\ \bibnamefont{Sedrakian}},\ }%
  \bibfield{title}{%
  \enquote{\bibinfo {title} {{Implications from GW170817 and I-Love-Q relations
  for relativistic hybrid stars}},}\ }%
  \bibfield{journal}{%
  \Doi{10.1103/PhysRevD.97.084038}{\bibinfo {journal} {Phys. Rev.}}\ }%
  \textbf{\bibinfo {volume} {D97}},\ \bibinfo {pages} {084038} (\bibinfo {year}
  {2018}),\ \Eprint{http://arxiv.org/abs/1712.00451}{arXiv:1712.00451
  [astro-ph.HE]}%
  \bibAnnoteFile{NoStop}{Paschalidis:2017qmb}%
\bibitem{Most:2018hfd}%
  \BibitemOpen
  \bibfield{author}{%
  \bibinfo {author} {\bibfnamefont{Elias~R.}\ \bibnamefont{Most}}, \bibinfo
  {author} {\bibfnamefont{Lukas~R.}\ \bibnamefont{Weih}}, \bibinfo {author}
  {\bibfnamefont{Luciano}\ \bibnamefont{Rezzolla}},\ and\ \bibinfo {author}
  {\bibfnamefont{Jürgen}\ \bibnamefont{Schaffner-Bielich}},\ }%
  \bibfield{title}{%
  \enquote{\bibinfo {title} {{New constraints on radii and tidal
  deformabilities of neutron stars from GW170817}},}\ }%
  \bibfield{journal}{%
  \Doi{10.1103/PhysRevLett.120.261103}{\bibinfo {journal} {Phys. Rev. Lett.}}\
  }%
  \textbf{\bibinfo {volume} {120}},\ \bibinfo {pages} {261103} (\bibinfo {year}
  {2018}),\ \Eprint{http://arxiv.org/abs/1803.00549}{arXiv:1803.00549 [gr-qc]}%
  \bibAnnoteFile{NoStop}{Most:2018hfd}%
\bibitem{Nandi:2018ami}%
  \BibitemOpen
  \bibfield{author}{%
  \bibinfo {author} {\bibfnamefont{Rana}\ \bibnamefont{Nandi}}, \bibinfo
  {author} {\bibfnamefont{Prasanta}\ \bibnamefont{Char}},\ and\ \bibinfo
  {author} {\bibfnamefont{Subrata}\ \bibnamefont{Pal}},\ }%
  \bibfield{title}{%
  \enquote{\bibinfo {title} {{Constraining the equation of state with
  gravitational wave observation}},}\ }%
   (\bibinfo {year} {2018}),\
  \Eprint{http://arxiv.org/abs/1809.07108}{arXiv:1809.07108 [astro-ph.HE]}%
  \bibAnnoteFile{NoStop}{Nandi:2018ami}%
\bibitem{Urbano:2018nrs}%
  \BibitemOpen
  \bibfield{author}{%
  \bibinfo {author} {\bibfnamefont{Alfredo}\ \bibnamefont{Urbano}}\ and\
  \bibinfo {author} {\bibfnamefont{Hardi}\ \bibnamefont{Veermäe}},\ }%
  \bibfield{title}{%
  \enquote{\bibinfo {title} {{On gravitational echoes from ultracompact exotic
  stars}},}\ }%
   (\bibinfo {year} {2018}),\
  \Eprint{http://arxiv.org/abs/1810.07137}{arXiv:1810.07137 [gr-qc]}%
  \bibAnnoteFile{NoStop}{Urbano:2018nrs}%
\bibitem{Demorest:2010bx}%
  \BibitemOpen
  \bibfield{author}{%
  \bibinfo {author} {\bibfnamefont{Paul}\ \bibnamefont{Demorest}}, \bibinfo
  {author} {\bibfnamefont{Tim}\ \bibnamefont{Pennucci}}, \bibinfo {author}
  {\bibfnamefont{Scott}\ \bibnamefont{Ransom}}, \bibinfo {author}
  {\bibfnamefont{Mallory}\ \bibnamefont{Roberts}},\ and\ \bibinfo {author}
  {\bibfnamefont{Jason}\ \bibnamefont{Hessels}},\ }%
  \bibfield{title}{%
  \enquote{\bibinfo {title} {{Shapiro Delay Measurement of A Two Solar Mass
  Neutron Star}},}\ }%
  \bibfield{journal}{%
  \Doi{10.1038/nature09466}{\bibinfo {journal} {Nature}}\ }%
  \textbf{\bibinfo {volume} {467}},\ \bibinfo {pages} {1081--1083} (\bibinfo
  {year} {2010}),\ \Eprint{http://arxiv.org/abs/1010.5788}{arXiv:1010.5788
  [astro-ph.HE]}%
  \bibAnnoteFile{NoStop}{Demorest:2010bx}%
\bibitem{Antoniadis:2013pzd}%
  \BibitemOpen
  \bibfield{author}{%
  \bibinfo {author} {\bibfnamefont{John}\ \bibnamefont{Antoniadis}}
  \emph{et~al.},\ }%
  \bibfield{title}{%
  \enquote{\bibinfo {title} {{A Massive Pulsar in a Compact Relativistic
  Binary}},}\ }%
  \bibfield{journal}{%
  \Doi{10.1126/science.1233232}{\bibinfo {journal} {Science}}\ }%
  \textbf{\bibinfo {volume} {340}},\ \bibinfo {pages} {6131} (\bibinfo {year}
  {2013}),\ \Eprint{http://arxiv.org/abs/1304.6875}{arXiv:1304.6875
  [astro-ph.HE]}%
  \bibAnnoteFile{NoStop}{Antoniadis:2013pzd}%
\bibitem{Chandrasekhar:1931ih}%
  \BibitemOpen
  \bibfield{author}{%
  \bibinfo {author} {\bibfnamefont{Subrahmanyan}\
  \bibnamefont{Chandrasekhar}},\ }%
  \bibfield{title}{%
  \enquote{\bibinfo {title} {{The maximum mass of ideal white dwarfs}},}\ }%
  \bibfield{journal}{%
  \Doi{10.1086/143324}{\bibinfo {journal} {Astrophys. J.}}\ }%
  \textbf{\bibinfo {volume} {74}},\ \bibinfo {pages} {81--82} (\bibinfo {year}
  {1931})%
  \bibAnnoteFile{NoStop}{Chandrasekhar:1931ih}%
\bibitem{Baym:1976yu}%
  \BibitemOpen
  \bibfield{author}{%
  \bibinfo {author} {\bibfnamefont{G.}~\bibnamefont{Baym}}\ and\ \bibinfo
  {author} {\bibfnamefont{S.~A.}\ \bibnamefont{Chin}},\ }%
  \bibfield{title}{%
  \enquote{\bibinfo {title} {{Can a Neutron Star Be a Giant MIT Bag?}}.}\ }%
  \bibfield{journal}{%
  \Doi{10.1016/0370-2693(76)90517-7}{\bibinfo {journal} {Phys. Lett.}}\ }%
  \textbf{\bibinfo {volume} {62B}},\ \bibinfo {pages} {241--244} (\bibinfo
  {year} {1976})%
  \bibAnnoteFile{NoStop}{Baym:1976yu}%
\bibitem{Drago:1995ah}%
  \BibitemOpen
  \bibfield{author}{%
  \bibinfo {author} {\bibfnamefont{A.}~\bibnamefont{Drago}}, \bibinfo {author}
  {\bibfnamefont{U.}~\bibnamefont{Tambini}},\ and\ \bibinfo {author}
  {\bibfnamefont{M.}~\bibnamefont{Hjorth-Jensen}},\ }%
  \bibfield{title}{%
  \enquote{\bibinfo {title} {{Neutron stars and massive quark matter}},}\ }%
  \bibfield{journal}{%
  \Doi{10.1016/0370-2693(96)00479-0}{\bibinfo {journal} {Phys. Lett.}}\ }%
  \textbf{\bibinfo {volume} {B380}},\ \bibinfo {pages} {13--17} (\bibinfo
  {year} {1996}),\
  \Eprint{http://arxiv.org/abs/nucl-th/9505037}{arXiv:nucl-th/9505037
  [nucl-th]}%
  \bibAnnoteFile{NoStop}{Drago:1995ah}%
\bibitem{RikovskaStone:2006ta}%
  \BibitemOpen
  \bibfield{author}{%
  \bibinfo {author} {\bibfnamefont{J.}~\bibnamefont{Rikovska-Stone}}, \bibinfo
  {author} {\bibfnamefont{Pierre A.~M.}\ \bibnamefont{Guichon}}, \bibinfo
  {author} {\bibfnamefont{Hrayr~H.}\ \bibnamefont{Matevosyan}},\ and\ \bibinfo
  {author} {\bibfnamefont{Anthony~William}\ \bibnamefont{Thomas}},\ }%
  \bibfield{title}{%
  \enquote{\bibinfo {title} {{Cold uniform matter and neutron stars in the
  quark-mesons-coupling model}},}\ }%
  \bibfield{journal}{%
  \Doi{10.1016/j.nuclphysa.2007.05.011}{\bibinfo {journal} {Nucl. Phys.}}\ }%
  \textbf{\bibinfo {volume} {A792}},\ \bibinfo {pages} {341--369} (\bibinfo
  {year} {2007}),\
  \Eprint{http://arxiv.org/abs/nucl-th/0611030}{arXiv:nucl-th/0611030
  [nucl-th]}%
  \bibAnnoteFile{NoStop}{RikovskaStone:2006ta}%
\bibitem{Stone:2010jt}%
  \BibitemOpen
  \bibfield{author}{%
  \bibinfo {author} {\bibfnamefont{J.~R.}\ \bibnamefont{Stone}}, \bibinfo
  {author} {\bibfnamefont{P.~A.~M.}\ \bibnamefont{Guichon}},\ and\ \bibinfo
  {author} {\bibfnamefont{A.~W.}\ \bibnamefont{Thomas}},\ }%
  \bibfield{title}{%
  \enquote{\bibinfo {title} {{Role of Hyperons in Neutron Stars}},}\ }%
   (\bibinfo {year} {2010}),\
  \Eprint{http://arxiv.org/abs/1012.2919}{arXiv:1012.2919 [nucl-th]}%
  \bibAnnoteFile{NoStop}{Stone:2010jt}%
\bibitem{Motta:2018rxp}%
  \BibitemOpen
  \bibfield{author}{%
  \bibinfo {author} {\bibfnamefont{T.~F.}\ \bibnamefont{Motta}}, \bibinfo
  {author} {\bibfnamefont{P.~A.~M.}\ \bibnamefont{Guichon}},\ and\ \bibinfo
  {author} {\bibfnamefont{A.~W.}\ \bibnamefont{Thomas}},\ }%
  \bibfield{title}{%
  \enquote{\bibinfo {title} {{Implications of Neutron Star Properties for the
  Existence of Light Dark Matter}},}\ }%
  \bibfield{journal}{%
  \Doi{10.1088/1361-6471/aab689}{\bibinfo {journal} {J. Phys.}}\ }%
  \textbf{\bibinfo {volume} {G45}},\ \bibinfo {pages} {05LT01} (\bibinfo {year}
  {2018}),\ \Eprint{http://arxiv.org/abs/1802.08427}{arXiv:1802.08427
  [nucl-th]}%
  \bibAnnoteFile{NoStop}{Motta:2018rxp}%
\bibitem{Deb:2018}%
  \BibitemOpen
  \bibfield{author}{%
  \bibinfo {author} {\bibfnamefont{Debabrata}\ \bibnamefont{Deb}}, \bibinfo
  {author} {\bibfnamefont{Sergei~V.}\ \bibnamefont{Ketov}}, \bibinfo {author}
  {\bibfnamefont{S.~K.}\ \bibnamefont{Maurya}}, \bibinfo {author}
  {\bibfnamefont{Maxim}\ \bibnamefont{Khlopov}}, \bibinfo {author}
  {\bibfnamefont{P.~H. R.~S.}\ \bibnamefont{Moraes}},\ and\ \bibinfo {author}
  {\bibfnamefont{Saibal}\ \bibnamefont{Ray}},\ }%
  \bibfield{title}{%
  \enquote{\bibinfo {title} {{Exploring physical features of anisotropic
  strange stars beyond standard maximum mass limit in
  $f\left(R,\mathcal{T}\right)$ gravity}},}\ }%
   (\bibinfo {year} {2018}),\
  \Eprint{http://arxiv.org/abs/1810.07678}{arXiv:1810.07678 [gr-qc]}%
  \bibAnnoteFile{NoStop}{Deb:2018}%
\bibitem{Abhishek:2018xml}%
  \BibitemOpen
  \bibfield{author}{%
  \bibinfo {author} {\bibfnamefont{Aman}\ \bibnamefont{Abhishek}}\ and\
  \bibinfo {author} {\bibfnamefont{Hiranmaya}\ \bibnamefont{Mishra}},\ }%
  \bibfield{title}{%
  \enquote{\bibinfo {title} {{Chiral symmetry breaking, color
  superconductivity, and the equation of state for magnetized strange quark
  matter}},}\ }%
   (\bibinfo {year} {2018}),\
  \Eprint{http://arxiv.org/abs/1810.09276}{arXiv:1810.09276 [hep-ph]}%
  \bibAnnoteFile{NoStop}{Abhishek:2018xml}%
\bibitem{Jokela:2018ers}%
  \BibitemOpen
  \bibfield{author}{%
  \bibinfo {author} {\bibfnamefont{Niko}\ \bibnamefont{Jokela}}, \bibinfo
  {author} {\bibfnamefont{Matti}\ \bibnamefont{Järvinen}},\ and\ \bibinfo
  {author} {\bibfnamefont{Jere}\ \bibnamefont{Remes}},\ }%
  \bibfield{title}{%
  \enquote{\bibinfo {title} {{Holographic QCD in the Veneziano limit and
  neutron stars}},}\ }%
   (\bibinfo {year} {2018}),\
  \Eprint{http://arxiv.org/abs/1809.07770}{arXiv:1809.07770 [hep-ph]}%
  \bibAnnoteFile{NoStop}{Jokela:2018ers}%
\bibitem{Lorce:2011kd}%
  \BibitemOpen
  \bibfield{author}{%
  \bibinfo {author} {\bibfnamefont{C.}~\bibnamefont{Lorcé}}\ and\ \bibinfo
  {author} {\bibfnamefont{B.}~\bibnamefont{Pasquini}},\ }%
  \bibfield{title}{%
  \enquote{\bibinfo {title} {{Quark Wigner Distributions and Orbital Angular
  Momentum}},}\ }%
  \bibfield{journal}{%
  \Doi{10.1103/PhysRevD.84.014015}{\bibinfo {journal} {Phys. Rev.}}\ }%
  \textbf{\bibinfo {volume} {D84}},\ \bibinfo {pages} {014015} (\bibinfo {year}
  {2011}),\ \Eprint{http://arxiv.org/abs/1106.0139}{arXiv:1106.0139 [hep-ph]}%
  \bibAnnoteFile{NoStop}{Lorce:2011kd}%
\bibitem{Lorce:2014mxa}%
  \BibitemOpen
  \bibfield{author}{%
  \bibinfo {author} {\bibfnamefont{C.}~\bibnamefont{Lorcé}},\ }%
  \bibfield{title}{%
  \enquote{\bibinfo {title} {{Spin–orbit correlations in the nucleon}},}\ }%
  \bibfield{journal}{%
  \Doi{10.1016/j.physletb.2014.06.068}{\bibinfo {journal} {Phys. Lett.}}\ }%
  \textbf{\bibinfo {volume} {B735}},\ \bibinfo {pages} {344--348} (\bibinfo
  {year} {2014}),\ \Eprint{http://arxiv.org/abs/1401.7784}{arXiv:1401.7784
  [hep-ph]}%
  \bibAnnoteFile{NoStop}{Lorce:2014mxa}%
\bibitem{Bhoonah:2017olu}%
  \BibitemOpen
  \bibfield{author}{%
  \bibinfo {author} {\bibfnamefont{Amit}\ \bibnamefont{Bhoonah}}\ and\ \bibinfo
  {author} {\bibfnamefont{Cédric}\ \bibnamefont{Lorcé}},\ }%
  \bibfield{title}{%
  \enquote{\bibinfo {title} {{Quark transverse spin–orbit correlations}},}\
  }%
  \bibfield{journal}{%
  \Doi{10.1016/j.physletb.2017.10.003}{\bibinfo {journal} {Phys. Lett.}}\ }%
  \textbf{\bibinfo {volume} {B774}},\ \bibinfo {pages} {435--440} (\bibinfo
  {year} {2017}),\ \Eprint{http://arxiv.org/abs/1703.08322}{arXiv:1703.08322
  [hep-ph]}%
  \bibAnnoteFile{NoStop}{Bhoonah:2017olu}%
\bibitem{Weinberg:1966jm}%
  \BibitemOpen
  \bibfield{author}{%
  \bibinfo {author} {\bibfnamefont{Steven}\ \bibnamefont{Weinberg}},\ }%
  \bibfield{title}{%
  \enquote{\bibinfo {title} {{Dynamics at infinite momentum}},}\ }%
  \bibfield{journal}{%
  \Doi{10.1103/PhysRev.150.1313}{\bibinfo {journal} {Phys. Rev.}}\ }%
  \textbf{\bibinfo {volume} {150}},\ \bibinfo {pages} {1313--1318} (\bibinfo
  {year} {1966})%
  \bibAnnoteFile{NoStop}{Weinberg:1966jm}%
\bibitem{Burkardt:2000za}%
  \BibitemOpen
  \bibfield{author}{%
  \bibinfo {author} {\bibfnamefont{Matthias}\ \bibnamefont{Burkardt}},\ }%
  \bibfield{title}{%
  \enquote{\bibinfo {title} {{Impact parameter dependent parton distributions
  and off forward parton distributions for zeta ---> 0}},}\ }%
  \bibfield{journal}{%
  \Doi{10.1103/PhysRevD.62.071503, 10.1103/PhysRevD.66.119903}{\bibinfo
  {journal} {Phys. Rev.}}\ }%
  \textbf{\bibinfo {volume} {D62}},\ \bibinfo {pages} {071503} (\bibinfo {year}
  {2000}),\ \bibinfo {note} {[Erratum: Phys. Rev.D66,119903(2002)]},\
  \Eprint{http://arxiv.org/abs/hep-ph/0005108}{arXiv:hep-ph/0005108 [hep-ph]}%
  \bibAnnoteFile{NoStop}{Burkardt:2000za}%
\bibitem{Belitsky:2005qn}%
  \BibitemOpen
  \bibfield{author}{%
  \bibinfo {author} {\bibfnamefont{A.~V.}\ \bibnamefont{Belitsky}}\ and\
  \bibinfo {author} {\bibfnamefont{A.~V.}\ \bibnamefont{Radyushkin}},\ }%
  \bibfield{title}{%
  \enquote{\bibinfo {title} {{Unraveling hadron structure with generalized
  parton distributions}},}\ }%
  \bibfield{journal}{%
  \Doi{10.1016/j.physrep.2005.06.002}{\bibinfo {journal} {Phys. Rept.}}\ }%
  \textbf{\bibinfo {volume} {418}},\ \bibinfo {pages} {1--387} (\bibinfo {year}
  {2005}),\ \Eprint{http://arxiv.org/abs/hep-ph/0504030}{arXiv:hep-ph/0504030
  [hep-ph]}%
  \bibAnnoteFile{NoStop}{Belitsky:2005qn}%
\bibitem{Dirac:1949cp}%
  \BibitemOpen
  \bibfield{author}{%
  \bibinfo {author} {\bibfnamefont{Paul A.~M.}\ \bibnamefont{Dirac}},\ }%
  \bibfield{title}{%
  \enquote{\bibinfo {title} {{Forms of Relativistic Dynamics}},}\ }%
  \bibfield{journal}{%
  \Doi{10.1103/RevModPhys.21.392}{\bibinfo {journal} {Rev. Mod. Phys.}}\ }%
  \textbf{\bibinfo {volume} {21}},\ \bibinfo {pages} {392--399} (\bibinfo
  {year} {1949})%
  \bibAnnoteFile{NoStop}{Dirac:1949cp}%
\bibitem{Brodsky:1997de}%
  \BibitemOpen
  \bibfield{author}{%
  \bibinfo {author} {\bibfnamefont{Stanley~J.}\ \bibnamefont{Brodsky}},
  \bibinfo {author} {\bibfnamefont{Hans-Christian}\ \bibnamefont{Pauli}},\ and\
  \bibinfo {author} {\bibfnamefont{Stephen~S.}\ \bibnamefont{Pinsky}},\ }%
  \bibfield{title}{%
  \enquote{\bibinfo {title} {{Quantum chromodynamics and other field theories
  on the light cone}},}\ }%
  \bibfield{journal}{%
  \Doi{10.1016/S0370-1573(97)00089-6}{\bibinfo {journal} {Phys. Rept.}}\ }%
  \textbf{\bibinfo {volume} {301}},\ \bibinfo {pages} {299--486} (\bibinfo
  {year} {1998}),\
  \Eprint{http://arxiv.org/abs/hep-ph/9705477}{arXiv:hep-ph/9705477 [hep-ph]}%
  \bibAnnoteFile{NoStop}{Brodsky:1997de}%
\bibitem{Susskind:1967rg}%
  \BibitemOpen
  \bibfield{author}{%
  \bibinfo {author} {\bibfnamefont{Leonard}\ \bibnamefont{Susskind}},\ }%
  \bibfield{title}{%
  \enquote{\bibinfo {title} {{Model of selfinduced strong interactions}},}\ }%
  \bibfield{journal}{%
  \Doi{10.1103/PhysRev.165.1535}{\bibinfo {journal} {Phys. Rev.}}\ }%
  \textbf{\bibinfo {volume} {165}},\ \bibinfo {pages} {1535--1546} (\bibinfo
  {year} {1968})%
  \bibAnnoteFile{NoStop}{Susskind:1967rg}%
\bibitem{Lorce:2011ni}%
  \BibitemOpen
  \bibfield{author}{%
  \bibinfo {author} {\bibfnamefont{Cédric}\ \bibnamefont{Lorcé}}, \bibinfo
  {author} {\bibfnamefont{Barbara}\ \bibnamefont{Pasquini}}, \bibinfo {author}
  {\bibfnamefont{Xiaonu}\ \bibnamefont{Xiong}},\ and\ \bibinfo {author}
  {\bibfnamefont{Feng}\ \bibnamefont{Yuan}},\ }%
  \bibfield{title}{%
  \enquote{\bibinfo {title} {{The quark orbital angular momentum from Wigner
  distributions and light-cone wave functions}},}\ }%
  \bibfield{journal}{%
  \Doi{10.1103/PhysRevD.85.114006}{\bibinfo {journal} {Phys. Rev.}}\ }%
  \textbf{\bibinfo {volume} {D85}},\ \bibinfo {pages} {114006} (\bibinfo {year}
  {2012}),\ \Eprint{http://arxiv.org/abs/1111.4827}{arXiv:1111.4827 [hep-ph]}%
  \bibAnnoteFile{NoStop}{Lorce:2011ni}%
\bibitem{Lorce:2012ce}%
  \BibitemOpen
  \bibfield{author}{%
  \bibinfo {author} {\bibfnamefont{Cédric}\ \bibnamefont{Lorcé}},\ }%
  \bibfield{title}{%
  \enquote{\bibinfo {title} {{Wilson lines and orbital angular momentum}},}\ }%
  \bibfield{journal}{%
  \Doi{10.1016/j.physletb.2013.01.007}{\bibinfo {journal} {Phys. Lett.}}\ }%
  \textbf{\bibinfo {volume} {B719}},\ \bibinfo {pages} {185--190} (\bibinfo
  {year} {2013}),\ \Eprint{http://arxiv.org/abs/1210.2581}{arXiv:1210.2581
  [hep-ph]}%
  \bibAnnoteFile{NoStop}{Lorce:2012ce}%
\bibitem{Lorce:2015sqe}%
  \BibitemOpen
  \bibfield{author}{%
  \bibinfo {author} {\bibfnamefont{C.}~\bibnamefont{Lorcé}}\ and\ \bibinfo
  {author} {\bibfnamefont{B.}~\bibnamefont{Pasquini}},\ }%
  \bibfield{title}{%
  \enquote{\bibinfo {title} {{Multipole decomposition of the nucleon transverse
  phase space}},}\ }%
  \bibfield{journal}{%
  \Doi{10.1103/PhysRevD.93.034040}{\bibinfo {journal} {Phys. Rev.}}\ }%
  \textbf{\bibinfo {volume} {D93}},\ \bibinfo {pages} {034040} (\bibinfo {year}
  {2016}),\ \Eprint{http://arxiv.org/abs/1512.06744}{arXiv:1512.06744
  [hep-ph]}%
  \bibAnnoteFile{NoStop}{Lorce:2015sqe}%
\bibitem{Abidin:2008sb}%
  \BibitemOpen
  \bibfield{author}{%
  \bibinfo {author} {\bibfnamefont{Zainul}\ \bibnamefont{Abidin}}\ and\
  \bibinfo {author} {\bibfnamefont{Carl~E.}\ \bibnamefont{Carlson}},\ }%
  \bibfield{title}{%
  \enquote{\bibinfo {title} {{Hadronic Momentum Densities in the Transverse
  Plane}},}\ }%
  \bibfield{journal}{%
  \Doi{10.1103/PhysRevD.78.071502}{\bibinfo {journal} {Phys. Rev.}}\ }%
  \textbf{\bibinfo {volume} {D78}},\ \bibinfo {pages} {071502} (\bibinfo {year}
  {2008}),\ \Eprint{http://arxiv.org/abs/0808.3097}{arXiv:0808.3097 [hep-ph]}%
  \bibAnnoteFile{NoStop}{Abidin:2008sb}%
\bibitem{Selyugin:2009ic}%
  \BibitemOpen
  \bibfield{author}{%
  \bibinfo {author} {\bibfnamefont{O.~V.}\ \bibnamefont{Selyugin}}\ and\
  \bibinfo {author} {\bibfnamefont{O.~V.}\ \bibnamefont{Teryaev}},\ }%
  \bibfield{title}{%
  \enquote{\bibinfo {title} {{Generalized Parton Distributions and Description
  of Electromagnetic and Graviton form factors of nucleon}},}\ }%
  \bibfield{journal}{%
  \Doi{10.1103/PhysRevD.79.033003}{\bibinfo {journal} {Phys. Rev.}}\ }%
  \textbf{\bibinfo {volume} {D79}},\ \bibinfo {pages} {033003} (\bibinfo {year}
  {2009}),\ \Eprint{http://arxiv.org/abs/0901.1786}{arXiv:0901.1786 [hep-ph]}%
  \bibAnnoteFile{NoStop}{Selyugin:2009ic}%
\bibitem{Son:2014sna}%
  \BibitemOpen
  \bibfield{author}{%
  \bibinfo {author} {\bibfnamefont{Hyeon-Dong}\ \bibnamefont{Son}}\ and\
  \bibinfo {author} {\bibfnamefont{Hyun-Chul}\ \bibnamefont{Kim}},\ }%
  \bibfield{title}{%
  \enquote{\bibinfo {title} {{Stability of the pion and the pattern of chiral
  symmetry breaking}},}\ }%
  \bibfield{journal}{%
  \Doi{10.1103/PhysRevD.90.111901}{\bibinfo {journal} {Phys. Rev.}}\ }%
  \textbf{\bibinfo {volume} {D90}},\ \bibinfo {pages} {111901} (\bibinfo {year}
  {2014}),\ \Eprint{http://arxiv.org/abs/1410.1420}{arXiv:1410.1420 [hep-ph]}%
  \bibAnnoteFile{NoStop}{Son:2014sna}%
\bibitem{Chakrabarti:2015lba}%
  \BibitemOpen
  \bibfield{author}{%
  \bibinfo {author} {\bibfnamefont{Dipankar}\ \bibnamefont{Chakrabarti}},
  \bibinfo {author} {\bibfnamefont{Chandan}\ \bibnamefont{Mondal}},\ and\
  \bibinfo {author} {\bibfnamefont{Asmita}\ \bibnamefont{Mukherjee}},\ }%
  \bibfield{title}{%
  \enquote{\bibinfo {title} {{Gravitational form factors and transverse spin
  sum rule in a light front quark-diquark model in AdS/QCD}},}\ }%
  \bibfield{journal}{%
  \Doi{10.1103/PhysRevD.91.114026}{\bibinfo {journal} {Phys. Rev.}}\ }%
  \textbf{\bibinfo {volume} {D91}},\ \bibinfo {pages} {114026} (\bibinfo {year}
  {2015}),\ \Eprint{http://arxiv.org/abs/1505.02013}{arXiv:1505.02013
  [hep-ph]}%
  \bibAnnoteFile{NoStop}{Chakrabarti:2015lba}%
\bibitem{Mondal:2016xsm}%
  \BibitemOpen
  \bibfield{author}{%
  \bibinfo {author} {\bibfnamefont{Chandan}\ \bibnamefont{Mondal}}, \bibinfo
  {author} {\bibfnamefont{Narinder}\ \bibnamefont{Kumar}}, \bibinfo {author}
  {\bibfnamefont{Harleen}\ \bibnamefont{Dahiya}},\ and\ \bibinfo {author}
  {\bibfnamefont{Dipankar}\ \bibnamefont{Chakrabarti}},\ }%
  \bibfield{title}{%
  \enquote{\bibinfo {title} {{Charge and longitudinal momentum distributions in
  transverse coordinate space}},}\ }%
  \bibfield{journal}{%
  \Doi{10.1103/PhysRevD.94.074028}{\bibinfo {journal} {Phys. Rev.}}\ }%
  \textbf{\bibinfo {volume} {D94}},\ \bibinfo {pages} {074028} (\bibinfo {year}
  {2016}),\ \Eprint{http://arxiv.org/abs/1608.01095}{arXiv:1608.01095
  [hep-ph]}%
  \bibAnnoteFile{NoStop}{Mondal:2016xsm}%
\bibitem{SattaryNikkhoo:2018odd}%
  \BibitemOpen
  \bibfield{author}{%
  \bibinfo {author} {\bibfnamefont{Negin}\ \bibnamefont{Sattary~Nikkhoo}}\ and\
  \bibinfo {author} {\bibfnamefont{Mohammad~Reza}\ \bibnamefont{Shojaei}},\ }%
  \bibfield{title}{%
  \enquote{\bibinfo {title} {{Electromagnetic and gravitational form factors by
  using the modified Gaussian ansatz for $H^q(x,t)$}},}\ }%
  \bibfield{journal}{%
  \Doi{10.1016/j.nuclphysa.2017.09.010}{\bibinfo {journal} {Nucl. Phys.}}\ }%
  \textbf{\bibinfo {volume} {A969}},\ \bibinfo {pages} {138--150} (\bibinfo
  {year} {2018})%
  \bibAnnoteFile{NoStop}{SattaryNikkhoo:2018odd}%
\bibitem{Kaur:2018ewq}%
  \BibitemOpen
  \bibfield{author}{%
  \bibinfo {author} {\bibfnamefont{Navdeep}\ \bibnamefont{Kaur}}, \bibinfo
  {author} {\bibfnamefont{Narinder}\ \bibnamefont{Kumar}}, \bibinfo {author}
  {\bibfnamefont{Chandan}\ \bibnamefont{Mondal}},\ and\ \bibinfo {author}
  {\bibfnamefont{Harleen}\ \bibnamefont{Dahiya}},\ }%
  \bibfield{title}{%
  \enquote{\bibinfo {title} {{Generalized Parton Distributions of Pion for
  Non-Zero Skewness in AdS/QCD}},}\ }%
  \bibfield{journal}{%
  \Doi{10.1016/j.nuclphysb.2018.07.003}{\bibinfo {journal} {Nucl. Phys.}}\ }%
  \textbf{\bibinfo {volume} {B934}},\ \bibinfo {pages} {80--95} (\bibinfo
  {year} {2018}),\ \Eprint{http://arxiv.org/abs/1807.01076}{arXiv:1807.01076
  [hep-ph]}%
  \bibAnnoteFile{NoStop}{Kaur:2018ewq}%
\bibitem{Adhikari:2016dir}%
  \BibitemOpen
  \bibfield{author}{%
  \bibinfo {author} {\bibfnamefont{Lekha}\ \bibnamefont{Adhikari}}\ and\
  \bibinfo {author} {\bibfnamefont{Matthias}\ \bibnamefont{Burkardt}},\ }%
  \bibfield{title}{%
  \enquote{\bibinfo {title} {{Angular Momentum Distribution in the Transverse
  Plane}},}\ }%
  \bibfield{journal}{%
  \Doi{10.1103/PhysRevD.94.114021}{\bibinfo {journal} {Phys. Rev.}}\ }%
  \textbf{\bibinfo {volume} {D94}},\ \bibinfo {pages} {114021} (\bibinfo {year}
  {2016}),\ \Eprint{http://arxiv.org/abs/1609.07099}{arXiv:1609.07099
  [hep-ph]}%
  \bibAnnoteFile{NoStop}{Adhikari:2016dir}%
\bibitem{Soper:1976bb}%
  \BibitemOpen
  \bibfield{author}{%
  \bibinfo {author} {\bibfnamefont{D.~E.}\ \bibnamefont{Soper}},\ }%
  \emph{\bibinfo {title} {{Classical Field Theory}}}\ (\bibinfo {publisher}
  {New York, Usa: Wiley (1976) 259p},\ \bibinfo {year} {1976})%
  \bibAnnoteFile{NoStop}{Soper:1976bb}%
\bibitem{Geracie:2016bkg}%
  \BibitemOpen
  \bibfield{author}{%
  \bibinfo {author} {\bibfnamefont{Michael}\ \bibnamefont{Geracie}},\ }%
  \emph{\bibinfo {title} {{Galilean Geometry in Condensed Matter Systems}}},\
  Ph.D. thesis,\ \bibinfo {school} {Chicago U.} (\bibinfo {year} {2016}),\
  \Eprint{http://arxiv.org/abs/1611.01198}{arXiv:1611.01198 [hep-th]}%
  \bibAnnoteFile{NoStop}{Geracie:2016bkg}%
\bibitem{Hawking:1973uf}%
  \BibitemOpen
  \bibfield{author}{%
  \bibinfo {author} {\bibfnamefont{S.~W.}\ \bibnamefont{Hawking}}\ and\
  \bibinfo {author} {\bibfnamefont{G.~F.~R.}\ \bibnamefont{Ellis}},\ }%
  \Doi{10.1017/CBO9780511524646}{\emph{\bibinfo {title} {{The Large Scale
  Structure of Space-Time}}}},\ Cambridge Monographs on Mathematical Physics\
  (\bibinfo {publisher} {Cambridge University Press},\ \bibinfo {year} {2011})\
  ISBN \bibinfo {isbn} {9780521200165, 9780521099066, 9780511826306,
  9780521099066}%
  \bibAnnoteFile{NoStop}{Hawking:1973uf}%
\bibitem{Husain:1995bf}%
  \BibitemOpen
  \bibfield{author}{%
  \bibinfo {author} {\bibfnamefont{Viqar}\ \bibnamefont{Husain}},\ }%
  \bibfield{title}{%
  \enquote{\bibinfo {title} {{Exact solutions for null fluid collapse}},}\ }%
  \bibfield{journal}{%
  \Doi{10.1103/PhysRevD.53.R1759}{\bibinfo {journal} {Phys. Rev.}}\ }%
  \textbf{\bibinfo {volume} {D53}},\ \bibinfo {pages} {1759--1762} (\bibinfo
  {year} {1996}),\
  \Eprint{http://arxiv.org/abs/gr-qc/9511011}{arXiv:gr-qc/9511011 [gr-qc]}%
  \bibAnnoteFile{NoStop}{Husain:1995bf}%
\bibitem{Wang:1998qx}%
  \BibitemOpen
  \bibfield{author}{%
  \bibinfo {author} {\bibfnamefont{Anzhong}\ \bibnamefont{Wang}}\ and\ \bibinfo
  {author} {\bibfnamefont{Yumei}\ \bibnamefont{Wu}},\ }%
  \bibfield{title}{%
  \enquote{\bibinfo {title} {{Generalized Vaidya solutions}},}\ }%
  \bibfield{journal}{%
  \Doi{10.1023/A:1018819521971}{\bibinfo {journal} {Gen. Rel. Grav.}}\ }%
  \textbf{\bibinfo {volume} {31}},\ \bibinfo {pages} {107} (\bibinfo {year}
  {1999}),\ \Eprint{http://arxiv.org/abs/gr-qc/9803038}{arXiv:gr-qc/9803038
  [gr-qc]}%
  \bibAnnoteFile{NoStop}{Wang:1998qx}%
\bibitem{Brodsky:2006in}%
  \BibitemOpen
  \bibfield{author}{%
  \bibinfo {author} {\bibfnamefont{S.~J.}\ \bibnamefont{Brodsky}}, \bibinfo
  {author} {\bibfnamefont{D.}~\bibnamefont{Chakrabarti}}, \bibinfo {author}
  {\bibfnamefont{A.}~\bibnamefont{Harindranath}}, \bibinfo {author}
  {\bibfnamefont{A.}~\bibnamefont{Mukherjee}},\ and\ \bibinfo {author}
  {\bibfnamefont{J.~P.}\ \bibnamefont{Vary}},\ }%
  \bibfield{title}{%
  \enquote{\bibinfo {title} {{Hadron optics: Diffraction patterns in deeply
  virtual Compton scattering}},}\ }%
  \bibfield{journal}{%
  \Doi{10.1016/j.physletb.2006.08.061}{\bibinfo {journal} {Phys. Lett.}}\ }%
  \textbf{\bibinfo {volume} {B641}},\ \bibinfo {pages} {440--446} (\bibinfo
  {year} {2006}),\
  \Eprint{http://arxiv.org/abs/hep-ph/0604262}{arXiv:hep-ph/0604262 [hep-ph]}%
  \bibAnnoteFile{NoStop}{Brodsky:2006in}%
\bibitem{Brodsky:2006ku}%
  \BibitemOpen
  \bibfield{author}{%
  \bibinfo {author} {\bibfnamefont{S.~J.}\ \bibnamefont{Brodsky}}, \bibinfo
  {author} {\bibfnamefont{D.}~\bibnamefont{Chakrabarti}}, \bibinfo {author}
  {\bibfnamefont{A.}~\bibnamefont{Harindranath}}, \bibinfo {author}
  {\bibfnamefont{A.}~\bibnamefont{Mukherjee}},\ and\ \bibinfo {author}
  {\bibfnamefont{J.~P.}\ \bibnamefont{Vary}},\ }%
  \bibfield{title}{%
  \enquote{\bibinfo {title} {{Hadron optics in three-dimensional invariant
  coordinate space from deeply virtual compton scattering}},}\ }%
  \bibfield{journal}{%
  \Doi{10.1103/PhysRevD.75.014003}{\bibinfo {journal} {Phys. Rev.}}\ }%
  \textbf{\bibinfo {volume} {D75}},\ \bibinfo {pages} {014003} (\bibinfo {year}
  {2007}),\ \Eprint{http://arxiv.org/abs/hep-ph/0611159}{arXiv:hep-ph/0611159
  [hep-ph]}%
  \bibAnnoteFile{NoStop}{Brodsky:2006ku}%
\bibitem{Diehl:2002he}%
  \BibitemOpen
  \bibfield{author}{%
  \bibinfo {author} {\bibfnamefont{M.}~\bibnamefont{Diehl}},\ }%
  \bibfield{title}{%
  \enquote{\bibinfo {title} {{Generalized parton distributions in impact
  parameter space}},}\ }%
  \bibfield{journal}{%
  \Doi{10.1007/s10052-002-1016-9}{\bibinfo {journal} {Eur. Phys. J.}}\ }%
  \textbf{\bibinfo {volume} {C25}},\ \bibinfo {pages} {223--232} (\bibinfo
  {year} {2002}),\ \bibinfo {note} {[Erratum: Eur. Phys. J.C31,277(2003)]},\
  \Eprint{http://arxiv.org/abs/hep-ph/0205208}{arXiv:hep-ph/0205208 [hep-ph]}%
  \bibAnnoteFile{NoStop}{Diehl:2002he}%
\bibitem{Gokhroo:1994fbj}%
  \BibitemOpen
  \bibfield{author}{%
  \bibinfo {author} {\bibfnamefont{M.~K.}\ \bibnamefont{Gokhroo}}\ and\
  \bibinfo {author} {\bibfnamefont{A.~L.}\ \bibnamefont{Mehra}},\ }%
  \bibfield{title}{%
  \enquote{\bibinfo {title} {{Anisotropic spheres with variable energy density
  in general relativity}},}\ }%
  \bibfield{journal}{%
  \Doi{10.1007/BF02088210}{\bibinfo {journal} {Gen. Rel. Grav.}}\ }%
  \textbf{\bibinfo {volume} {26}},\ \bibinfo {pages} {75--84} (\bibinfo {year}
  {1994})%
  \bibAnnoteFile{NoStop}{Gokhroo:1994fbj}%
\bibitem{Marchand:2011}%
  \BibitemOpen
  \bibfield{author}{%
  \bibinfo {author} {\bibfnamefont{A.}~\bibnamefont{Marchand}}, \bibinfo
  {author} {\bibfnamefont{J.H.}\ \bibnamefont{Weijs}}, \bibinfo {author}
  {\bibfnamefont{J.H.}\ \bibnamefont{Snoeijer}},\ and\ \bibinfo {author}
  {\bibfnamefont{B.}~\bibnamefont{Andreotti}},\ }%
  \bibfield{title}{%
  \enquote{\bibinfo {title} {{Why is surface tension a force parallel to the
  interface?}}.}\ }%
  \bibfield{journal}{%
  \Doi{10.1119/1.3619866}{\bibinfo {journal} {Am. J. Phys.}}\ }%
  \textbf{\bibinfo {volume} {79}},\ \bibinfo {pages} {999--1008} (\bibinfo
  {year} {2011})%
  \bibAnnoteFile{NoStop}{Marchand:2011}%
\bibitem{Weijs:2011}%
  \BibitemOpen
  \bibfield{author}{%
  \bibinfo {author} {\bibfnamefont{J.H.}\ \bibnamefont{Weijs}}, \bibinfo
  {author} {\bibfnamefont{A.}~\bibnamefont{Marchand}}, \bibinfo {author}
  {\bibfnamefont{B.}~\bibnamefont{Andreotti}}, \bibinfo {author}
  {\bibfnamefont{D.}~\bibnamefont{Lohse}},\ and\ \bibinfo {author}
  {\bibfnamefont{J.H.}\ \bibnamefont{Snoeijer}},\ }%
  \bibfield{title}{%
  \enquote{\bibinfo {title} {{Origin of line tension for a Lennard-Jones
  nanodroplet}},}\ }%
  \bibfield{journal}{%
  \Doi{10.1063/1.3546008}{\bibinfo {journal} {Phys. Fluids.}}\ }%
  \textbf{\bibinfo {volume} {23}},\ \bibinfo {pages} {022001--1--11} (\bibinfo
  {year} {2011})%
  \bibAnnoteFile{NoStop}{Weijs:2011}%
\bibitem{bakker:1928}%
  \BibitemOpen
  \bibfield{author}{%
  \bibinfo {author} {\bibfnamefont{G.}~\bibnamefont{Bakker}},\ }%
  \emph{\bibinfo {title} {Kapillarit{\"a}t und Oberfl{\"a}chenspannung}},\
  Handbuch der Experimentalphysik\ (\bibinfo {publisher} {Akademische
  Verlagsgesellschaft},\ \bibinfo {year} {1928})\
  \url{https://books.google.com/?id=T6haAQAACAAJ}%
  \bibAnnoteFile{NoStop}{bakker:1928}%
\bibitem{Kirkwood:1949}%
  \BibitemOpen
  \bibfield{author}{%
  \bibinfo {author} {\bibfnamefont{John~G.}\ \bibnamefont{Kirkwood}}\ and\
  \bibinfo {author} {\bibfnamefont{Frank~P.}\ \bibnamefont{Buff}},\ }%
  \bibfield{title}{%
  \enquote{\bibinfo {title} {{The Statistical Mechanical Theory of Surface
  Tension}},}\ }%
  \bibfield{journal}{%
  \Doi{10.1063/1.1747248}{\bibinfo {journal} {J. Chem. Phys.}}\ }%
  \textbf{\bibinfo {volume} {17}},\ \bibinfo {pages} {338--343} (\bibinfo
  {year} {1949})%
  \bibAnnoteFile{NoStop}{Kirkwood:1949}%
\bibitem{Thomson:1958}%
  \BibitemOpen
  \bibfield{author}{%
  \bibinfo {author} {\bibfnamefont{W.~(Lord~Kelvin)}\ \bibnamefont{Thomson}},\
  }%
  \bibfield{title}{%
  \enquote{\bibinfo {title} {{On the Thermal Effects of drawing out a Film of
  Liquid}},}\ }%
  \bibfield{journal}{%
  \Doi{10.1098/rspl.1857.0061}{\bibinfo {journal} {Roy. Soc. Proc.}}\ }%
  \textbf{\bibinfo {volume} {9}},\ \bibinfo {pages} {255--6} (\bibinfo {year}
  {1858})%
  \bibAnnoteFile{NoStop}{Thomson:1958}%
\bibitem{Laue:1911lrk}%
  \BibitemOpen
  \bibfield{author}{%
  \bibinfo {author} {\bibfnamefont{M.}~\bibnamefont{Laue}},\ }%
  \bibfield{title}{%
  \enquote{\bibinfo {title} {{Zur Dynamik der Relativitätstheorie}},}\ }%
  \bibfield{journal}{%
  \Doi{10.1002/andp.19113400808}{\bibinfo {journal} {Annalen Phys.}}\ }%
  \textbf{\bibinfo {volume} {340}},\ \bibinfo {pages} {524--542} (\bibinfo
  {year} {1911})%
  \bibAnnoteFile{NoStop}{Laue:1911lrk}%
\bibitem{Herrera:1992lwz}%
  \BibitemOpen
  \bibfield{author}{%
  \bibinfo {author} {\bibfnamefont{L.}~\bibnamefont{Herrera}},\ }%
  \bibfield{title}{%
  \enquote{\bibinfo {title} {{Cracking of self-gravitating compact objects}},}\
  }%
  \bibfield{journal}{%
  \Doi{10.1016/0375-9601(92)90036-L}{\bibinfo {journal} {Phys. Lett.}}\ }%
  \textbf{\bibinfo {volume} {A165}},\ \bibinfo {pages} {206--210} (\bibinfo
  {year} {1992})%
  \bibAnnoteFile{NoStop}{Herrera:1992lwz}%
\bibitem{Abreu:2007ew}%
  \BibitemOpen
  \bibfield{author}{%
  \bibinfo {author} {\bibfnamefont{H.}~\bibnamefont{Abreu}}, \bibinfo {author}
  {\bibfnamefont{H.}~\bibnamefont{Hernandez}},\ and\ \bibinfo {author}
  {\bibfnamefont{L.~A.}\ \bibnamefont{Nunez}},\ }%
  \bibfield{title}{%
  \enquote{\bibinfo {title} {{Sound Speeds, Cracking and Stability of
  Self-Gravitating Anisotropic Compact Objects}},}\ }%
  \bibfield{journal}{%
  \Doi{10.1088/0264-9381/24/18/005}{\bibinfo {journal} {Class. Quant. Grav.}}\
  }%
  \textbf{\bibinfo {volume} {24}},\ \bibinfo {pages} {4631--4646} (\bibinfo
  {year} {2007}),\ \Eprint{http://arxiv.org/abs/0706.3452}{arXiv:0706.3452
  [gr-qc]}%
  \bibAnnoteFile{NoStop}{Abreu:2007ew}%
\bibitem{Matveev:1973ra}%
  \BibitemOpen
  \bibfield{author}{%
  \bibinfo {author} {\bibfnamefont{V.~A.}\ \bibnamefont{Matveev}}, \bibinfo
  {author} {\bibfnamefont{R.~M.}\ \bibnamefont{Muradian}},\ and\ \bibinfo
  {author} {\bibfnamefont{A.~N.}\ \bibnamefont{Tavkhelidze}},\ }%
  \bibfield{title}{%
  \enquote{\bibinfo {title} {{Automodellism in the large - angle elastic
  scattering and structure of hadrons}},}\ }%
  \bibfield{journal}{%
  \Doi{10.1007/BF02728133}{\bibinfo {journal} {Lett. Nuovo Cim.}}\ }%
  \textbf{\bibinfo {volume} {7}},\ \bibinfo {pages} {719--723} (\bibinfo {year}
  {1973})%
  \bibAnnoteFile{NoStop}{Matveev:1973ra}%
\bibitem{Brodsky:1973kr}%
  \BibitemOpen
  \bibfield{author}{%
  \bibinfo {author} {\bibfnamefont{Stanley~J.}\ \bibnamefont{Brodsky}}\ and\
  \bibinfo {author} {\bibfnamefont{Glennys~R.}\ \bibnamefont{Farrar}},\ }%
  \bibfield{title}{%
  \enquote{\bibinfo {title} {{Scaling Laws at Large Transverse Momentum}},}\ }%
  \bibfield{journal}{%
  \Doi{10.1103/PhysRevLett.31.1153}{\bibinfo {journal} {Phys. Rev. Lett.}}\ }%
  \textbf{\bibinfo {volume} {31}},\ \bibinfo {pages} {1153--1156} (\bibinfo
  {year} {1973})%
  \bibAnnoteFile{NoStop}{Brodsky:1973kr}%
\bibitem{Brodsky:1974vy}%
  \BibitemOpen
  \bibfield{author}{%
  \bibinfo {author} {\bibfnamefont{Stanley~J.}\ \bibnamefont{Brodsky}}\ and\
  \bibinfo {author} {\bibfnamefont{Glennys~R.}\ \bibnamefont{Farrar}},\ }%
  \bibfield{title}{%
  \enquote{\bibinfo {title} {{Scaling Laws for Large Momentum Transfer
  Processes}},}\ }%
  \bibfield{journal}{%
  \Doi{10.1103/PhysRevD.11.1309}{\bibinfo {journal} {Phys. Rev.}}\ }%
  \textbf{\bibinfo {volume} {D11}},\ \bibinfo {pages} {1309} (\bibinfo {year}
  {1975})%
  \bibAnnoteFile{NoStop}{Brodsky:1974vy}%
\bibitem{Lepage:1980fj}%
  \BibitemOpen
  \bibfield{author}{%
  \bibinfo {author} {\bibfnamefont{G.~Peter}\ \bibnamefont{Lepage}}\ and\
  \bibinfo {author} {\bibfnamefont{Stanley~J.}\ \bibnamefont{Brodsky}},\ }%
  \bibfield{title}{%
  \enquote{\bibinfo {title} {{Exclusive Processes in Perturbative Quantum
  Chromodynamics}},}\ }%
  \bibfield{journal}{%
  \Doi{10.1103/PhysRevD.22.2157}{\bibinfo {journal} {Phys. Rev.}}\ }%
  \textbf{\bibinfo {volume} {D22}},\ \bibinfo {pages} {2157} (\bibinfo {year}
  {1980})%
  \bibAnnoteFile{NoStop}{Lepage:1980fj}%
\bibitem{Wald:1984rg}%
  \BibitemOpen
  \bibfield{author}{%
  \bibinfo {author} {\bibfnamefont{Robert~M.}\ \bibnamefont{Wald}},\ }%
  \Doi{10.7208/chicago/9780226870373.001.0001}{\emph{\bibinfo {title} {{General
  Relativity}}}}\ (\bibinfo {publisher} {Chicago Univ. Pr.},\ \bibinfo
  {address} {Chicago, USA},\ \bibinfo {year} {1984})%
  \bibAnnoteFile{NoStop}{Wald:1984rg}%
\bibitem{Poisson:2009pwt}%
  \BibitemOpen
  \bibfield{author}{%
  \bibinfo {author} {\bibfnamefont{Eric}\ \bibnamefont{Poisson}},\ }%
  \Doi{10.1017/CBO9780511606601}{\emph{\bibinfo {title} {{A Relativist's
  Toolkit: The Mathematics of Black-Hole Mechanics}}}}\ (\bibinfo {publisher}
  {Cambridge University Press},\ \bibinfo {year} {2009})%
  \bibAnnoteFile{NoStop}{Poisson:2009pwt}%
\bibitem{Senovilla:2014gza}%
  \BibitemOpen
  \bibfield{author}{%
  \bibinfo {author} {\bibfnamefont{José M.~M.}\ \bibnamefont{Senovilla}}\ and\
  \bibinfo {author} {\bibfnamefont{David}\ \bibnamefont{Garfinkle}},\ }%
  \bibfield{title}{%
  \enquote{\bibinfo {title} {{The 1965 Penrose singularity theorem}},}\ }%
  \bibfield{journal}{%
  \Doi{10.1088/0264-9381/32/12/124008}{\bibinfo {journal} {Class. Quant.
  Grav.}}\ }%
  \textbf{\bibinfo {volume} {32}},\ \bibinfo {pages} {124008} (\bibinfo {year}
  {2015}),\ \Eprint{http://arxiv.org/abs/1410.5226}{arXiv:1410.5226 [gr-qc]}%
  \bibAnnoteFile{NoStop}{Senovilla:2014gza}%
\bibitem{Curiel:2014zba}%
  \BibitemOpen
  \bibfield{author}{%
  \bibinfo {author} {\bibfnamefont{Erik}\ \bibnamefont{Curiel}},\ }%
  \bibfield{title}{%
  \enquote{\bibinfo {title} {{A Primer on Energy Conditions}},}\ }%
  \bibfield{journal}{%
  \Doi{10.1007/978-1-4939-3210-8_3}{\bibinfo {journal} {Einstein Stud.}}\ }%
  \textbf{\bibinfo {volume} {13}},\ \bibinfo {pages} {43--104} (\bibinfo {year}
  {2017}),\ \Eprint{http://arxiv.org/abs/1405.0403}{arXiv:1405.0403
  [physics.hist-ph]}%
  \bibAnnoteFile{NoStop}{Curiel:2014zba}%
\bibitem{Martin-Moruno:2017exc}%
  \BibitemOpen
  \bibfield{author}{%
  \bibinfo {author} {\bibfnamefont{Prado}\ \bibnamefont{Martin-Moruno}}\ and\
  \bibinfo {author} {\bibfnamefont{Matt}\ \bibnamefont{Visser}},\ }%
  \bibfield{title}{%
  \enquote{\bibinfo {title} {{Classical and semi-classical energy
  conditions}},}\ }%
  \bibfield{journal}{%
  \Doi{10.1007/978-3-319-55182-1_9}{\bibinfo {journal} {Fundam. Theor. Phys.}}\
  }%
  \textbf{\bibinfo {volume} {189}},\ \bibinfo {pages} {193--213} (\bibinfo
  {year} {2017}),\ \Eprint{http://arxiv.org/abs/1702.05915}{arXiv:1702.05915
  [gr-qc]}%
  \bibAnnoteFile{NoStop}{Martin-Moruno:2017exc}%
\bibitem{Maeda:2018hqu}%
  \BibitemOpen
  \bibfield{author}{%
  \bibinfo {author} {\bibfnamefont{Hideki}\ \bibnamefont{Maeda}}\ and\ \bibinfo
  {author} {\bibfnamefont{Cristian}\ \bibnamefont{Martinez}},\ }%
  \bibfield{title}{%
  \enquote{\bibinfo {title} {{Energy conditions in arbitrary dimensions}},}\ }%
   (\bibinfo {year} {2018}),\
  \Eprint{http://arxiv.org/abs/1810.02487}{arXiv:1810.02487 [gr-qc]}%
  \bibAnnoteFile{NoStop}{Maeda:2018hqu}%
\bibitem{Ford:1978qya}%
  \BibitemOpen
  \bibfield{author}{%
  \bibinfo {author} {\bibfnamefont{L.~H.}\ \bibnamefont{Ford}},\ }%
  \bibfield{title}{%
  \enquote{\bibinfo {title} {{Quantum Coherence Effects and the Second Law of
  Thermodynamics}},}\ }%
  \bibfield{journal}{%
  \Doi{10.1098/rspa.1978.0197}{\bibinfo {journal} {Proc. Roy. Soc. Lond.}}\ }%
  \textbf{\bibinfo {volume} {A364}},\ \bibinfo {pages} {227--236} (\bibinfo
  {year} {1978})%
  \bibAnnoteFile{NoStop}{Ford:1978qya}%
\bibitem{Fewster:2012yh}%
  \BibitemOpen
  \bibfield{author}{%
  \bibinfo {author} {\bibfnamefont{Christopher~J.}\ \bibnamefont{Fewster}},\ }%
  \bibfield{title}{%
  \enquote{\bibinfo {title} {{Lectures on quantum energy inequalities}},}\ }%
   (\bibinfo {year} {2012}),\
  \Eprint{http://arxiv.org/abs/1208.5399}{arXiv:1208.5399 [gr-qc]}%
  \bibAnnoteFile{NoStop}{Fewster:2012yh}%
\bibitem{Fewster:2017wmn}%
  \BibitemOpen
  \bibfield{author}{%
  \bibinfo {author} {\bibfnamefont{Christopher~J.}\ \bibnamefont{Fewster}},\ }%
  \bibfield{title}{%
  \enquote{\bibinfo {title} {{Quantum Energy Inequalities}},}\ }%
  \bibfield{journal}{%
  \Doi{10.1007/978-3-319-55182-1_10}{\bibinfo {journal} {Fundam. Theor.
  Phys.}}\ }%
  \textbf{\bibinfo {volume} {189}},\ \bibinfo {pages} {215--254} (\bibinfo
  {year} {2017})%
  \bibAnnoteFile{NoStop}{Fewster:2017wmn}%
\bibitem{Fewster:2018pey}%
  \BibitemOpen
  \bibfield{author}{%
  \bibinfo {author} {\bibfnamefont{Christopher~J.}\ \bibnamefont{Fewster}}\
  and\ \bibinfo {author} {\bibfnamefont{Eleni-Alexandra}\
  \bibnamefont{Kontou}},\ }%
  \bibfield{title}{%
  \enquote{\bibinfo {title} {{Quantum strong energy inequalities}},}\ }%
   (\bibinfo {year} {2018}),\
  \Eprint{http://arxiv.org/abs/1809.05047}{arXiv:1809.05047 [gr-qc]}%
  \bibAnnoteFile{NoStop}{Fewster:2018pey}%
\end{thebibliography}%

\end{document}